\documentclass[fleqn,usenatbib,onecolumn]{mnras}

\usepackage{newtxtext,newtxmath}

\usepackage[T1]{fontenc}
\usepackage{ae,aecompl}

\usepackage[utf8]{inputenc}

\usepackage{graphicx}	
\usepackage{amsmath}	
\usepackage{amssymb}	

\usepackage{multirow}
\usepackage{multicol}
\usepackage{booktabs}
\usepackage{siunitx}
\usepackage{xcolor}

\usepackage{braket}

\usepackage[caption=false]{subfig}

\newcommand{\Mpc}{\mathrm{Mpc}}

\newcommand{\bk}{\mathbf{k}}
\newcommand{\bn}{\mathbf{n}}
\newcommand{\hbn}{\hat{\mathbf{n}}}
\newcommand{\FoG}{\textrm{FoG}}

\renewcommand{\vec}[1]{\boldsymbol{#1}}
\newcommand{\mat}[1]{\textbf{\textsf{#1}}}

\newcommand{\D}[1]{\ensuremath{\text{d}#1}}

\title[Cosmology with the power spectrum and LCF]{Constraining the growth rate of structure with phase correlations}

\author[Byun, Franco, Howlett, Bonvin \& Obreschkow]{
Joyce Byun,$^{1}$\thanks{E-mail: joyce.byun@unige.ch}
Felipe Oliveira Franco,$^{1}$
Cullan Howlett,$^{2}$
Camille Bonvin$^{1}$ and
\newauthor
Danail Obreschkow$^{3}$
\\
$^{1}$D\'epartement de Physique Th\'eorique and Center for Astroparticle Physics (CAP), University of Geneva, 24 quai Ernest Ansermet, \protect\\CH-1211 Geneva, Switzerland\\
$^{2}$School of Mathematics and Physics, The University of Queensland, Brisbane, QLD 4072, Australia\\
$^{3}$International Centre for Radio Astronomy Research (ICRAR), University of Western Australia, 35 Stirling Highway, Crawley WA \protect\\6009, Australia
}

\date{Accepted XXX. Received YYY; in original form ZZZ}

\pubyear{2020}

\begin{document}
\label{firstpage}
\pagerange{\pageref{firstpage}--\pageref{lastpage}}
\maketitle
\begin{abstract}
We show that correlations between the phases of the galaxy density field in redshift space provide additional information about the growth rate of large-scale structure that is complementary to the power spectrum multipoles. In particular, we consider the multipoles of the line correlation function (LCF), which correlates phases between three collinear points, and use the Fisher forecasting method to show that the LCF multipoles can break the degeneracy between the measurement of the growth rate of structure $f$ and the amplitude of perturbations $\sigma_8$ that is present in the power spectrum multipoles at large scales. This leads to an improvement in the measurement of $f$ and $\sigma_8$ by up to 220 per cent for $k_{\rm max} = 0.15 \, h\mathrm{Mpc}^{-1}$ and up to 50 per cent for $k_{\rm max} = 0.30 \, h\mathrm{Mpc}^{-1}$ at redshift $z=0.25$, with respect to power spectrum measurements alone for the upcoming generation of galaxy surveys like DESI and Euclid. The average improvements in the constraints on $f$ and $\sigma_8$ for $k_{\rm max} = 0.15 \, h\mathrm{Mpc}^{-1}$ are $\sim 90$ per cent for the DESI BGS sample with mean redshift $\overline{z}=0.25$, $\sim 40$ per cent for the DESI ELG sample with $\overline{z}=1.25$, and $\sim 40$ per cent for the Euclid H$\alpha$ galaxies with $\overline{z}=1.3$. For $k_{\rm max} = 0.30 \, h\mathrm{Mpc}^{-1}$, the average improvements are $\sim 40$ per cent for the DESI BGS sample and $\sim 20$ per cent for both the DESI ELG and Euclid H$\alpha$ galaxies.
\end{abstract}




\section{Introduction}

Galaxy clustering is a powerful cosmological probe that provides a way of measuring the growth of structure and constraining cosmological parameters. The standard way of extracting information from galaxy clustering is to measure the two-point correlation function, or its Fourier-space counterpart, the power spectrum. Since observations are performed in redshift-space, the power spectrum is sensitive not only to the matter density distribution, but also to the peculiar velocities of galaxies, through the so-called redshift-space distortions (RSD). These distortions generate anisotropies in the power spectrum, in the form of a quadrupole and hexadecapole. Measuring these multipoles along with the monopole provides a measurement of three key variables in our Universe: the growth rate of structure $f$, the amplitude of fluctuations $\sigma_8$, and the linear galaxy bias $b_1$ \citep{Kaiser:1987qv}. However, these quantities are not measured independently: at large scales the power spectrum is sensitive to $f\sigma_8$ and $b_1\sigma_8$. Although in theory these degeneracies can be overcome by including information from non-linear scales, in practice additional data like the Cosmic Microwave Background (CMB) or gravitational lensing are used.

Instead, in this paper we show that by using Fourier phase information it is possible to break the degeneracies and constrain $f, b_1$ and $\sigma_8$ separately, without relying on external data sets. This method has the strong advantage of being model-independent, as it does not depend on a model for how the linear growth factor and growth rate evolve with redshift. This is not the case when probes at different redshifts are combined, for example the power spectrum and the CMB, where translating the measured primordial amplitude of perturbations $A_s$ into a measurement of $\sigma_8$ requires assuming a cosmological model, like $\Lambda$CDM and general relativity (GR), to describe the growth of density perturbations \citep{Alam:2016hwk}. Similarly, lensing data provide a measurement of $S_8\equiv \sigma_8\sqrt{\Omega_m/0.3}$, which requires a knowledge of the matter density parameter $\Omega_m$ to extract $\sigma_8$ \citep{Abbott:2017wau,Hildebrandt:2016iqg,Hildebrandt:2018yau,Hikage:2018qbn,Hamana:2019etx}. Aiming for a more model-independent determination of $f, b_1$ and $\sigma_8$ is particularly important in the current situation where tensions exist between different data sets \citep{Verde:2019ivm}.

Various estimators have been proposed in the past to measure information contained in Fourier phases $\epsilon(\bk) \equiv \delta(\bk)/|\delta(\bk)|$, where $\delta(\bk)$ is the density field. Here we focus on the line correlation function (LCF) that measures a type of three-point correlation function restricted to phases along a line \citep{Obreschkow:2012yb}. Geometrically, the LCF can be interpreted as a quantification of the amount of filamentary structure in the cosmic web on different scales (see \citealt{Obreschkow:2012yb} for details). The LCF has been modelled in real space using both perturbation theory and effective non-linear kernels and was shown to agree well with measurements from N-body simulations over a wide range of scales \citep{Wolstenhulme:2014cla}. \cite{Eggemeier:2016asq} demonstrated that the LCF used in combination with the power spectrum improves the constraints on cosmological parameters by up to a factor of 2 compared with the power spectrum alone. \cite{Ali:2018sdk} then extended these forecasts to models beyond $\Lambda$CDM, including warm dark matter models and modified gravity models, and showed that the LCF also improves parameter constraints in these cases. 

These previous works have shown the potential of the LCF to improve parameter constraints in real space. The goal of this work is to extend these analyses to redshift space, where observations are made. In \cite{Eggemeier:2015ifa}, the LCF was modified to be more sensitive to RSD, and the resulting anisotropic LCF estimator was applied to toy mocks. More recently, \cite{Franco:2018yag} used tree-level perturbation theory to model the LCF multipoles in redshift space. They showed that even though in principle an infinite number of multipoles are generated by RSD, nearly all of the information about the velocity phase correlations is contained in the monopole, quadrupole and hexadecapole. In this work we explore how these multipoles may be used in combination with the power spectrum multipoles to break the degeneracy between $f, b_1$ and $\sigma_8$, with data from the upcoming generation of large-scale structure surveys like the Dark Energy Spectroscopic Instrument (DESI; \citealt{Aghamousa:2016zmz}) and Euclid \citep{Laureijs2011}.

We note that this study of the information content of the LCF multipoles is within the larger context of how higher-order statistics can complement and strengthen cosmological constraints from power spectrum analyses. In real-space, \cite{Byun:2017fkz} showed that even though the LCF does not constrain $\Lambda$CDM parameters as well as the full bispectrum or the modal bispectrum, it still provides an efficient way to compress information and gives improvements in cosmological parameter constraints ranging from a few per cent to 70 per cent, depending on the parameter. In redshift space, analyses of the Baryon Oscillation Spectroscopic Survey (BOSS) have combined the bispectrum monopole with power spectrum multipoles \citep{Gil-Marin:2016wya} and detected the bispectrum quadrupole \citep{Sugiyama:2018yzo}, while for future surveys, \cite{Gagrani:2016rfy}, \cite{Yankelevich:2018uaz} and \cite{Gualdi:2020ymf} have explored the additional benefit of using the anisotropic bispectrum beyond the monopole for cosmological constraints. We defer a comparison of the information in the redshift-space LCF and redshift-space bispectrum to a future work.

The main results of this work are Fisher forecasts for how well the LCF multipoles may further improve constraints on $f$ and $\sigma_8$ from the power spectrum multipoles that will be measured by upcoming surveys such as DESI and Euclid. To provide guidance for our forecasts, we compare our theoretical model for the power spectrum and LCF multipoles with simulations and find that the model reproduces well the measured power spectrum multipoles up to $k_{\rm max} = 0.3\,h \Mpc^{-1}$ and the LCF multipoles down to $r_{\rm min} = 20\,h^{-1}\Mpc$. We also use perturbation theory to calculate the full covariance of the power spectrum and LCF multipoles, and we compare it with the covariance measured from 500 \textsc{l-picola} simulations at redshift $z=0$. We find that using the theoretical covariance matrix underestimates the forecasted errors on cosmological parameters by 20 per cent at most, compared with constraints obtained using the simulated covariance matrix, and we find a similar level of agreement when comparing constraints with measured or predicted values of the Fingers-of-God velocity dispersions. We also show that neglecting the cross-covariance between the power spectrum and LCF changes the resulting constraints by less than 10 per cent. To estimate the future constraints that may be achievable with a joint power spectrum and LCF multipoles analysis, at present we must rely more heavily on our theoretical models, but the encouraging results of these checks support the extension of our forecasting pipeline to upcoming surveys.

For upcoming surveys like DESI and Euclid, we calculate Fisher forecasted constraints for $f$ and $\sigma_8$ in each redshift bin and compare constraints from only the power spectrum multipoles with constraints from the combined power spectrum and LCF multipoles. Our forecast estimates that the LCF multipoles could strengthen the constraints on $f$ and $\sigma_8$ from the power spectrum multipoles by up to 220 per cent if scales up to $k_{\rm max}=0.15\,h\Mpc^{-1}$ are included. If $k_{\rm max}=0.3\,h\Mpc^{-1}$, the improvement from the LCF multipoles is smaller but still significant, up to 50 per cent. If CMB measurements from the Planck survey \citep{Aghanim:2018eyx} are included through a prior on $\sigma_8$, the improvement brought by the LCF multipoles becomes marginal, less than $\sim 10$ per cent. However, we note again that the advantage of using the LCF multipoles instead of external CMB data is that translating the CMB constraint on $A_s$ into a constraint on $\sigma_8$ depends on an assumed cosmological model, such as $\Lambda$CDM and general relativity.

The outline for this paper is as follows. In Section~\ref{sec:modeling}, we summarise the theoretical models we use for the halo power spectrum and LCF multipoles. Section~\ref{sec:sims} describes the simulation data and the power spectrum and LCF multipole estimators. Section~\ref{sec:theoretical_covariance} summarises the theoretical models we use for the covariance matrices, while the detailed derivations are in Appendices \ref{app:Qn_cov} and \ref{app:PlQn_crosscov}. In Section~\ref{sec:compare}, we compare our model predictions with measurements from simulations, and in Section~\ref{sec:fisher} we validate our forecasting pipeline with Fisher forecasts based on the simulation boxes. We present our forecasts for surveys like DESI and Euclid in Section~\ref{sec:surveys}, and we end with our conclusions in Section~\ref{sec:conclusions}.


\section{Theoretical modeling}
\label{sec:modeling}

In this section, we specify the models we use for the power spectrum and line correlation function (LCF) multipoles in redshift space. Galaxy clustering observables measure the correlations between galaxy overdensities,
\begin{equation}
	\Delta(\mathbf{x})\equiv\frac{N(\mathbf{x})-\bar N}{\bar N},
\end{equation}
where $N(\mathbf{x})$ is the number density of galaxies at position $\mathbf{x}$ and $\bar N$ is the average number density. The LCF is defined as the correlation between the phases of $\Delta$ at three collinear points,
\begin{align}
	\ell\left(\mathbf{r}\right) & \equiv V^3 \left(\frac{r^{3}}{V}\right)^{3/2}\Braket{\epsilon\left(\mathbf{x}+\mathbf{r}\right)\epsilon\left(\mathbf{x}\right)\epsilon\left(\mathbf{x}-\mathbf{r}\right)} \nonumber\\
	& = \frac{V^{3}}{\left(2\pi\right)^{9}}\left(\frac{r^{3}}{V}\right)^{3/2}\iiintop_{k_{1},k_{2},k_{3}\leq\frac{2\pi}{r}}\D{^3 k_1}\D{^3 k_2}\D{^3 k_3}e^{i\mathbf{x}\cdot\left(\mathbf{k}_{1}+\mathbf{k}_{2}+\mathbf{k}_{3}\right)}e^{i\mathbf{r}\cdot\left(\mathbf{k}_{1}-\mathbf{k}_{2}\right)}\Braket{\epsilon\left(\mathbf{k}_{1}\right)\,\epsilon\left(\mathbf{k}_{2}\right)\,\epsilon\left(\mathbf{k}_{3}\right)}, 
	\label{eq:LCF definition fourier}
\end{align}
where $\epsilon(\mathbf{x})$ is the inverse Fourier transform of $\epsilon(\mathbf{k}) \equiv \frac{\Delta(\mathbf{k})}{|\Delta(\mathbf{k})|}$ and $\Delta(\mathbf{k})$ is the Fourier transform of $\Delta(\mathbf{x})$.

In principle the 3-point correlation function of phases depends on the bispectrum and all higher-order cumulants, but in the mildly non-Gaussian regime, it can be expressed in terms of the power spectrum and bispectrum of $\Delta$ using the Edgeworth expansion as \citep{Matsubara:2003te,Wolstenhulme:2014cla}
\begin{equation}
	\Braket{\epsilon\left(\mathbf{k}_{1}\right)\,\epsilon\left(\mathbf{k}_{2}\right)\,\epsilon\left(\mathbf{k}_{3}\right)} = \frac{(2\pi)^3}{V} \left( \frac{\sqrt{\pi}}{2}\right)^3 \frac{B(\mathbf{k}_{1},\mathbf{k}_{2},\mathbf{k}_{3})}{\sqrt{V P(\mathbf{k}_{1})P(\mathbf{k}_{2})P(\mathbf{k}_{3})}} \delta_D(\mathbf{k}_{1}+\mathbf{k}_{2}+\mathbf{k}_{3}). 
	\label{eq:3eps}
\end{equation}
In~\cite{Franco:2018yag}, the ratio in eq.~\eqref{eq:3eps} was calculated using tree-level perturbation theory, where two types of contributions arise: an intrinsic contribution due to the fact that the density, velocity and bias are non-linear, and a mapping contribution generated by the non-linear mapping between real space and redshift space. In this paper, we modify the intrinsic contribution in the bispectrum and use a non-linear model for the power spectrum to better describe the LCF down to smaller scales where tree-level perturbation theory breaks down. In the following subsections, we summarise how we model these quantities using established models in the literature.

The LCF multipoles are then given by
\begin{equation}
	Q_n(r) = \frac{2n+1}{2} \int_{-1}^1 d\nu\, \ell (r,\nu) L_n(\nu)\, , \label{eq:Qn}
\end{equation}
where $\nu$ depends on the orientation of the line with respect to the direction of observation, $\nu \equiv \hat{\mathbf{r}} \cdot \hat{\bn}$, and $L_n$ is the Legendre polynomial of degree $n$.


\subsection{Power spectrum modeling}

For the redshift-space power spectrum, we use a variant of the model from \cite{Vlah:2013lia}, as implemented in \cite{Howlett:2019bky}. To summarise, the model accounts for the large-scale coherent motions of galaxies towards overdensities (the Kaiser effect), random non-linear motions of these galaxies within virialised halos, and the non-linear and non-local bias of galaxies with respect to the underlying matter field. It does this using a distribution function approach, where different contributions to the full anisotropic power spectrum are broken down into terms containing unique powers of the mass-weighted velocity (momentum) field. Each of these contributions is then evaluated using standard (one-loop) perturbation theory for biased tracers.

For consistency with the Fingers-of-God modeling used in the bispectrum and LCF, we remove the terms dependent on the velocity dispersion from within the various components of the model and replace them with a single, global damping term, $D^{P}_{\FoG}$. We adopt three different forms of the damping term, which are described in Section~\ref{sec:fog}.

Following this, we write the anisotropic power spectrum as
\begin{equation}
	P(k,\mu) = D^{P}_{\FoG}(k,\mu,\sigma_P)\biggl[P_{00}(k) + \mu^{2}(2P_{01}(k) + P_{02}(k,\mu)) + P_{11}(k,\mu) + \mu^{4}(P_{12}(k,\mu) + 1/4P_{22}(k,\mu))\biggl], 
	\label{eq:Pvlah}
\end{equation}
where $\mu=\hat{\bk}\cdot\hat{\bn}$ and 
\begin{align}
    P_{00} & = b_{1}^{2}P_{\delta \delta} + 2b_{1}(b_{2}K_{00} + b_{s}K^{s}_{00} + b_{3nl}\sigma^{2}_{3}P_{m}) + 1/2b_{2}^{2}K_{01} + 1/2b^{2}_{s}K^{s}_{01} + b_{2}b_{s}K^{s}_{02}\, , \\
    P_{01} & = fb_{1}(P_{\delta \theta} + 2b_{1}I_{10} + 6k^{2}P_{m}b_{1}J_{10} - b_{2}K_{11} - b_{s}K^{s}_{11}) - f(b_{2}K_{10} + b_{s}K^{s}_{10} + b_{3nl}\sigma^{2}_{3}P_{m})\, , \\
    P_{02} & = f^{2}b_{1}(I_{02} + \mu^{2}I_{20} + 2k^{2}P_{m}(J_{02} + \mu^{2}J_{20})) - f^{2}(b_{2}(K_{20}+\mu^{2}K_{30}) + b_{s}(K^{s}_{20}+\mu^{2}K^{s}_{30}))\, , \\
    P_{11} & = f^{2}(\mu^{2}(P_{\theta \theta} + 4b_{1}I_{22} + b_{1}^{2}I_{13} + 12k^{2}P_{m}b_{1}J_{10}) + b_{1}^{2}I_{31})\, , \\ 
    P_{12} & = f^{3}(I_{12} + \mu^{2}I_{21} - b_{1}(I_{03} + \mu^{2}I_{30}) + 2k^{2}P_{m}(J_{02} + \mu^{2}J_{20}))\, , \\
    P_{22} & = 1/4f^{4}(I_{23} + 2\mu^{2}I_{32} + \mu^{4}I_{33})\, .
\end{align}
Here $f$ is the growth rate of structure, and $b_{1}$, $b_{2}$, $b_{s}$ and $b_{3nl}$ are the galaxy bias parameters under a Eulerian bias expansion relating the galaxy overdensity $\delta_{g}$ and matter overdensity $\delta$,
\begin{equation}
	\delta_{g}(\vec{x}) = b_{1}\delta(\vec{x}) + \frac{b_{2}}{2}(\delta^{2}(\vec{x})-\langle \delta \rangle) + \frac{b_{s}}{2}(s^{2}(\vec{x})-\langle s \rangle) + \frac{b_{3nl}}{6}\delta^{3}(\vec{x}).
\end{equation}
In deriving the power spectrum model, the above bias expansion is renormalised following \cite{Saito:2014qha} and from which we also adopt $b_{s}=-4/7(b_{1}-1)$ and $b_{3nl}=32/315(b_{1}-1)$. The terms $I_{ij}$, $J_{ij}$, $K_{ij}$ and $\sigma_{3}$ are integrals over the matter power spectrum $P_{m}$ and are presented in the appendices of \cite{Vlah:2012ni} and \cite{Howlett:2019bky}. $s(\boldsymbol{x})$ is the Fourier transform of $s(\boldsymbol{k})$, which is also presented in \cite{Vlah:2013lia}. Finally, $P_{\delta \delta}$, $P_{\delta \theta}$ and $P_{\theta \theta}$ are the non-linear density-density, density-velocity divergence and velocity divergence-velocity divergence power spectra. In \cite{Vlah:2013lia} these were computed using one-loop standard perturbation theory, but in this work we use the non-linear real-space matter power spectrum measured from the simulations in Section~\ref{sec:sims} for $P_{\delta \delta}$ and to compute the non-linear RSD terms. As such we also set $P_{m}=P_{\delta\delta}$ in the above equations. For $P_{\delta \theta}$ and $P_{\theta \theta}$, we use the fitting formulae from \cite{Jennings:2012ej}, again with the measured simulation matter power spectrum as input. Although the use of the simulated non-linear power spectrum in the above model is not strictly consistent from a theoretical point of view, this gave the best fit to the simulation redshift-space power spectrum and is suitable for the forecasts in this work.


\subsection{Bispectrum modeling}

As our model for the redshift-space halo bispectrum, we have implemented the fitting formula in \cite{Gil-Marin:2014pva} that extends the tree-level expression for the bispectrum to smaller scales by modifying the perturbation theory kernels that describe the non-linear density and peculiar velocity.

The standard perturbation theory expression for the tree-level halo bispectrum in redshift space is 
\begin{align}
B_{\rm tree}(\bk_1,\bk_2) &= 2 Z_1(\bk_1)  Z_1(\bk_2) Z_2(\bk_1,\bk_2)P_m(k_1)P_m(k_2)  + 2\;\mathrm{cyclic\;permutations}\,,\label{eq:B}\\
	Z_1(\bk_i) &\equiv b_1 + f \mu_i^2, \label{eq:Z1}\\
	Z_2(\bk_1,\bk_2) &\equiv b_1 F_2(\bk_1,\bk_2) + f \mu^2 G_2(\bk_1,\bk_2) + \frac{b_2}{2} + \frac{b_{s^2}}{2} S_2(\bk_1,\bk_2)+ \frac{b_1f \mu k}{2} \left( \frac{\mu_1}{k_1} + \frac{\mu_2}{k_2} \right) 
	+ \frac{f^2 \mu k}{2} \mu_1 \mu_2 \left( \frac{\mu_2}{k_1} + \frac{\mu_1}{k_2} \right),\label{eq:Z2}
\end{align}
where $\mu_i\equiv \hat{\bk}_i \cdot \hat{\bn}$, $\mu \equiv (\mu_1k_1+\mu_2k_2)/k$, and $k^2 \equiv (\bk_1 + \bk_2)^2$. $F_2$ and $G_2$ are the non-linear density and velocity kernels at second-order in perturbation theory \citep{Bernardeau:2001qr}. The first four terms in eq.~\eqref{eq:Z2} give rise to the intrinsic kernel of the 3-point phase correlations, called $W_2^{\rm int}$ in \cite{Franco:2018yag}, while the last two terms give rise to the mapping kernel, $W_2^{\rm map}$. Here we modify the intrinsic part following \cite{Gil-Marin:2014pva}, by replacing the $F_2$ and $G_2$ kernels in the above expression with the fitted effective kernels $F_2^{\mathrm{eff}}$ and $G_2^{\mathrm{eff}}$ given in eqs.~(2.19), (2.20), (5.1), (5.2), and Appendix A of that paper. Each effective kernel depends on nine free parameters that are fitted using a subset of triangle configurations of the matter bispectrum monopole measured from simulations at $0 \leq z \leq 1.5$. As shown in~\cite{Gil-Marin:2014pva}, these parameters depend only weakly on cosmology, and therefore we do not vary them in our Fisher forecasts.

Similarly to the power spectrum model in eq.~\eqref{eq:Pvlah}, we also multiply the bispectrum by a global damping factor $D^B_{\FoG}$ to include the Fingers-of-God effect. The three different forms of this damping factor that we implement in this work are described in Section~\ref{sec:fog}. Thus the bispectrum model we use is
\begin{align}
B(\bk_1,\bk_2) &= D_{\FoG}^B(\bk_1,\bk_2,\sigma_B)
	\biggl[ 2 Z_1(\bk_1)  Z_1(\bk_2) Z_2^{\rm eff}(\bk_1,\bk_2)P_m(k_1)P_m(k_2)  + 2\;\mathrm{cyclic\;permutations}\biggl], \label{eq:Bgm}\\
	Z_2^{\rm eff}(\bk_1,\bk_2) &\equiv b_1 F_2^{\rm eff}(\bk_1,\bk_2) + f \mu^2 G_2^{\rm eff}(\bk_1,\bk_2) + \frac{b_2}{2} + \frac{b_{s^2}}{2} S_2(\bk_1,\bk_2)+ \frac{b_1f \mu k}{2} \left( \frac{\mu_1}{k_1} + \frac{\mu_2}{k_2} \right) 
	+ \frac{f^2 \mu k}{2} \mu_1 \mu_2 \left( \frac{\mu_2}{k_1} + \frac{\mu_1}{k_2} \right).
\end{align}
Strictly speaking, for the tree-level halo bispectrum in perturbation theory, the matter power spectrum $P_m$ in eq.~\eqref{eq:B} is the \textit{linear} matter power spectrum, since non-linearities are encoded in $Z_2$. However, as in \cite{Gil-Marin:2014pva}, we will use the average real-space \textit{non-linear} matter power spectrum measured from simulations for $P_m$ in eq.~\eqref{eq:Bgm}.


\subsection{Fingers-of-God}
\label{sec:fog}

In both the power spectrum and bispectrum models summarised above, we multiplied a phenomenological damping factor to account for the Fingers-of-God (FoG) effect on small scales. Damping factors for the power spectrum and bispectrum have taken different analytic forms in the literature, each with free parameters representing the velocity dispersions of galaxies. Previous works that have jointly analysed simulation power spectra and bispectra include a damping factor for each \citep{Gil-Marin:2014pva,Hashimoto:2017klo}, and we follow the same procedure to create a damping factor for the LCF by appropriately combining damping factors that have been used previously for the power spectrum and bispectrum.

We explore three choices for the functional forms of the damping factors. The first is a Gaussian function \citep{Hashimoto:2017klo,Yankelevich:2018uaz},
\begin{align}
	D_{\FoG}^P(k,\mu,\sigma_P) &= \exp \left( - \frac{1}{2} k^2 \mu^2 \sigma_P^2 \right), 
	&D_{\FoG}^B(k_i,\mu_i,\sigma_B) &= \exp \left( - \frac{1}{2} [ k_1^2 \mu_1^2 + k_2^2 \mu_2^2 + k_3^2 \mu_3^2 ] \sigma_B^2 \right),
	\label{eq:gaussianfog}
\end{align}
while the second is from \cite{Gil-Marin:2014pva},
\begin{align}
	D_{\FoG}^P(k,\mu,\sigma_P) &= \left( 1 + \frac{1}{2} k^2 \mu^2 \sigma_P^2 \right)^{-2},
	&D_{\FoG}^B(k_i,\mu_i,\sigma_B) &= \left( 1 + \frac{1}{2} [ k_1^2 \mu_1^2 + k_2^2 \mu_2^2 + k_3^2 \mu_3^2 ]^2 \sigma_B^2 \right)^{-2},
	\label{eq:gmfog}
\end{align}
and the third form is a Lorentzian function \citep{Hashimoto:2017klo},
\begin{align}
	D_{\FoG}^P(k,\mu,\sigma_P) &= \left( 1 + \frac{1}{2} k^2 \mu^2 \sigma_P^2 \right)^{-1},
	&D_{\FoG}^B(k_i,\mu_i,\sigma_B) &= \left( 1 + \frac{1}{2} [ k_1^2 \mu_1^2 + k_2^2 \mu_2^2 + k_3^2 \mu_3^2 ] \sigma_B^2 \right)^{-1}.
	\label{eq:lorentzianfog}
\end{align}

The velocity dispersions $\sigma_P$ and $\sigma_B$ generally depend on the halo populations and will evolve with redshift, so we do not adopt the exact values of the velocity dispersions from the previously mentioned papers. Instead, we either fit their values to the \textsc{l-picola} simulations at $z=0$ that we use in this work, or for the survey forecasts at higher redshifts, we use the linear prediction for the velocity dispersion, given by 
\begin{equation}
	\sigma_P^2 = \frac{2f^2}{3} \int \frac{\D{^3 k}}{(2\pi)^3} \, \frac{P_m(k)}{k^2}, 
	\label{eq:sigmav2}
\end{equation}
where, consistently with the rest of our analysis, we use the non-linear real-space matter power spectrum for $P_m$.

In perturbation theory, assuming that the power spectrum and bispectrum are treated consistently, one might expect that the velocity dispersions $\sigma_P$ and $\sigma_B$ are the same \citep{Hashimoto:2017klo}. However, in this work we conservatively treat $\sigma_P$ and $\sigma_B$ as two separate nuisance parameters that reflect our lack of knowledge about clustering on very small scales. For this reason, we allow the damping factors for the power spectrum and bispectrum to have independently varying velocity dispersions and constrain both at the same time. For the survey forecasts, we choose the fiducial values such that $\sigma_B = \sigma_P$, but treat $\sigma_P$ and $\sigma_B$ as two independently varying nuisance parameters.


\subsection{Shot noise}
\label{sec:shot noise}

Lastly, our models of the halo power spectrum and LCF must include the effect of shot noise. To calculate the shot noise effect on the LCF, we include Poissonian shot noise in the power spectrum and bispectrum in eq.~\eqref{eq:3eps} by adding the following $P_{\rm noise}$ and $B_{\rm noise}$ terms to the halo power spectrum and bispectrum, respectively \citep{Eggemeier:2016asq}: 
\begin{align}
	P_{\rm noise} &= \frac{1}{\overline{n}}\,,
	&B_{\rm noise}(\textbf{k}_1,\textbf{k}_2,\textbf{k}_3) &= \frac{1}{\overline{n}}\left[ P(\textbf{k}_1) + P(\textbf{k}_2) + P(\textbf{k}_3)\right] + \frac{1}{\overline{n}^2}\,,
	\label{eq:PBnoise}
\end{align}
where $P(\textbf{k}_i)$ here is the redshift-space halo power spectrum without shot noise. Replacing $P$ and $B$ in eq.~\eqref{eq:3eps} by $P+P_{\rm noise}$ and $B+B_{\rm noise}$ gives
\begin{equation}
	\Braket{\epsilon\left(\mathbf{k}_{1}\right)\epsilon\left(\mathbf{k}_{2}\right)\epsilon\left(\mathbf{k}_{3}\right)} = \frac{(2\pi)^3}{V} \left( \frac{\sqrt{\pi}}{2}\right)^3 \sqrt{\frac{\nu_{\rm eff}(\bk_1)\nu_{\rm eff}(\bk_2)\nu_{\rm eff}(\bk_3)}{V P(\mathbf{k}_{1})P(\mathbf{k}_{2})P(\mathbf{k}_{3})}}
	\left[B(\mathbf{k}_{1},\mathbf{k}_{2},\mathbf{k}_{3})+\frac{1}{\overline{n}}\left[ P(\textbf{k}_1) + P(\textbf{k}_2) + P(\textbf{k}_3)\right] + \frac{1}{\overline{n}^2}\right] \delta_D(\mathbf{k}_{1}+\mathbf{k}_{2}+\mathbf{k}_{3})\,,
	\label{eq:3epsshot}
\end{equation}
where
\begin{equation}
	\nu_{\rm eff}(\bk)\equiv\frac{\overline{n}P(\bk)}{1+\overline{n}P(\bk)}\,.
	\label{eq:nu eff}
\end{equation}

Shot noise therefore has two effects on the LCF. First, it suppresses the signal coming from the bispectrum, i.e.\ the term proportional to the bispectrum $B$, due to the shot noise suppression factor $\nu_{\rm eff}$. This term is called the effective LCF in \cite{Eggemeier:2016asq}. Second, shot noise adds a noise contribution to the signal, made of the last two terms in eq.~\eqref{eq:3epsshot}. The predicted effective LCF and total LCF are plotted as dashed and solid curves in Figure~\ref{fig:lcffits}, where the theoretical LCF multipoles are compared with measurements from simulations. 

As in the case of the bispectrum, the shot noise contribution to the LCF depends on cosmology. As discussed in \cite{Yankelevich:2018uaz}, there are then two possible ways to treat the shot noise terms when performing the forecasts. The first option is to conservatively assume that the shot noise modeling is not accurate enough to extract information from its cosmology dependence, and therefore not to vary this term in the Fisher forecast derivatives. In this case, the observable is the effective LCF, as in \cite{Eggemeier:2016asq}. The second option is to assume that the cosmology-dependence of the shot noise is accurately modeled by the form of $B_{\rm noise}$, such that we can use the cosmology-dependence of the shot noise as a source of information about the parameters of interest. In this case, the observable is the total LCF. In Section~\ref{sec:fisher}, we compare Fisher forecasts for both options, to demonstrate how much constraining power can be gained through the cosmology-dependence of the shot noise.


\subsection{Summary of theoretical modeling}
\label{sec:model_summary}

We use the halo power spectrum model in eq.~\eqref{eq:Pvlah}, which is originally from \cite{Vlah:2013lia} but modified according to \cite{Howlett:2019bky}, and further modified in this work to accommodate a global Fingers-of-God damping factor. Including the additive shot noise term in eq.~\eqref{eq:PBnoise}, our theoretical model for the power spectrum multipoles is then given by
\begin{equation}
	P_\ell(k) \equiv \frac{2n+1}{2} \int_{-1}^1 \D{\mu}\, \left[ P(k,\mu) + P_{\rm noise} \right] L_n(\mu).
\end{equation}

Our model for the LCF multipoles is built from a combination of eq.~\eqref{eq:Pvlah} and the bispectrum model from \cite{Gil-Marin:2014pva} in eq.~\eqref{eq:Bgm}. To obtain our predictions for the LCF multipoles, we combine eqs.~\eqref{eq:LCF definition fourier}, \eqref{eq:Qn} and \eqref{eq:3epsshot}. This results in an expression for $Q_n$ that contains a 7-dimensional integral. To simplify this expression, we first note that $\ell(r,\nu)$ depends only on the angle $\nu=\hat{\mathbf{r}}\cdot\hat\bn$ and not on the angle $\phi$, which describes a rotation of $\mathbf{r}$ around $\hat\bn$. We can therefore average the multipoles in eq.~\eqref{eq:Qn} over $\phi$. We then rewrite the exponential in eq.~\eqref{eq:LCF definition fourier} and the Legendre polynomial in eq.~\eqref{eq:Qn} using
\begin{equation}
	e^{i\mathbf{r}\cdot\boldsymbol{\kappa}_{1}}=4\pi\sum_{l=0}^{\infty}\sum_{m=-l}^{l}i^{l}j_{l}\left(\kappa_1 r\right)Y_{lm}\left(\hat{\boldsymbol{\kappa}}_1\right)Y_{lm}^{*}\left(\hat{\mathbf{r}}\right),
	\label{eq:exp as Ylm}
\end{equation}
where $\boldsymbol{\kappa}_{1}\equiv \mathbf{k}_{1}-\mathbf{k}_{2}$, and 
\begin{equation}
	L_{n}\left(\nu\right)=\frac{4\pi}{2n+1}\sum_{p=-n}^{n}Y_{np}\left(\hat{\mathbf{r}}\right)Y_{np}^*\left(\hat{\mathbf{n}}\right).
	\label{eq:L as Ylm}
\end{equation}
Using these substitutions and noting the orthonormality of the spherical harmonic functions,
\begin{equation}
	\int d\Omega_{\hat{\mathbf{r}}} \,Y_{lm}\left(\hat{\mathbf{r}}\right)Y_{np}^{*}\left(\hat{\mathbf{r}}\right) = \delta_{nl}^{K}\delta_{pm}^{K}\,,
	\label{eq:Ylm orth}
\end{equation}
we obtain our theoretical prediction for the LCF multipoles as
\begin{align}
	Q_n(r) &= \left(2n+1\right)i^{n}\frac{V^2}{(2\pi)^6}\left(\frac{r^{3}}{V}\right)^{3/2} \left( \frac{\sqrt{\pi}}{2}\right)^3 \iintop_{k_{1},k_{2},|\mathbf{k}_{1}+\mathbf{k}_{2}|\leq\frac{2\pi}{r}} \D{^3 k_1}\D{^3 k_2}\, j_{n}\left(\kappa_1r\right)L_{n}\left(\hat{\boldsymbol{\kappa}}_1\cdot\hat{\mathbf{n}}\right) \nonumber \\
	&\times \sqrt{\frac{\nu_{\rm eff}(\bk_1)\nu_{\rm eff}(\bk_2)\nu_{\rm eff}(\bk_3)}{V P(\mathbf{k}_{1})P(\mathbf{k}_{2})P(\mathbf{k}_{3})}} \left[B(\mathbf{k}_{1},\mathbf{k}_{2},\mathbf{k}_{3})+\frac{1}{\overline{n}}\left[ P(\textbf{k}_1) + P(\textbf{k}_2) + P(\textbf{k}_3)\right] + \frac{1}{\overline{n}^2}\right].
\end{align}

We note the LCF modeling in this work is comprised of a combination of existing models in the literature for the power spectrum and bispectrum which are themselves derived from different assumptions and frameworks. Thus they are not fully consistent from a perturbation theory point of view. However, for the purposes of the first forecast using the redshift-space LCF that is in this work, we implement this model and check how it compares with simulations in Section~\ref{sec:compare} before proceeding with the Fisher forecasts.

We have previously mentioned that the LCF multipoles aid in breaking parameter degeneracies, and here we discuss this point in more detail. \cite{Franco:2018yag} derived explicit expressions for the multipoles of the LCF using tree-level perturbation theory and found that they depend on different combinations of the growth rate $f$ and the amplitude of perturbations $\sigma_8$. For example, eq.~(40) in that work showed that the intrinsic part of the monopole takes the simple form
\begin{align}
    Q^{\rm int}_0(r)=&\frac{r^{9/2}}{8\pi\sqrt{2}}\int_{0}^{2\pi/r}\hspace{-0.2cm}dk_{1}k_{1}^{2}\int_{0}^{2\pi/r}\hspace{-0.2cm}dk_{2}k_{2}^{2}\int_{-1}^{\alpha_{\text{cut}}}\hspace{-0.1cm}d\alpha \sqrt{\frac{P_{L}\left(\left|\mathbf{k}_{1}+\mathbf{k}_{2}\right|\right)P_{L}\left(k_{1}\right)}{P_{L}\left(k_{2}\right)}} \Bigg\{F_{2}(-\bk_1-\bk_2,\bk_1) \nonumber\\
   +&\left(F_{2}(-\bk_1-\bk_2,\bk_1)-G_2(-\bk_1-\bk_2,\bk_1)\right)\left(\frac{\arctan\sqrt{f/b_1}}{\sqrt{f/b_1}}-1\right)+\left(\frac{b_2}{2b_1}+\frac{b_{s^2}}{2b_1}S_2(-\bk_1-\bk_2,\bk_1)\right)\left(\frac{\arctan\sqrt{f/b_1}}{\sqrt{f/b_1}}\right)\Bigg\}\sum_{i=1}^3j_0(\kappa_i r)\, ,
   \label{Qmono}
\end{align}
where $\alpha\equiv\hat{\bk}_1\cdot\hat{\bk}_2$, $\alpha_{\rm cut}\equiv\min\{1,\max\{-1,\big[(2\pi/r)^2-k_1^2-k_2^2\big]/[2k_1k_2]\}\}$ enforces the condition $|\bk_1 + \bk_2| \leq 2\pi/r$,  $\boldsymbol{\kappa}_{1}\equiv\mathbf{k}_{1}-\mathbf{k}_{2}$, $\boldsymbol{\kappa}_{2}\equiv\mathbf{k}_{1}+2\mathbf{k}_{2}$ and $\boldsymbol{\kappa}_{3}\equiv-2\mathbf{k}_{1}-\mathbf{k}_{2}$. The monopole is directly proportional to $\sigma_8$ through the ratio of power spectra in the first line of eq.~\eqref{Qmono}, but it depends on the growth rate $f$ through the arctangents and the square roots in the second line of eq.~\eqref{Qmono}. This explicitly shows that the monopole of the LCF can be used in conjunction with the power spectrum multipoles to break the degeneracy between the measurements of $f$ and $\sigma_8$. Similar expressions have been derived for the intrinsic and mapping parts of the other multipoles in~\cite{Franco:2018yag}. Going beyond tree-level perturbation theory, eq.~\eqref{Qmono} must be modified; in particular, the integrals over $\bk_1, \bk_2$ and $\nu$ in eq.~\eqref{eq:Qn} can no longer be reduced to three integrals over $k_1, k_2$ and $\alpha$, as in eq.~\eqref{Qmono}. However, the fact that the $Q_n$ multipoles can break the degeneracy between $f$ and $\sigma_8$ remains valid.

Finally, we end this section on the theoretical modeling of the power spectrum and LCF by discussing the broader modelling assumptions in this work. Traditional analyses of the redshift-space power spectrum typically use a fixed template for the linear power spectrum to measure $f\sigma_8$, which is then used to constrain potential deviations from the assumed cosmology that generated the fixed template, e.g. $\Lambda$CDM and GR. We note that CMB measurements strongly constrain the shape of the linear power spectrum at early times, independently of late-time acceleration and structure growth (as discussed in Section 3.3 of \cite{Jelic-Cizmek:2020pkh}), such that the galaxy power spectrum multipoles in the \textit{linear} regime can subsequently constrain $f\sigma_8$ in a model-independent way, without assuming that the linear growth factor or the growth rate match that of $\Lambda$CDM and GR. However, in this work, we use the \textit{non-linear} power spectrum and LCF that depend on the non-linear matter power spectrum and the perturbation theory kernels, which we fix to those predicted by $\Lambda$CDM and GR. This means that our forecasted constraints on $f$ and $\sigma_8$ represent a consistency test of $\Lambda$CDM and GR. We apply this test at many individual redshift bins, allowing for the redshift evolution of the growth of structure to deviate from $\Lambda$CDM and GR in a model-independent way.

However, it is important to note that these modelling assumptions are not a prerequisite for using the LCF for cosmological constraints, and the joint analysis of the power spectrum and LCF can also provide constraints on other non-$\Lambda$CDM or non-GR models as long as theoretical predictions for the power spectrum and LCF within these scenarios are available \citep{Ali:2018sdk}. For example, to constrain modified gravity models where the perturbation theory kernels are changed or the growth rate is scale-dependent, the power spectrum and LCF could be used to simultaneously constrain $f$, $\sigma_8$ and the additional free parameters that are specific to the model.


\section{Simulations and estimators}
\label{sec:sims}

In this work, we rely on simulations to check multiple components of our forecasting pipeline. Here, we describe the simulation data that we use, as well as the power spectrum and LCF multipole estimators. The LCF multipole estimator in eq.~\eqref{eq:Qn_estimator} is presented  and implemented for the first time in this work.


\subsection{Simulations}
\label{subsec:sims}

The simulations are the same as those of \cite{Ali:2018sdk} and consist of 500 unique dark matter realisations generated using the \textsc{l-picola} approximate simulation code \citep{Tassev:2013pn,Howlett:2014opa,Howlett:2015hfa}. The authors of these previous works have demonstrated that the COLA algorithm is able to reproduce well the two- and three-point clustering, and the LCF, for scales $k<0.3 \,h\Mpc^{-1}$ and $r>10 \,h^{-1}\mathrm{Mpc}$, which are the scales of interest to this work and next generation surveys.

The simulations were generated using $256^{3}$ particles in a box of length $512\,h^{-1}\mathrm{Mpc}$. This corresponds to a mass resolution of $6.7\times 10^{11}h^{-1}M_{\odot}$. Halos were then identified in each of the realisations using a Friends-of-Friends algorithm \citep{Davis:1985rj} limited to a minimum of 10 particles per halo. We do not expect the \textsc{l-picola} simulations to accurately reproduce the mass function of halos down to these scales, nor the internal properties of the halos. We also do not extract subhalos from the simulations. Nonetheless, they are adequate for checking the theoretical modelling presented in Section~\ref{sec:modeling}. The average halo number density in the simulations is $\overline{n} = 3.57 \times 10^{-4}\,h^{3}\mathrm{Mpc^{-3}}$. Details of how measurements of the power spectra and LCF were made from the simulated halo catalogs are given below.


\subsection{Power spectrum estimator}

The estimator for the bin-averaged isotropic power spectrum is
\begin{align}
	\hat{P}\left(k\right)=\frac{1}{V}\int_{k} \D{^3 q}
	\frac{\Delta\left(\mathbf{q}\right)\Delta\left(-\mathbf{q}\right)}{V_{P}\left(k\right)}\,,
\end{align}
where the integrals run over the interval $q\in\left[k-\Delta k/2,k+\Delta k/2\right]$, $\Delta k$ is the bin width, and $V_{P} \equiv 4\pi k^{2}\Delta k$ is the volume of modes in a spherical shell. Similarly, assuming the global plane-parallel approximation, one can define an estimator for the power spectrum multipoles as
\begin{align}
	\hat{P}_{n}\left(k\right)=\frac{2n+1}{V}\int_{k}\D{^3 q}\frac{\Delta\left(\mathbf{q}\right)\Delta\left(-\mathbf{q}\right)}{V_{P}\left(k\right)}L_{n}\left(\hat{\mathbf{q}}\cdot\hat{\mathbf{n}}\right).
	\label{eq:angular power spectrum estimator}
\end{align}


\subsection{LCF estimator}
 
The LCF multipoles are defined by eq.~\eqref{eq:Qn}. To derive an estimator for $Q_n$ we first use the fact that $\ell (r,\nu)$ depends only on $\nu \equiv \hat{\mathbf{r}} \cdot \hat{\bn}$, and not on the angle $\phi$ which describes a rotation of $\mathbf{r}$ around $\hat{\bn}$. We can therefore average the multipoles over $\phi$ to obtain
\begin{equation}
	Q_n(r) = \frac{2n+1}{4\pi} \int_0^{2\pi} \D{\phi} \int^{1}_{-1} \D{\nu}\,\ell(r,\nu) L_n(\nu).  \label{eq:Qest}
\end{equation}
The estimator for $\ell(\mathbf{r})$ is given by
\begin{equation}
	\ell(\mathbf{r}) = \frac{V^2}{(2\pi)^6} \left(\frac{r^{3}}{V}\right)^{3/2} \iintop_{k_{1},k_{2},|\mathbf{k}_{1}+\mathbf{k}_{2}|\leq\frac{2\pi}{r}} 
	\D{^3 k_1}\D{^3 k_2} \,
	e^{i\mathbf{r}\cdot\left(\mathbf{k}_{1}-\mathbf{k}_{2}\right)}\Braket{\epsilon(-\mathbf{k}_{1}-\mathbf{k}_{2})\epsilon(\mathbf{k}_{1})\epsilon(\mathbf{k}_{2})}. \label{eq:lcontinuous}
\end{equation}
The exponential in eq.~\eqref{eq:lcontinuous} and the Legendre polynomial in eq.~\eqref{eq:Qest} can be written in terms of spherical harmonics using the identities in eqs.~\eqref{eq:exp as Ylm} and \eqref{eq:L as Ylm}. Then the orthonormality of the spherical harmonic functions in eq.~\eqref{eq:Ylm orth} gives the final expression for the estimator,
\begin{align}
	\hat{Q}_n(r) = \left(2n+1\right)i^{n}\frac{V^2}{(2\pi)^6}\left(\frac{r^{3}}{V}\right)^{3/2} \iintop_{k_{1},k_{2},|\mathbf{k}_{1}+\mathbf{k}_{2}|\leq\frac{2\pi}{r}} \D{^3 k_1}\D{^3 k_2}\, j_{n}\left(\kappa_1r\right)L_{n}\left(\hat{\boldsymbol{\kappa}}_1\cdot\hat{\mathbf{n}}\right)\Braket{\epsilon(-\mathbf{k}_{1}-\mathbf{k}_{2})\epsilon(\mathbf{k}_{1})\epsilon(\mathbf{k}_{2})}. 
	\label{eq:Qn_estimator}
\end{align}

\subsection{Estimating means and covariance matrix}

We estimate the mean of the power spectrum and LCF multipoles as
\begin{equation}
	\overline{X}_i = \frac{1}{N_{\rm sims}} \sum_{n=1}^{N_{\rm sims}} \hat{X}_i^{(n)},
\end{equation}
while the covariance matrix is estimated as
\begin{equation}
	\hat{\mat{C}}_{*,ij} = \frac{1}{N_{\rm sims}-1} \sum_{n=1}^{N_{\rm sims}} 
	\left( \hat{X}_i^{(n)} - \overline{X}_i \right)
	\left( \hat{X}_j^{(n)} - \overline{X}_j \right).
\end{equation}
$N_{\rm sims} = 500$ is the total number of independent realisations of the \textsc{l-picola} simulations, and $\hat{X}_i^{(n)}$ is the measured data vector in the $n$-th realisation.

For calculating signal-to-noise ratios and the Fisher forecasts that follow, we require estimates of the inverse covariance matrix. Taking the direct inverse of $\hat{\mat{C}}_*$ results in a biased estimate of the true inverse covariance. To remedy this, we apply an approximate correction by multiplying the inverse of $\hat{\mat{C}}_*$ with the Anderson-Hartlap factor \citep{Anderson2003,Hartlap2007} and estimate the inverse covariance as
\begin{equation}
	\hat{\mat{C}}^{-1} = \frac{N_{\rm sims} - N_{\rm bins} - 2}{N_{\rm sims} - 1} \hat{\mat{C}}_*^{-1},
\end{equation}
where $N_{\rm bins}$ is the length of the data vector. 


\section{Theoretical covariance matrices}
\label{sec:theoretical_covariance}

We first review expressions for the leading-order power spectrum multipoles covariance. Then we present for the first time the leading-order contributions to the LCF multipoles covariance and the cross-covariance between the power spectrum multipoles and the LCF multipoles. In this section we briefly summarise the equations and defer the detailed derivations to Appendices~\ref{app:Qn_cov} and \ref{app:PlQn_crosscov}.

For the DESI and Euclid galaxy surveys that we consider in this work, the redshift bins have widths of $\Delta z \geq 0.1$, so we expect that the covariance between different redshift bins is negligible. All covariance expressions we present are therefore implicitly for one fixed redshift bin.


\subsection{Power spectrum covariance}

The Gaussian covariance for the power spectrum multipoles is
\begin{align} 
	{\rm cov}\big[P_{n_1}(k_i),P_{n_2}(k_j)\big]=	 \delta_{ij} \frac{(2n_1+1)(2n_2+1)}{V_P(k_i)} 
	\frac{(2\pi)^3}{V} 
	\int_{-1}^{1}d\mu \left[ P(k_i,\mu)+ \frac{1}{\overline{n}} \right]^2L_{n_1}(\mu)L_{n_2}(\mu),
	\label{eq:covP}
\end{align}
where $P(k_i,\mu)$ is given by eq.~\eqref{eq:Pvlah}.


\subsection{LCF covariance}

The Gaussian covariance for the LCF multipoles is
\begin{align}
	\text{cov}\left[Q_{n_{1}}\left(r_{i}\right),Q_{n_{2}}\left(r_{j}\right)\right]  =& \delta^K_{n_1n_2} \frac{\left(2n_{1}+1\right)(-1)^{n_{1}}\left(r_{i}r_{j}\right)^{9/2}}{4\pi^{4}V}\int_{0}^{2\pi/R}dk_{1}k_{1}^{2}\int_{0}^{2\pi/R}dk_{2}k_{2}^{2}\int_{-1}^{\alpha_{\text{cut}}}d\alpha j_{n_{1}}\left(\kappa_{1}r_{i}\right)\nonumber \\
 	& \times\left[j_{n_{1}}\left(\kappa_{1}r_{j}\right)+2j_{n_{1}}\left(\kappa_{2}r_{j}\right)L_{n_{1}}\left(\hat{\boldsymbol{\kappa}}_{1}\cdot\hat{\boldsymbol{\kappa}}_{2}\right)\right],
 	\label{eq:covQn}
	\end{align}
with $R\equiv\max\left(r_{i},r_{j}\right)$, $\alpha\equiv\hat{\bk}_1\cdot\hat{\bk}_2$, $\boldsymbol{\kappa}_{1}\equiv \mathbf{k}_{1}-\mathbf{k}_{2}$, and $\boldsymbol{\kappa}_{2}\equiv \mathbf{k}_{1}+2\mathbf{k}_{2}$. $\alpha_{\text{cut}}=\min\{ 1,\max\{ -1,[\left(2\pi/R\right)^{2}-k_{1}^{2}-k_{2}^{2}]/[2k_{1}k_{2}]\} \}$ is imposed to keep $\left|\mathbf{k}_{1}+\mathbf{k}_{2}\right|\leq2\pi/R$ satisfied. The argument of the Legendre polynomial is given by
\begin{equation}
	\hat{\boldsymbol{\kappa}}_{1}\cdot\hat{\boldsymbol{\kappa}}_{2}=\frac{k_1^2+k_1k_2\alpha-2k_2^2}{\sqrt{\big(k_1^2-2k_1k_2\alpha+k_2^2\big)\big(k_1^2+4k_1k_2\alpha+4k_2^2\big)}}\, .
\end{equation}
The covariance between different multipoles of the LCF is zero. A complete derivation of this covariance is presented in Appendix~\ref{app:Qn_cov}. 

We note that this covariance in eq.~\eqref{eq:covQn} is independent of both cosmology and shot noise, like the Gaussian covariance of the real-space LCF. This is due to the fact that the Gaussian covariance is proportional to products of two-point functions of the phases, 
\begin{equation}
	\langle\epsilon(\bk_1)\epsilon(\bk_2) \rangle=\frac{(2\pi)^3}{V}\delta_D(\bk_1+\bk_2)\,,
\end{equation}
which do not depend on the power spectrum and are therefore unaffected by shot noise and cosmology. The only dependence it has on the survey of interest is through the survey volume. Cosmology dependence would appear in the LCF covariance through additional non-Gaussian terms that we do not include here, but we expect that these contributions would be small, because in Section \ref{sec:theory_cov} and Figure \ref{fig:halo_theorycov} we find that the Gaussian covariance is a good approximation for the LCF scales of interest in this work.


\subsection{Power spectrum and LCF cross-covariance}

The covariance between the power spectrum and LCF multipoles is intrinsically non-Gaussian because it is a 5-point correlator. As shown in Appendix~\ref{app:PlQn_crosscov}, there are two contributions: one containing the product of the bispectrum and power spectrum and one containing a connected 5-point correlator term,
\begin{equation}
	\text{cov}\left[P_{n_{1}}\left(k_{i}\right),Q_{n_{2}}\left(r_{j}\right)\right]=\frac{V}{\left(2\pi\right)^{3}}\left(\frac{\sqrt{\pi}}{2}\right)^{3}\left(\frac{r_{j}^{3}}{V}\right)^{3/2}\left[\mathcal{C}^{\left(n_{1},n_{2}\right)}_{PB}+\mathcal{C}^{\left(n_{1},n_{2}\right)}_{P_{5}}\right].
	\label{eq:covPQ}
\end{equation}
The expression for $\mathcal{C}^{\left(n_{1},n_{2}\right)}_{PB}$ derived in Appendix~\ref{app:PlQn_crosscov}, and modified to include shot noise, is 
\begin{align}
	\mathcal{C}^{\left(n_{1},n_{2}\right)}_{PB} & = 2\left(2n_{1}+1\right)\left(2n_{2}+1\right)i^{n_2}
	\left(\frac{2}{\sqrt{\pi}}\right)^3
	\int_{k_{i}}\frac{\D{^3 k_2}}{V_{P}\left(k_{i}\right)}
	\left[P\left(\mathbf{k}_{2}\right) + \frac{1}{\overline{n}}\right]
	L_{n_{1}}\left(\hat{\mathbf{k}}_{2}\cdot\hat{\bn}\right)\Theta\left(1-\frac{k_{2}r_{j}}{2\pi}\right) \nonumber \\
	&\times\underset{k_{1},\left|\mathbf{k}_{1}+\mathbf{k}_{2}\right|\leq
	\frac{2\pi}{r_{j}}
	}{\int}\D{^3 k_1}
	\Braket{\epsilon\left(\mathbf{k}_{2}\right)\,\epsilon\left(-\mathbf{k}_{1}-\mathbf{k}_{2}\right)\,\epsilon\left(\mathbf{k}_{1}\right)} 
	\left[j_{n_{2}}\left(\kappa_{1}r_{j}\right)L_{n_{2}}\left(\hat{\boldsymbol{\kappa}}_{1}\cdot\hat{\bn}\right)+j_{n_{2}}\left(\kappa_{2}r_{j}\right)L_{n_{2}}\left(\hat{\boldsymbol{\kappa}}_{2}\cdot\hat{\bn}\right)+j_{n_{2}}\left(\kappa_{3}r_{j}\right)L_{n_{2}}\left(\hat{\boldsymbol{\kappa}}_{3}\cdot\hat{\bn}\right)\right],
	\label{eq:CPBfinal}
\end{align}
where for the 3-point phase correlations we use eq.~\eqref{eq:3epsshot}. The expression for $\mathcal{C}^{\left(n_{1},n_{2}\right)}_{P^5}$ is in eq.~\eqref{eq:CP5}. $\boldsymbol{\kappa}_{1}$ and $\boldsymbol{\kappa}_{2}$ are defined in the same way as before, while $\boldsymbol{\kappa}_{3}\equiv-2\mathbf{k}_{1}-\mathbf{k}_{2}$.

For combined power spectrum and bispectrum galaxy clustering analyses, the contribution from $\mathcal{C}^{\left(n_{1},n_{2}\right)}_{P^5}$ is expected to be subdominant on large scales \citep{Sefusatti:2006pa}. \cite{Sugiyama:2019ike} showed that indeed this is the case, while the relevance of the $\mathcal{C}^{\left(n_{1},n_{2}\right)}_{P^5}$ term on intermediate and smaller scales has not yet been fully confirmed. This term is also more difficult to compute, so in our theoretical model for the cross-covariance we do not include it. We assess the impact of this assumption to the power spectrum--LCF cross-covariance in Section~\ref{sec:theory_cov} by comparing the forecasts obtained using the full measured covariance from simulations with the forecasts obtained using the theoretical covariance matrix. We will see that the impact of ignoring the cross-covariance between the power spectrum and the LCF is similarly minimal in both cases, implying that it is a good approximation to neglect the $\mathcal{C}^{\left(n_{1},n_{2}\right)}_{P^5}$ term or even the cross-covariance $\text{cov}\left[P_{n_{1}}\left(k_{i}\right),Q_{n_{2}}\left(r_{j}\right)\right]$ entirely.


\section{Comparing models and simulations}
\label{sec:compare}

In this section we compare theoretical predictions for the power spectrum and LCF multipoles, and their covariance, with measurements from 500 \textsc{l-picola} simulations. This comparison, described in detail below, validates our implementation of the theoretical predictions and estimators, and gives valuable guidance on the modelling choices, range of scales and fiducial values to use in the Fisher forecasts that we present later in Sections~\ref{sec:fisher} and \ref{sec:surveys}.


\subsection{Mean power spectrum and LCF at fiducial cosmology}
\label{sec:means}

In Section~\ref{sec:modeling}, we presented the models that we use for the halo power spectrum and LCF multipoles. Here we compare these models to the simulations by fitting them to the mean of 500 \textsc{l-picola} simulation measurements obtained using the estimators in Section~\ref{sec:sims}. This comparison checks that our theoretical predictions and measurements are in good agreement before proceeding to the Fisher forecasts in the rest of this work.

In the ideal scenario, we would perform a fully joint fit to both the $P_\ell$ and $Q_n$ multipoles at the same time. However, the LCF predictions are computationally very demanding, so for each model we perform the fit in two stages. In the first stage, we fit the $P_\ell$ multipoles by running an MCMC chain with a Gaussian likelihood and the $P_\ell$ covariance matrix measured from simulations. We fit the shot noise-corrected $P_0$, while $P_2$ and $P_4$ do not have any shot noise contribution in our model. We include all $k$ bins up to some $k_{\rm max}$ and consider the 3-dimensional parameter space of $b_1$, $b_2$ and $\sigma_P$. We leave out $\sigma_B$ because it does not appear in the power spectrum, and for simplicity we fix the values of $f$ and $\sigma_8$ to the values that correspond to the known $\Lambda$CDM cosmology of the simulations. Then in the second stage, we fix $b_1$, $b_2$ and $\sigma_P$ to their best-fit values from the power spectrum fit, and only fit the LCF data for the last remaining parameter, $\sigma_B$. We note that we fit the total LCF, which includes both the effective and shot noise terms discussed in Section~\ref{sec:shot noise}. The LCF fitting procedure is not performed using MCMC chains because the LCF predictions are computed too slowly; instead, we maximize the likelihood interpolated over a grid in the 1-dimensional $\sigma_B$ parameter space. Assuming that the likelihood is approximately Gaussian, the error $\sigma_\theta$ on the parameter $\theta$ is estimated as
\begin{align}
	\frac{1}{\sigma_\theta^2} = \frac{1}{2} \frac{\partial^2 \chi^2}{\partial \theta^2},
\end{align}
where the second derivative is evaluated numerically at the maximum likelihood point. 

The results from this fitting procedure with $k_{\rm max} = 0.30  \, h\Mpc^{-1}$ are shown in Table~\ref{tab:fits} for all three models, where the only difference between them is whether the analytic form of the Fingers-of-God damping factor is a Gaussian, the same as in \cite{Gil-Marin:2014pva}, or a Lorentzian. For the power spectrum, the fitted $k$ bins are centered at $k_i = k_{\rm f} (i + \frac{1}{2})$, where $i=1, 2, \ldots, 24$, the fundamental wavenumber is $k_{\rm f} \equiv 2 \pi / L$, and the bin width is $\Delta k = k_{\rm f}$. For the LCF, the fitted $r$ bins are linearly spaced in $\ln(r)$ such that $\ln(r_j) = \ln (10) + (j-1) \left[ \ln(100)-\ln(10)\right]/19$, for $j=7, 8, \ldots, 20$.

\begin{figure*}
\includegraphics[width=\textwidth]{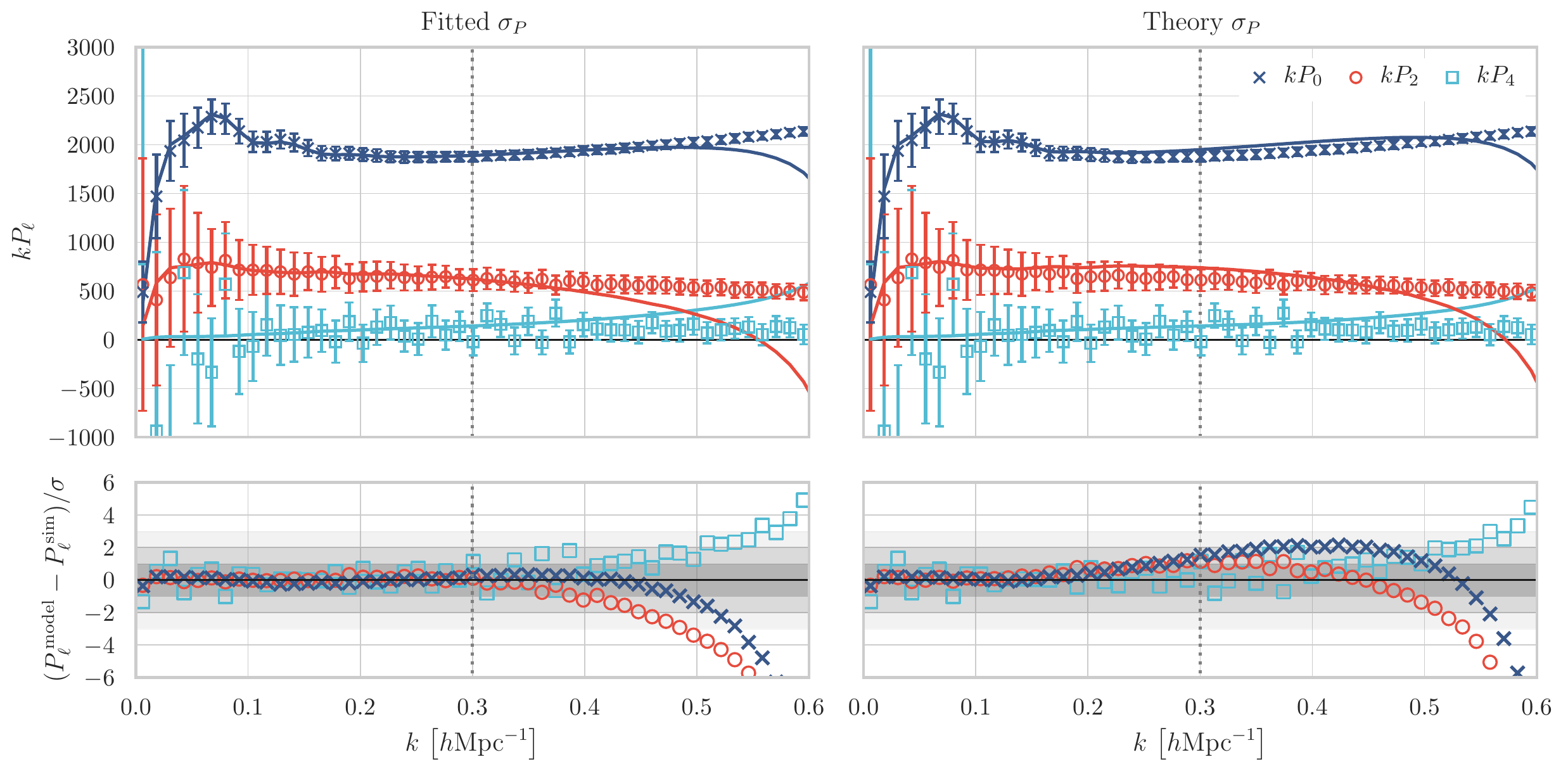}
\caption{The best-fit power spectrum multipoles with Lorentzian FoG are compared to the mean of 500 \textsc{l-picola} simulations. The shot noise has been corrected for in the measurements and models of the $P_0$ shown here. The gray dotted vertical lines indicate the $k_{\rm max} = 0.3 \,h\Mpc^{-1}$ that was used in the fit, and the solid lines are the best-fit model predictions. The model predictions in the left panel use the best-fit values from Table~\ref{tab:fits}, while the model in the right panel is different only in setting the value of $\sigma_P$ to the theory prediction, while $b_1$ and $b_2$ are still the best-fit values as in the left panel. The measurements shown in both panels are identical.}
\label{fig:pkfits}
\end{figure*}

For the power spectrum multipoles, we find that all three models perform well up to $k_{\rm max} = 0.30  \, h\Mpc^{-1}$, but the Lorentzian form returns the lowest value of the minimum $\chi^2$.\footnote{We note that it is not straightforward to compute corresponding values of reduced $\chi^2$ as $\chi_{\rm red}^2 = \chi^2/N_{\rm dof}$, because for models that are nonlinear in the parameters, as we have in our case, the number of degrees of freedom is not simply $N_{\rm dof} = N_{\rm data} - N_{\rm parameters}$, where $N_{\rm data}$ is the number of data bins and $N_{\rm parameters}$ is the number of model parameters \citep{Andrae:2010gh}. Additionally, the fact that we are fitting the mean of 500 simulations means that the fitted data has very little statistical noise. This, combined with the simple assumption for $N_{\rm dof}$, results in $\chi_{\rm red}^2$ values that are very small (much less than 1). However, here we only use the minimum $\chi^2$ values to compare between models with the same data bins and free parameters, so the absolute magnitude of $\chi_{\rm red}^2$ is not especially important.} The best-fit $P_\ell$ with Lorentzian FoG are compared with the data in the left column of Figure~\ref{fig:pkfits}, where we show that the model for the monopole and quadrupole is consistent with the measurements to within $1\sigma$ up to $k_{\rm max} = 0.3 \,h\Mpc^{-1}$. The hexadecapole is at most 2$\sigma$ away from the measurements for the same $k$ bins. 

The choice of $k_{\rm max} = 0.3 \,h\Mpc^{-1}$ is our optimistic case, but in later sections we also explore a more conservative forecast by using $k_{\rm max} = 0.15 \,h\Mpc^{-1}$. However, we do not show fits separately for this lower $k_{\rm max}$ because the MCMC analysis showed that $b_2$ cannot be constrained independently from the other parameters using this limited range of scales. The MCMC chains resulted in a bimodal posterior, with one of the modes being consistent with the best-fit from $k_{\rm max} = 0.3 \,h\Mpc^{-1}$.

\begin{table*}
\centering
\begin{tabular}{lccccccc} \hline
&  \multicolumn{5}{c}{Power spectrum} &  \multicolumn{2}{c}{LCF}\\
Model & $k_{\rm max}$ & $b_{1}$ & $b_{2}$ & $\sigma_P$ & min $\chi^2(P_\ell)$ &  $\sigma_B$ & min $\chi^2(Q_n)$ \\
\cmidrule(lr){1-1} \cmidrule(lr){2-6} \cmidrule(lr){7-8}
\\[-1.0em]
Vlah + GM fit + Gaussian FoG & 0.30 & $1.642\pm 0.023$ & $0.406^{+0.063}_{-0.080}$ & $4.18\pm 0.10$ & 11.5 & $2.73\pm 0.90$ & 4.3 \\
\\[-1.0em]
Vlah + GM fit + GM FoG & 0.30 & $1.650\pm 0.023$ & $0.406^{+0.062}_{-0.078}$ & $3.169\pm 0.088$ & 9.2 & $8.11 \pm 2.54$ & 5.7 \\
\\[-1.0em]
Vlah + GM fit + Lorentzian FoG & 0.30 & $1.657\pm 0.023$ & $0.408^{+0.060}_{-0.078}$ & $4.81\pm 0.15$ & 9.1 & $3.17\pm 1.01$ & 4.3 \\
\\[-1.0em]
\hline
\end{tabular}
\caption{Constraints on the galaxy bias parameters $b_{1}$ and $b_{2}$ and the Fingers-of-God velocity dispersions $\sigma_P$ and $\sigma_B$ from fitting halo power spectrum and LCF multipoles to the mean of 500 \textsc{l-picola} simulations.}
\label{tab:fits}
\end{table*}

\begin{figure*}
\includegraphics[width=\textwidth]{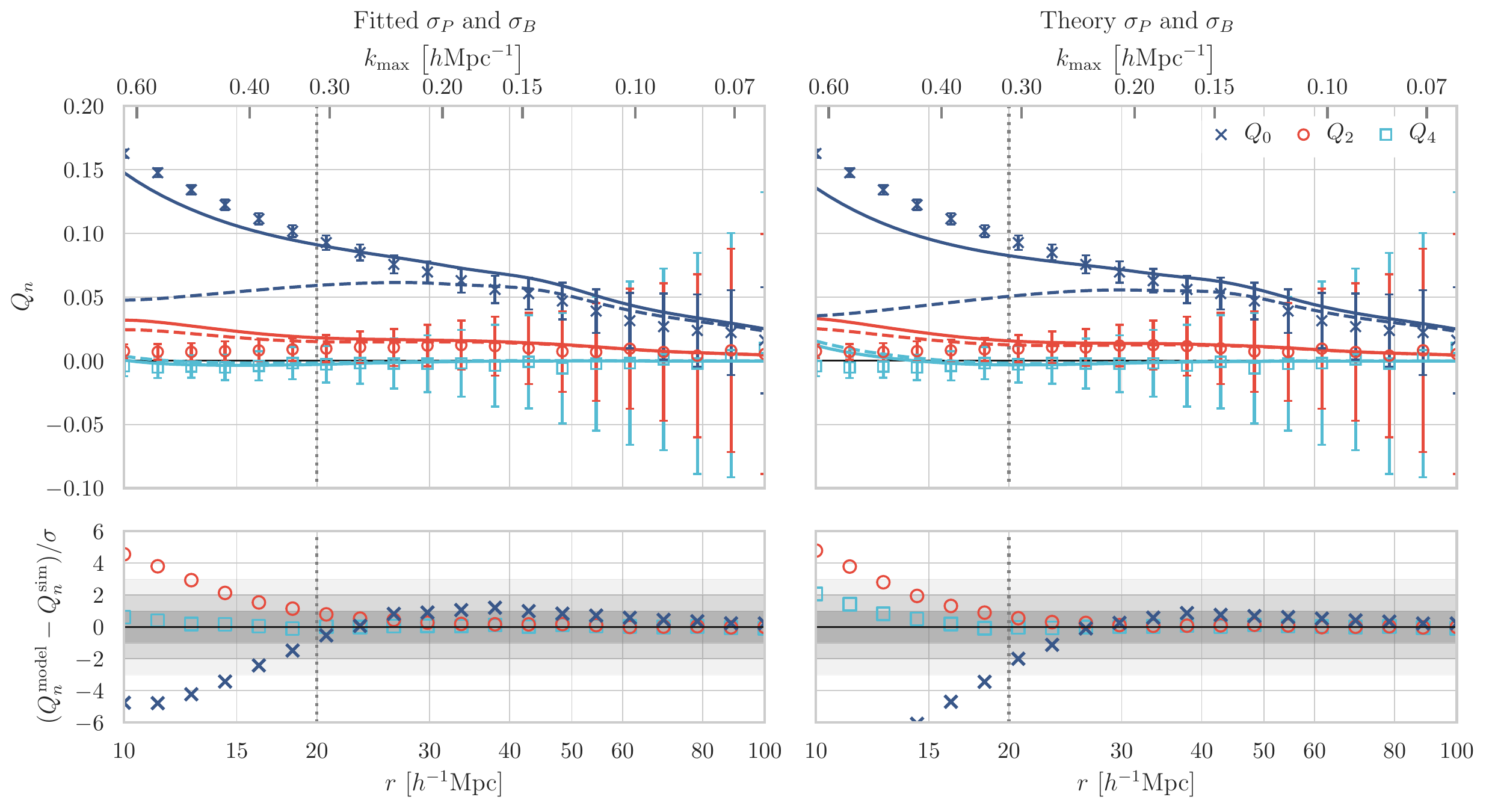}
\caption{The best-fit LCF multipoles with Lorentzian FoG are compared to the mean of 500 \textsc{l-picola} simulations. The gray dotted vertical lines indicate the $r_{\rm min} = 20 \, h^{-1}\Mpc$ that was used in the fit, and the solid lines are the best-fit model predictions. The effective LCF is shown in dashed curves, while the total LCF (the sum of the effective and shot noise LCF terms) is shown as a solid curve. The model predictions in the left panel use the best-fit values from Table~\ref{tab:fits}, while the model in the right panel is different only in setting the value of $\sigma_P=\sigma_B$ to the theory prediction. $b_1$ and $b_2$ in the right panel are the same best-fit values as in the left panel. The measurements shown in both panels are identical. The ticks for $k_{\rm max}$ at the top of both panels indicate the maximum wavenumber that is integrated (summed) over to obtain the predicted (measured) LCF at a given $r$.}
\label{fig:lcffits}
\end{figure*}

Once the power spectrum fits are complete, we fit the halo LCF multipoles using each model. For this we fix the parameter values for $b_1$, $b_2$ and $\sigma_P$ to the best-fit values from the power spectrum, and subsequently only fit the LCF multipoles for $\sigma_B$ using the $r_{\rm min}$ value that corresponds to $k_{\rm max} = 0.3 \,h\Mpc^{-1}$. 

As seen in eq.~\eqref{eq:LCF definition fourier}, the LCF integrates modes up to $k=2\pi/r$, and indeed as shown in~\cite{Wolstenhulme:2014cla}, the integral is dominated by this upper bound. Therefore, to trust the LCF at a scale $r_{\rm min}$ we need to trust the modelling of the power spectrum and bispectrum up to $k_{\rm max}=2\pi/r_{\rm min}$. We use this to determine the cut-off scale, i.e.~$r_{\min} = 21 \, h^{-1}\Mpc$ if $k_{\rm max}=0.30\,h\Mpc^{-1}$. Similarly, for the more conservative forecast with $k_{\rm max}=0.15\,h\Mpc^{-1}$, we use $r_{\min} = 43 \, h^{-1}\Mpc$. We note, however, that the cutoff for $\mathbf{k}_i$ in the integrals of eq.~\eqref{eq:LCF definition fourier}, and the choice of corresponding $k_{\rm max}$ for a chosen $r_{\rm min}$, sometimes differ in previous works on the LCF. For example, in the Fisher forecasts of \cite{Eggemeier:2016asq} and \cite{Byun:2017fkz}, though the cutoff was also $k_i \leq 2\pi/r$ as we use here, the LCF bins down to $r_{\rm min} = 10 \,h^{-1}\Mpc$ were combined with power spectrum bins up to $k_{\rm max} \approx 0.30\,h\Mpc^{-1}$. Both works used $k_{\rm max} = \pi/r_{\rm min}$, which comes from arguing that a density perturbation with wavelength $\lambda=2\pi/k$ corresponds to an overdensity of size $r=\lambda/2=\pi/k$. In this work, our different way of choosing $k_{\rm max}$ and our more conservative choices for $r_{\rm min}$ ensure that both the power spectrum and the LCF only have access to Fourier modes up to $k_{\rm max} = 0.30\,h\Mpc^{-1}$ in both the theoretical modeling and the estimation of these observables.

The resulting best-fit values for $\sigma_B$ are shown in Table \ref{tab:fits}, where we find that the Gaussian and Lorentzian FoG give the same minimum $\chi^2$ value that is lower than from the Gil-Marin FoG. The best-fit model using the Lorentzian FoG is shown in the left column of Figure~\ref{fig:lcffits}, where we see that the model for the LCF multipoles is in good agreement with the simulations down to $r_{\rm min} \approx 20\,h^{-1}\Mpc$: the quadrupole and hexadecapole are within 1$\sigma$ agreement with the measurements, while the monopole is at most 1.2$\sigma$ away from the measurements. 

The panels on the right sides of Figures~\ref{fig:pkfits} and \ref{fig:lcffits} are only different from the left panels in that, rather than using the best-fit values of $\sigma_P$ and $\sigma_B$, they use the linear theory prediction in eq.~\eqref{eq:sigmav2}, which for $z=0$ is $\sigma_P = \sigma_B = 4.5 \, h^{-1}\Mpc$. This changes the power spectrum at the highest $k$-bins, while the LCF monopole is $2\sigma$ away from the measurement in the smallest $r \approx 20 \,h^{-1}\Mpc$ bin. In Section~\ref{subsec:theory_veldisp}, we explore in more detail the impact of using the theoretical velocity dispersions in the Fisher forecasts.

\begin{figure*}
     \centering
     \subfloat{\includegraphics[width=0.5\textwidth]{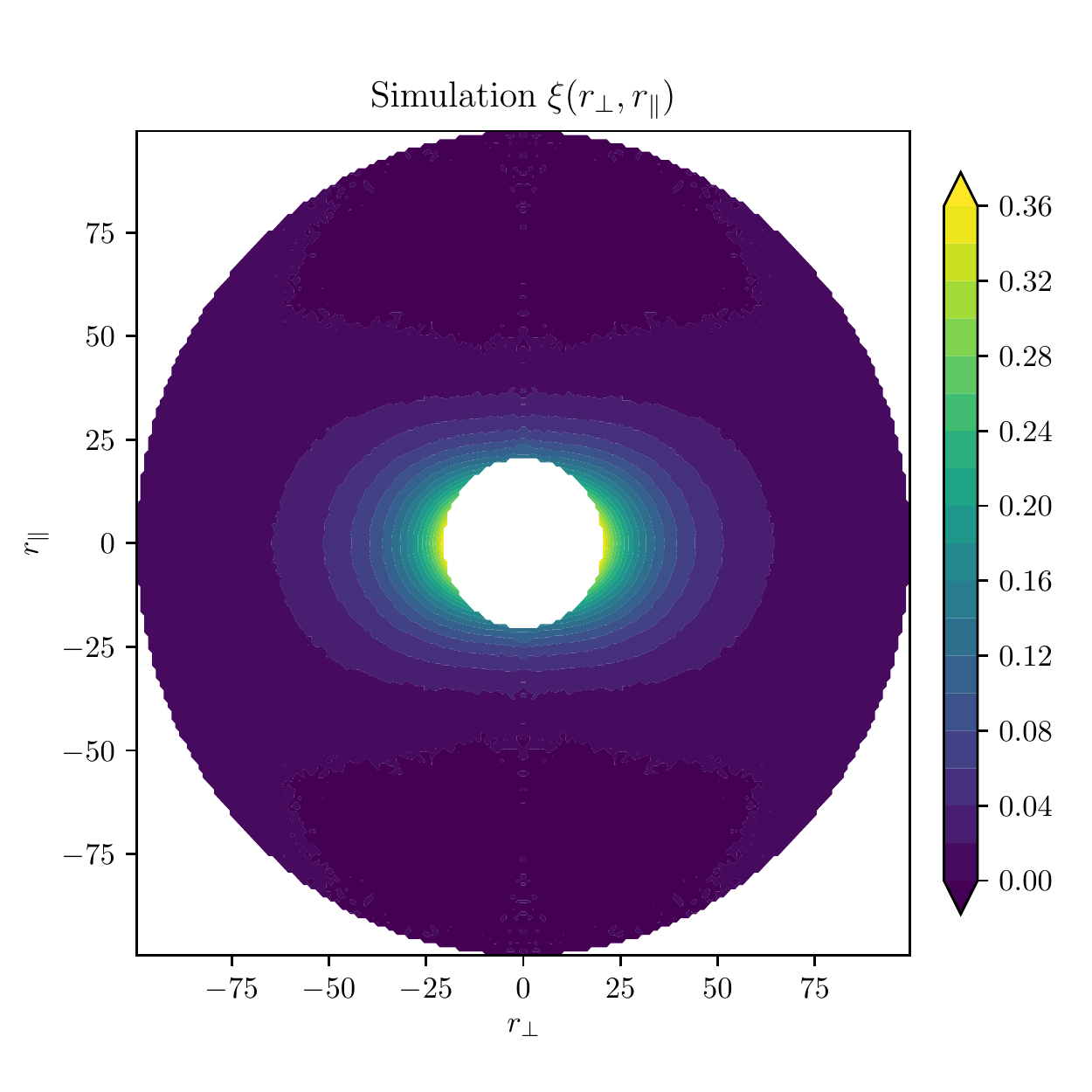}}
     \subfloat{\includegraphics[width=0.5\textwidth]{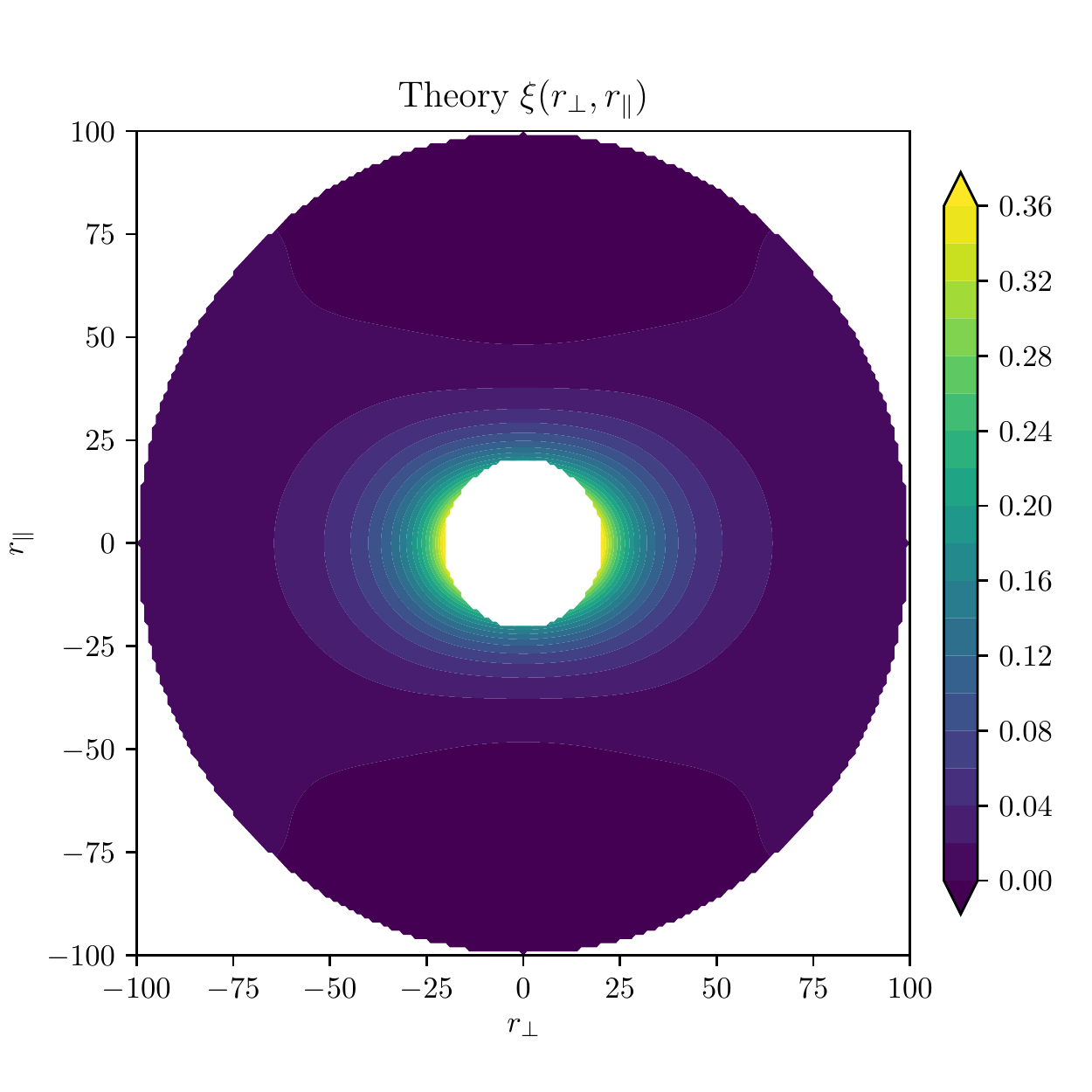}}
     \\
     \vspace{-2\baselineskip}
     \subfloat{\includegraphics[width=0.5\textwidth]{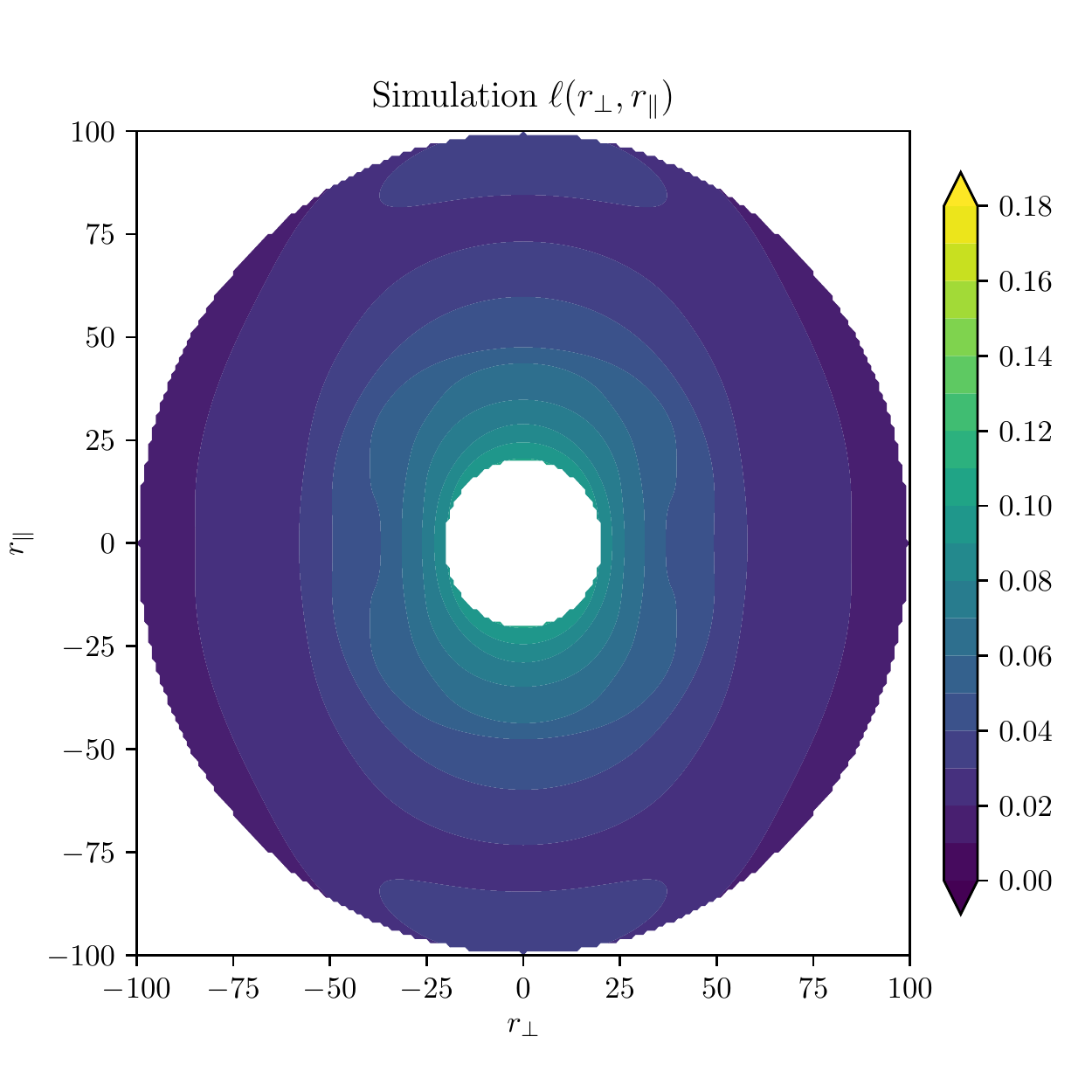}}
     \subfloat{\includegraphics[width=0.5\textwidth]{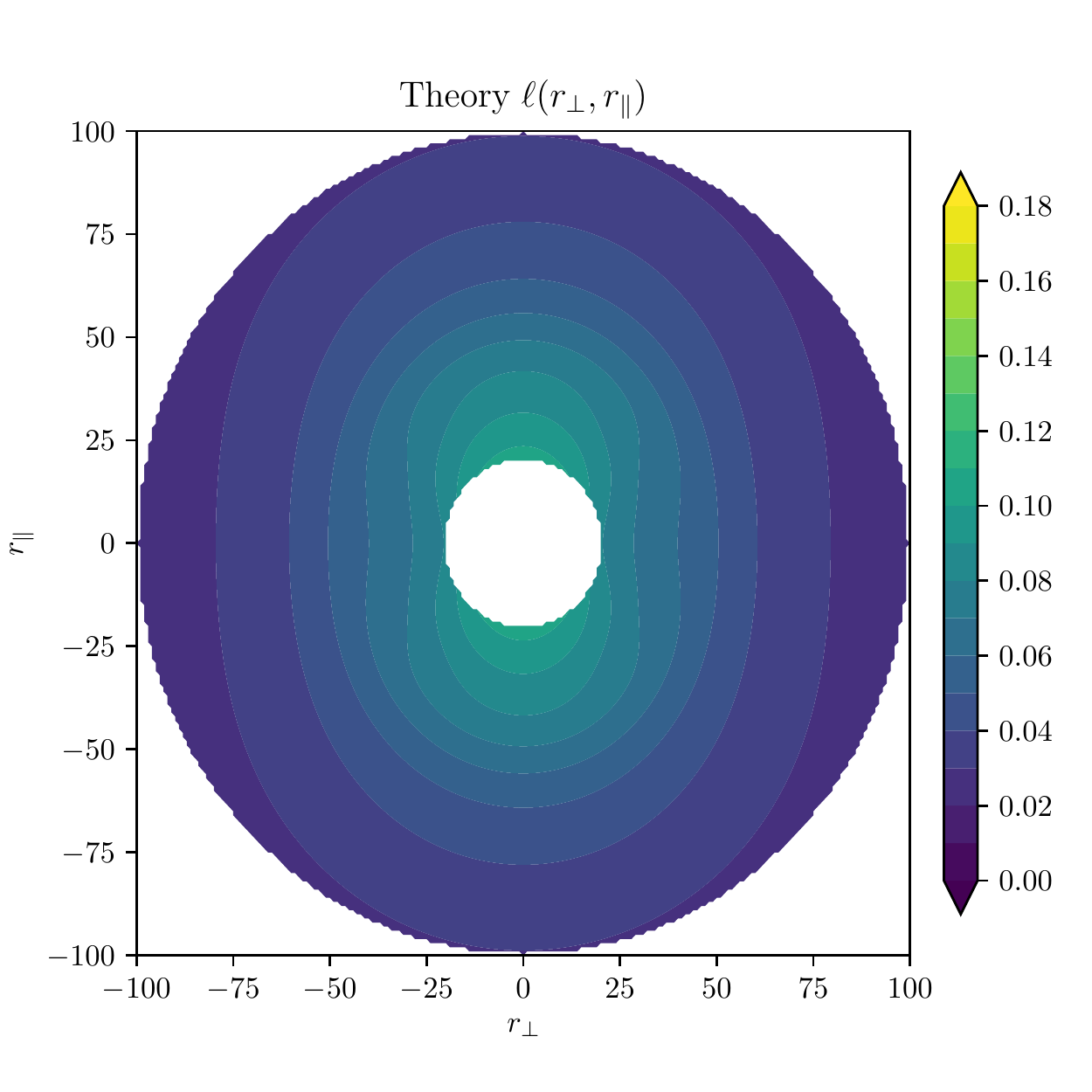}}
     \caption{Visualization of the 2-point correlation function (top row) and LCF (bottom row) from simulations (left column) and theory predictions (right column). Only scales with $20 < r < 100 \,h^{-1}\Mpc$ are shown. The 2-point correlation functions in the top panels both show the mild squashing along the line of sight that is characteristic of linear redshift-space distortions, but no visible signatures of the Fingers-of-God effect, which would typically be on smaller scales (inside the white circles). The LCF in the bottom panels show that peculiar velocities cause the phases to be slightly more correlated when the three points are along the line of sight.}
     \label{fig:onions}
\end{figure*}

Given these results from the power spectrum and LCF fits, we choose to use the model with Lorentzian FoG for our Fisher forecasts, since it has a lower minimum $\chi^2$ value for both the power spectrum and LCF, while also providing best-fit $\sigma_P$ and $\sigma_B$ that are more in agreement. In the following sections, we present the forecasts for both an optimistic $k_{\rm max} = 0.30 \,h\Mpc^{-1}$ ($r_{\rm min}=20\,h^{-1}\Mpc$) and a conservative $k_{\rm max} = 0.15 \,h\Mpc^{-1}$ ($r_{\rm min}=40\,h^{-1}\Mpc$). When the fiducial values of the parameters are set to their best-fit values, we will take these to be the best-fit values from the full range of scales, i.e.\ up to $k_{\rm max} = 0.3 \,h\Mpc^{-1}$, even for the forecasts where only the $k$ bins up to $k_{\rm max} = 0.15 \,h\Mpc^{-1}$ are used.

Lastly, in Figure \ref{fig:onions} we show the two-dimensional 2-point correlation function and LCF from both simulations and the best-fit model with Lorentzian FoG. In all panels, we only show contours in the regions where $r \geq 20 \, h^{-1}\Mpc$ to correspond to the range of scales that we use in the fits. The simulation 2PCF is the average of the 2PCF measured in 500 \textsc{l-picola} simulations using the direct pair counting method implemented in \texttt{nbodykit} \citep{Hand:2017pqn}, while the theoretical 2PCF is calculated by taking the inverse Fourier transform of the best-fit anisotropic power spectrum. In both the simulation and theoretical 2PCF panels, we see the characteristic Kaiser squashing along the line of sight that is indicative of linear redshift-space distortions. The non-linear Fingers-of-God effect resulting from virialized subhalos typically appears as a strong elongation along the line of sight, but we do not see it here because the halos in the \textsc{l-picola} simulations are larger than $6.7\times 10^{12} \,h^{-1}M_{\odot}$ and do not include subhalos. 

In the bottom panels of Figure \ref{fig:onions}, the simulation LCF is reconstructed from the average $Q_n$ measured in the simulations, while the theoretical LCF is reconstructed from the best-fit $Q_n$. In both the measured and theoretical LCF panels, we see a slight elongation along the line of sight which comes from the positive quadrupole $Q_2$. This higher correlation along the line of sight is a general feature of RSD in the LCF on these scales: it is present even for tree-level matter densities in redshift-space when we do not include halo biasing, shot noise, or the Fingers-of-God effect. The counter-intuitive enhancement of the LCF along the line of sight is qualitatively explained by the \textit{inverse} relationship between the LCF amplitude and the number density of filaments, as discussed Section 3.4 of \cite{Obreschkow:2012yb} and Section 3.3 of \cite{Eggemeier:2015ifa}. Briefly, the addition of spatially uncorrelated filamentary structure aligned with a particular direction reduces the LCF signal along the same direction due to the increased random phase noise from the different filaments. In the case of RSD, the Kaiser effect works in the opposite way: it reduces the apparent density of filamentary structure along the line of sight, and the resulting decreased phase noise boosts the LCF along the line of sight.


\subsection{Covariance matrices at fiducial cosmology}
\label{sec:cov}

Here we present the covariance matrix measured from simulations and compare it with the theoretical covariance matrix in Section~\ref{sec:theoretical_covariance}, evaluated at the best-fit model from the previous subsection.

\begin{figure*}
\includegraphics[width=\textwidth]{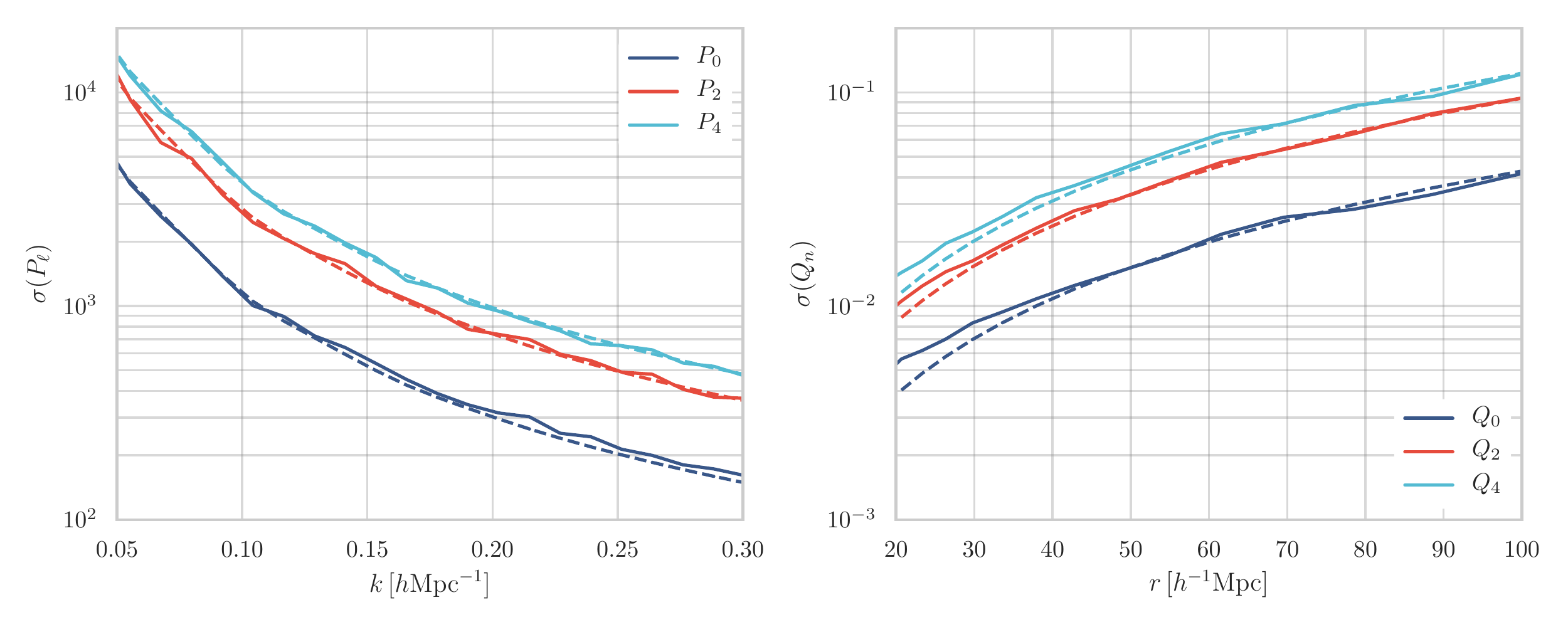}
\caption{The standard deviations from the simulations (solid lines) and theoretical predictions (dashed lines) are compared for the power spectrum (left panel) and LCF multipoles (right panel). The agreement between theory and simulations is better than $\approx 10$ per cent for all scales shown, except for the $r < 30 \, h^{-1}\Mpc$ bins of the LCF multipoles in the right panel.}
\label{fig:var}
\end{figure*}

Figure~\ref{fig:var} compares the standard deviations of the power spectrum and LCF multipoles measured from simulations (solid lines) with theoretical predictions (dashed lines). We find that the predictions for the power spectrum standard deviations match the measured ones to within approximately 10 per cent in all $k$ bins. For the LCF multipoles, the bins with $r > 30 \, h^{-1}\Mpc$ have standard deviations that are also predicted to within 10 per cent, but this grows up to 30 per cent for the smallest $r$ bin shown in Figure~\ref{fig:var}.

\begin{figure}
\centering
\includegraphics[width=0.5\textwidth, height=0.5\textwidth]{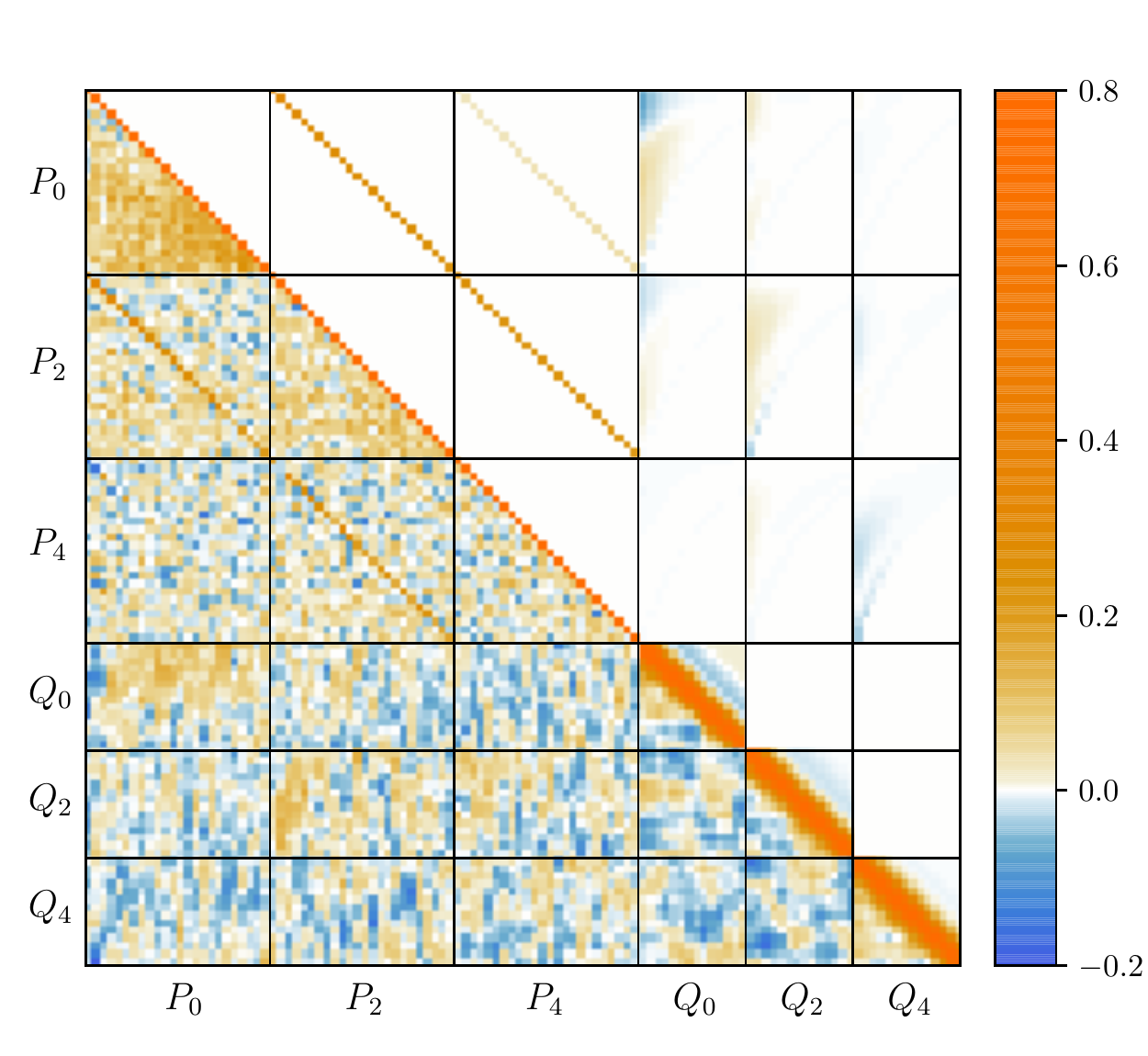}
\caption{Correlation coefficients for the simulation covariance (lower left) compared to the theoretical covariance (upper right). Bins for $k_{\rm max}=0.30 \,h\Mpc^{-1}$ are shown, and the grid of black vertical and horizontal lines delineate the different multipoles. Starting from the upper left corner of each matrix and going down or to the right, $k$ increases for $P_\ell$ and $r$ increases for $Q_n$.}
\label{fig:cov}
\end{figure}

In Figure~\ref{fig:cov} we compare the correlation coefficients of the simulation and theoretical covariance matrices,
\begin{equation}
	C_{ij}\equiv\frac{{\rm cov}[X_i,X_j]}{\sqrt{{\rm cov}[X_i,X_i]{\rm cov}[X_j, X_j]}}\, ,
\end{equation}
where $X_i$ are elements of the data vector containing $P_0(k),P_2(k), P_4(k), Q_0(r), Q_2(r)$ and $Q_4(r)$. We note that the theoretical covariance here is not internally fully consistent, since for the power spectrum multipoles and the LCF multipoles we include only the Gaussian contribution, while we include a non-Gaussian contribution for the cross-covariance between the power spectrum and the LCF multipoles. Still, Figure~\ref{fig:cov} shows that the theoretical covariance matrix captures some of the most prominent features in the simulation covariance matrix: for example, the covariance between different $P_\ell$ multipoles in the same $k$ bins, the covariance between neighboring $r$ bins of $Q_n$, and the cross-covariance between $P_0$ and $Q_0$.

\begin{figure*}
\centering
\includegraphics[width=\textwidth]{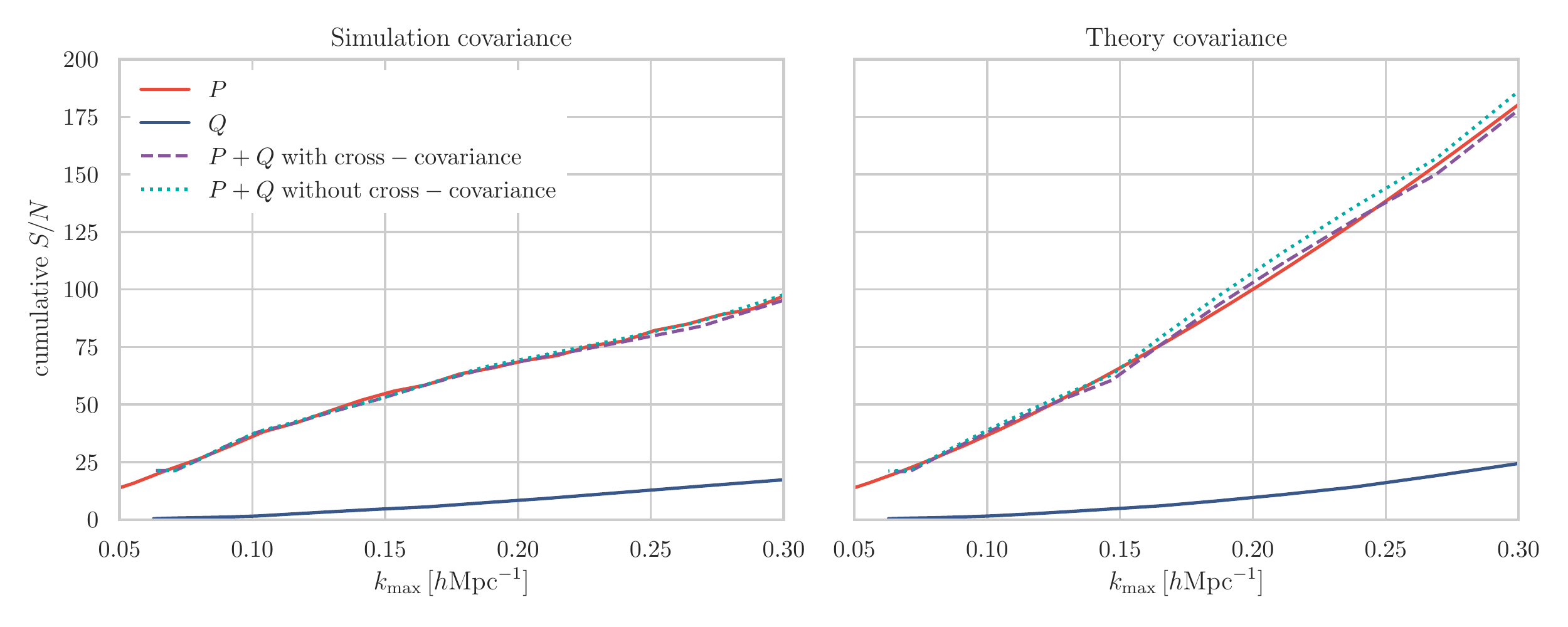}
\caption{The cumulative signal-to-noise ratio from the power spectrum (solid red lines), LCF (solid navy lines), their combination with cross-covariance (dashed purple lines), and their combination without cross-covariance (dotted teal lines) are compared. The left panel uses the simulated covariance matrix, while the right panel uses the theoretically predicted covariance matrix. The signal in both panels is the measured mean from the simulations, so the differences between the two panels is due only to the differences in the covariance matrices.}
\label{fig:SN}
\end{figure*}

In Figure~\ref{fig:SN}, we show the cumulative signal-to-noise ratio as a function of $k_{\rm max}$ for the power spectrum, LCF, and their combination. Each panel is computed using either the simulation covariance matrix on the left or the theoretically predicted covariance matrix on the right. In all cases, the signal is fixed to the average measurement from the simulations, so differences between the two panels are due only to differences in the covariance matrices. We see that the total signal-to-noise from the $P+Q$ combination begins to be overestimated by the theoretical covariance by more than 10 per cent for $k_{\rm max} \gtrsim 0.12 \, h\Mpc^{-1}$. We have also checked the importance of the cross-covariance between the power spectrum and LCF by comparing the signal-to-noise ratio using the full covariance matrix with the total signal-to-noise ratio without cross-covariance, $\frac{S}{N}(P+Q) = \sqrt{\frac{S}{N}(P)^2 + \frac{S}{N}(Q)^2}$. This comparison shows that the signal-to-noise does not appear to be sensitive to the cross-covariance between the power spectrum and LCF; in both panels of Figure~\ref{fig:SN}, neglecting the cross-covariance changes the total signal-to-noise for the combined probes by less than 5 per cent for all $k_{\rm max}$ shown.

Interestingly, the LCF multipoles do not appear to contribute noticeably to the total signal-to-noise, but we expect that how the available signal-to-noise is translated into parameter constraints depends on the modelling of the signal and the parameters under consideration \citep{Byun:2017fkz}. In the next section, we will present the results of Fisher forecasts based on the simulation data that explore the parameter constraints in more detail.


\section{Forecasts based on simulations}
\label{sec:fisher}

We present power spectrum and LCF Fisher forecasts which are based on our simulation data. We first present our benchmark forecast in Section~\ref{sec:benchmark}, which relies most heavily on the simulations and least on theoretical assumptions. Then in Sections~\ref{sec:theory_cov} and \ref{subsec:theory_veldisp}, we test the modelling of the covariance matrix and the velocity dispersions by comparing forecasts with more theoretical modelling to the benchmark forecast. We end this section by discussing how we obtain predictions for the galaxy bias that we will use for the survey forecasts in Section~\ref{sec:surveys}.

\subsection{Benchmark forecast}
\label{sec:benchmark}

\begin{table}
\centering
\begin{tabular}{cS[table-format=1.4]S[table-format=1.4]cS[table-format=1.4]S[table-format=1.4]c}
\toprule
\multicolumn{7}{c}{$k_{\rm max} = 0.15 \,h\Mpc^{-1}$} \\
\midrule
& {$P$} & \multicolumn{2}{c}{$P+Q$} 
	& {$\mathrm{Planck}+P$} & \multicolumn{2}{c}{$\mathrm{Planck}+P+Q$} \\ 
\cmidrule(lr){2-2} \cmidrule(lr){3-4} \cmidrule(lr){5-5} \cmidrule(lr){6-7}
{$b_1$} & 0.676 & 0.286 & (137\%) 
	& 0.0164 & 0.0160 & (2.1\%) \\
{$b_2$} & 0.571 & 0.271 & (10\%) 
	& 0.0754 & 0.0725 & (4.0\%) \\
{$f$} & 0.218 & 0.103 & (113\%) 
	& 0.0332 & 0.0327 & (1.3\%) \\
{$\sigma_8$} & 0.331 & 0.141 & (134\%) 
	& 0.00600 & 0.00599 & (0.1\%) \\
\midrule
\multicolumn{7}{c}{$k_{\rm max} = 0.30 \,h\Mpc^{-1}$} \\
\midrule
& {$P$} & \multicolumn{2}{c}{$P+Q$} 
	& {$\mathrm{Planck}+P$} & \multicolumn{2}{c}{$\mathrm{Planck}+P+Q$} \\ 
\cmidrule(lr){2-2} \cmidrule(lr){3-4} \cmidrule(lr){5-5} \cmidrule(lr){6-7}
{$b_1$} & 0.120 & 0.0779 & (54\%) 
	& 0.0139 & 0.0135 & (2.8\%) \\
{$b_2$} & 0.0962 & 0.0683 & (41\%) 
	& 0.0220 & 0.0216 & (1.7\%) \\
{$f$} & 0.0458 & 0.0340 & (35\%) 
	& 0.0171 & 0.0167 & (2.1\%) \\
{$\sigma_8$} & 0.0577 & 0.0382 & (51\%) 
	& 0.00597 & 0.00593 & (0.7\%) \\
\bottomrule
\end{tabular}
\caption{Fisher forecasted constraints on $b_1$, $b_2$, $f$ and $\sigma_8$ from the benchmark forecast for both $k_{\rm max} = 0.15 \,h\Mpc^{-1}$ and $0.30 \,h\Mpc^{-1}$. All constraints are marginalised over the $\sigma_P$ and $\sigma_B$ velocity dispersions. The per cent values in parentheses are the reduction in the forecasted errors as a result of including the LCF multipoles. The last three columns show the constraints when a Planck prior on $\sigma_8$ is included.}
\label{tab:fisher}
\end{table}

\begin{figure*}
    \centering
    \subfloat{\includegraphics[width=0.5\textwidth]{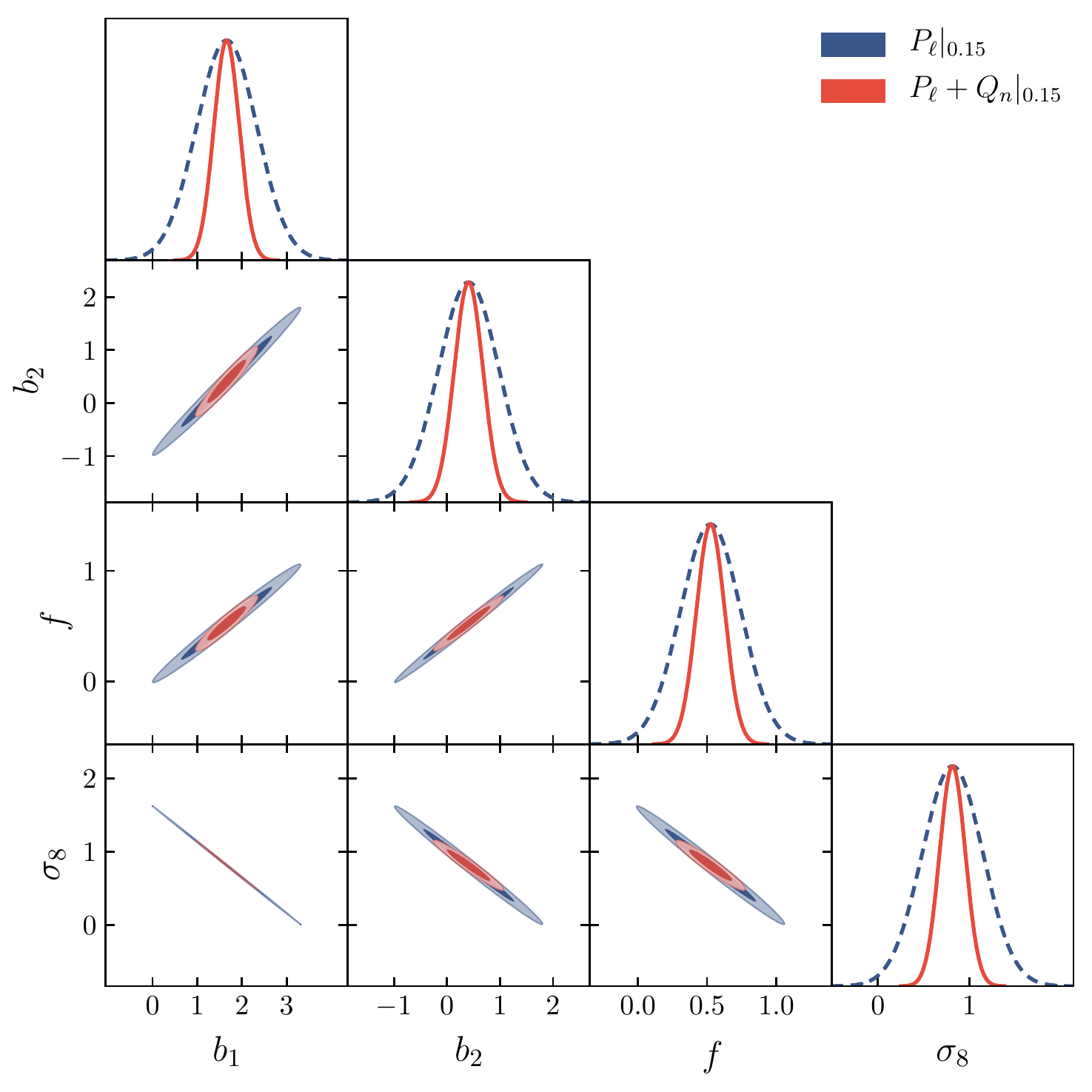}}
    \subfloat{\includegraphics[width=0.5\textwidth]{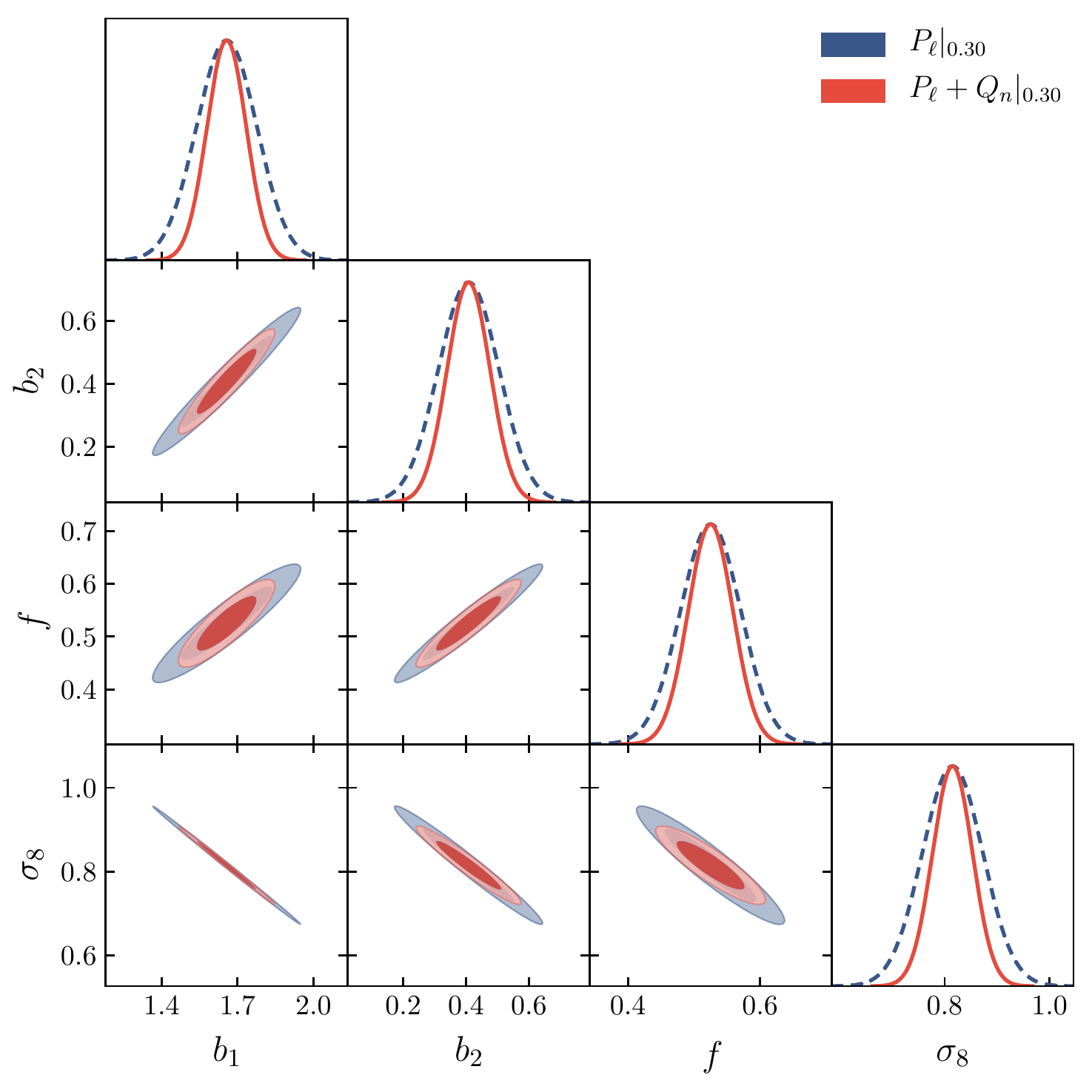}}
    \caption{Constraints from the benchmark forecast using the power spectrum only (blue dashed) and the power spectrum combined with the LCF (red solid) for $k_{\rm max} = 0.15 \,h\Mpc^{-1}$ (left) and $0.30 \,h\Mpc^{-1}$ (right). These constraints are the same as those in Table~\ref{tab:fisher} (without Planck). All constraints are marginalised over the $\sigma_P$ and $\sigma_B$ velocity dispersions.}
    \label{fig:halo_fisher_woPlanck}
\end{figure*}

In this work, we use the Fisher forecasting method to estimate parameter constraints \citep{Tegmark:1996bz}. Assuming the likelihood is a multivariate Gaussian and the data covariance does not vary with the parameters of interest, the Fisher information matrix is
\begin{equation}
	\mat{F} = \frac{\partial\vec{D}}{\partial\vec{\theta}}^T \cdot \mat{C}^{-1} \cdot \frac{\partial\vec{D}}{\partial\vec{\theta}},
\end{equation}
where our data vector $\vec{D}$ is comprised of the power spectrum and LCF multipoles, $\vec{\theta}$ are the parameters of interest, and $\mat{C}$ is the data covariance matrix. Both the partial derivatives and the data covariance are evaluated at a fiducial cosmology.\footnote{Our fiducial flat $\Lambda$CDM cosmology is given by $\Omega_{\rm b} = 0.0486$, $\Omega_{\rm m} = 0.3089$, $\sigma_8 = 0.8159$, $n_{\rm s} = 0.9667$, and $H_0 = 67.74 \,{\rm km}\,{\rm s}^{-1} \Mpc^{-1}$.} Then $\mat{F}^{-1}$ is the forecasted parameter covariance matrix.

For the data vector, we use the same data bins that were used in Section~\ref{sec:means}. This means that for the power spectrum forecast, we include $P_0$, $P_2$ and $P_4$, with all $k$-modes between $k_{\rm min}=0.018\,h\Mpc^{-1}$ and $k_{\rm max}=0.15 \,h\Mpc^{-1}$ or $k_{\rm max}=0.3 \,h\Mpc^{-1}$. Equivalently, for the LCF forecast, we include all separations between $r_{\rm min}=40\,h^{-1}\Mpc$ or $r_{\rm min}=20\,h^{-1}\Mpc$ and $r_{\rm max}=100\,h^{-1}\Mpc$.

We calculate all of our Fisher matrices for six parameters, $\vec{\theta} = (f,\sigma_8,b_1,b_2,\sigma_P,\sigma_B)$, where the $\sigma_P$ and $\sigma_B$ velocity dispersions are considered nuisance parameters and marginalised over. Fiducial values of these parameters are evaluated at the known cosmology or best-fit parameter values discussed in Section~\ref{sec:means}. All other cosmological parameters are kept fixed to their fiducial values, which match those of the simulations, since our goal is to determine how the LCF multipoles can break degeneracies between $f$, $\sigma_8$, $b_1$ and $b_2$. In $\Lambda$CDM, $f$ is fully determined by $\Omega_m$. Therefore, by keeping $\Omega_m$ fixed and varying $f$, we promote $f$ to a free parameter that is used to test models beyond $\Lambda$CDM. We evaluate the derivatives of the power spectrum and LCF multipoles with respect to the parameters numerically. 

The covariance matrix measured from simulations is representative of a survey at redshift $z=0$ that has volume $V_{\rm sim} = 0.13 \, h^{-3} \mathrm{Gpc}^3$, but upcoming surveys will be much larger than this by covering large sky areas in multiple redshift bins. Therefore, for the forecasts in this section, we rescale the simulation covariance to a comoving volume of $V = 3 \, h^{-3} \mathrm{Gpc}^3$, which approximates more closely the effective volume of a single redshift slice of width $\Delta z = 0.1$ from upcoming surveys at $z \approx 0.75$. More explicitly, the inverse covariance matrix we use in the benchmark forecast is $\mat{C}^{-1} = \hat{\mat{C}}^{-1} \, V/V_{\rm sim}$, where $\hat{\mat{C}}^{-1}$ is the inverse covariance matrix that we estimate from the simulations.

The results from the benchmark forecast for both $k_{\rm max} = 0.15 \,h\Mpc^{-1}$ and $0.30 \,h\Mpc^{-1}$ are shown in Figure~\ref{fig:halo_fisher_woPlanck} and Table~\ref{tab:fisher}. The per cent values in the table show the amount of improvement in the forecasted error that is brought by the LCF multipoles. We see that the constraints from the LCF give significant improvements to the power spectrum for both $k_{\rm max}$, but the improvement is significantly larger for $k_{\rm max} = 0.15 \,h\Mpc^{-1}$. This is due to the fact that if $k_{\rm max}$ is smaller, the power spectrum is more degenerate in the parameters, so the LCF has a larger opportunity to help by breaking some of these degeneracies. In the case where $k_{\rm max}$ is higher, the mild non-linearities in the power spectrum at smaller scales also help in breaking the degeneracies, so as a consequence, the LCF gives less improvement. Still, for both $k_{\rm max}$ values, adding the LCF multipoles provides noticeable improvements in constraining these parameters.

\begin{figure*}
\centering
\includegraphics[width=\textwidth]{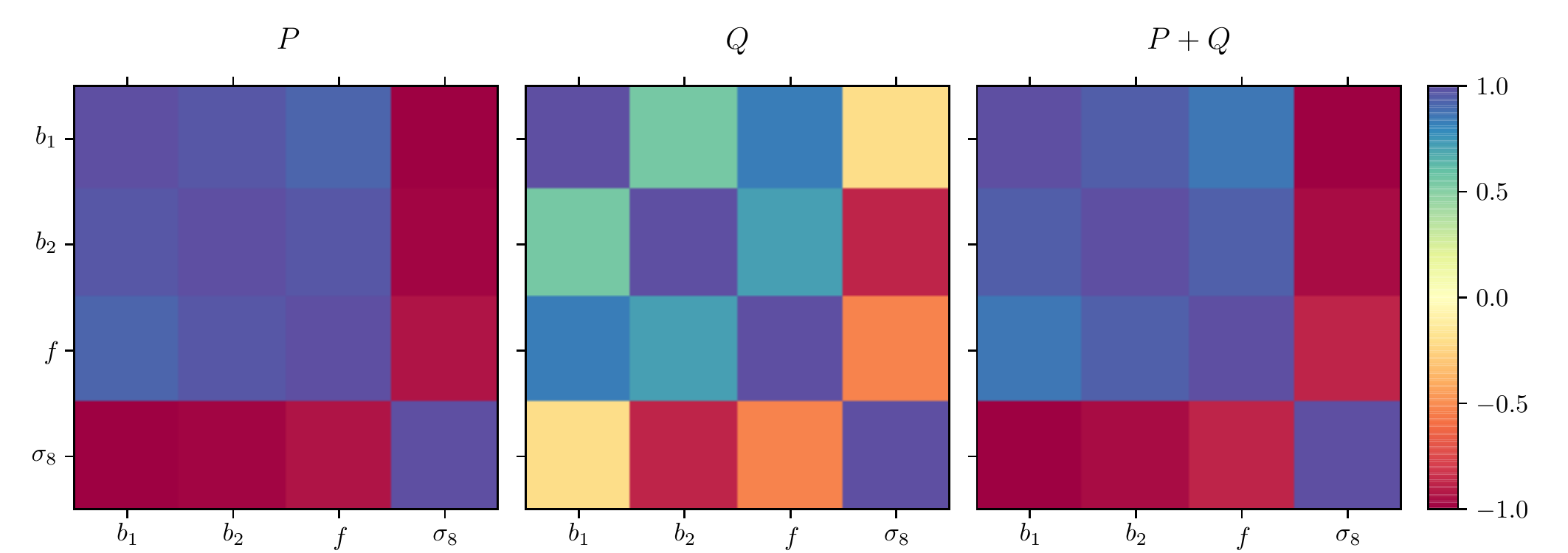}
\caption{Correlation coefficients of the parameter covariance matrix for the power spectrum (left), LCF (center) and their combination (right) for $k_{\rm max} = 0.30 \,h\Mpc^{-1}$. Comparing the left and center matrices shows that the LCF multipoles exhibit very different degeneracies between parameters, but because the LCF constraint is much weaker overall, the final joint constraint in the right panel has correlation coefficients that are similar to those from the power spectrum except the correlation coefficients are slightly closer to zero.}
\label{fig:halo_fisher_corrmats_kmax0.3}
\end{figure*}

In Figure \ref{fig:halo_fisher_corrmats_kmax0.3}, we show the correlation coefficients of the parameter covariance matrix from the power spectrum, LCF and their combination for $k_{\rm max} = 0.3 \,h\Mpc^{-1}$. Compared to the power spectrum (left panel), which has strong degeneracies for all pairs of parameters, the LCF (central panel) exhibits a very different pattern of correlations. In particular, as expected, $\sigma_8$ is much less degenerate with $f$ and $b_1$ in the LCF than in the power spectrum, showing that the LCF contains new, complementary information with respect to the power spectrum. The amount of new information is, however, lessened by the fact that the overall constraining power of the power spectrum is much stronger than that of the LCF. As a consequence, the correlation coefficients from the joint constraint (right panel) are more similar to the power spectrum than to the LCF, except now with less severe degeneracies between parameters. Another way to see that the main advantage of the LCF is to break degeneracies is by forecasting the constraints on $(f,\sigma_8, \sigma_P, \sigma_B)$ while fixing the values of $b_1$ and $b_2$. In this case, we find that the improvement from the LCF is negligible---less than 3 per cent for both $k_{\rm max}$ values. This is due to the fact that when $b_1$ and $b_2$ are fixed, the power spectrum multipoles are sensitive to different combinations of $\sigma_8$ and $f\sigma_8$. As a consequence, the power spectrum multipoles alone can break the degeneracy between $f$ and $\sigma_8$, and there is little benefit to including the LCF multipoles.

\begin{figure}
\centering	
\includegraphics[width=\textwidth]{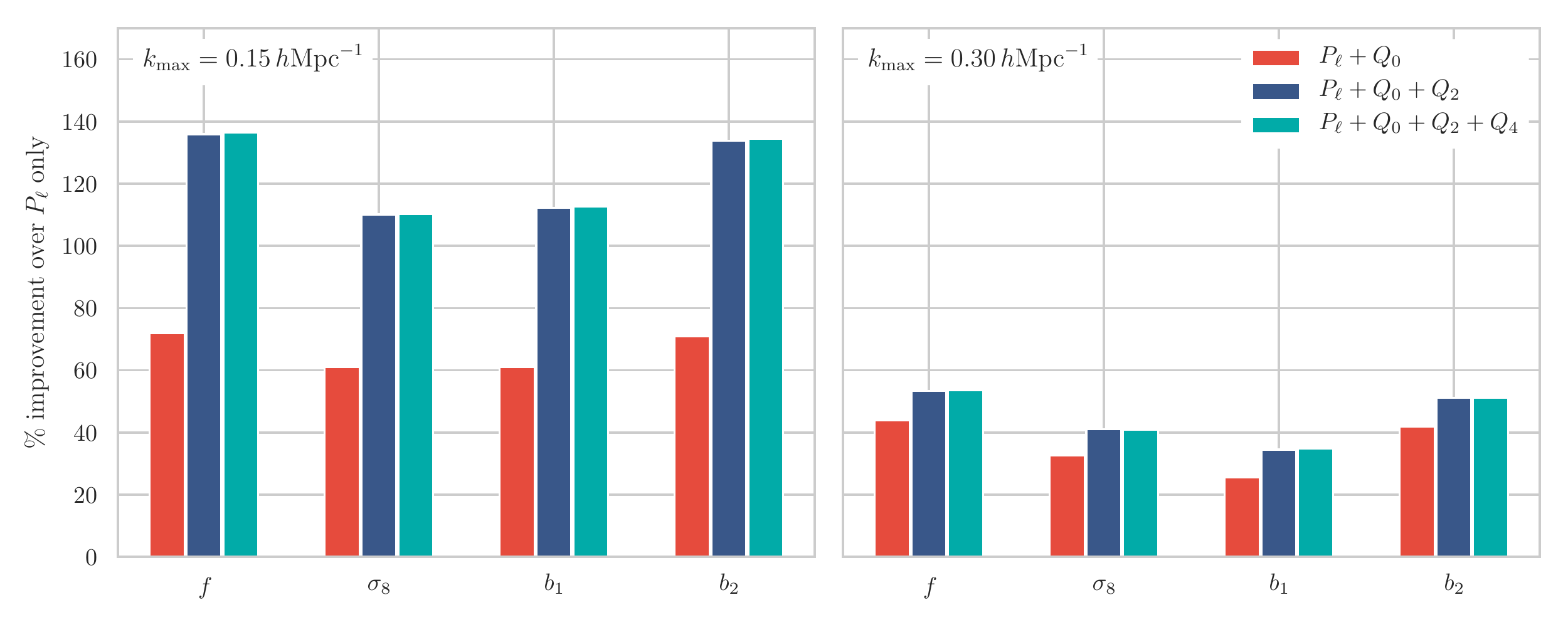}
\caption{Percentage improvement in the forecasted constraints from including LCF multipoles one by one, relative to the $P_\ell$-only forecast, when $k_{\rm max} = 0.15 \,h\Mpc^{-1}$ (left panel) or $0.30 \,h\Mpc^{-1}$ (right panel). For each parameter, the bars show the improvement in the forecasted constraint from adding only the monopole $Q_0$ (red bar on the left), adding the monopole $Q_0$ and quadrupole $Q_2$ (blue bar in the middle), and adding all three LCF multipoles $Q_0$, $Q_2$ and $Q_4$ (turquoise bar on the right), to the $P_\ell$ multipoles. Most of the constraining power of the LCF comes from the monopole $Q_0$ and the quadrupole $Q_2$, while adding the hexadecapole $Q_4$ provides less than 1 per cent of additional improvement.}
\label{fig:halo_fisher_Qn_contribs}
\end{figure}

What impact does the cross-covariance between the power spectrum and LCF multipoles have on the parameter constraints? In Section~\ref{sec:cov}, we saw that the impact of the cross-covariance on the signal-to-noise ratio was small. Similarly, the impact of the cross-covariance on the benchmark forecast is also small. For both $k_{\rm max}$ values, we confirm that neglecting the cross-covariance between the power spectrum and the LCF changes the forecasted constraints minimally, by less than 7 per cent. This confirms the expectation that an estimator targeted at measuring the phases of the density and velocity fields is minimally correlated with the power spectrum, which is only sensitive to the amplitude of these fields.  As a consequence, the information in the LCF multipoles is minimally redundant with that in the power spectrum multipoles.

\begin{figure*}
\centering
\includegraphics[width=0.5\textwidth]{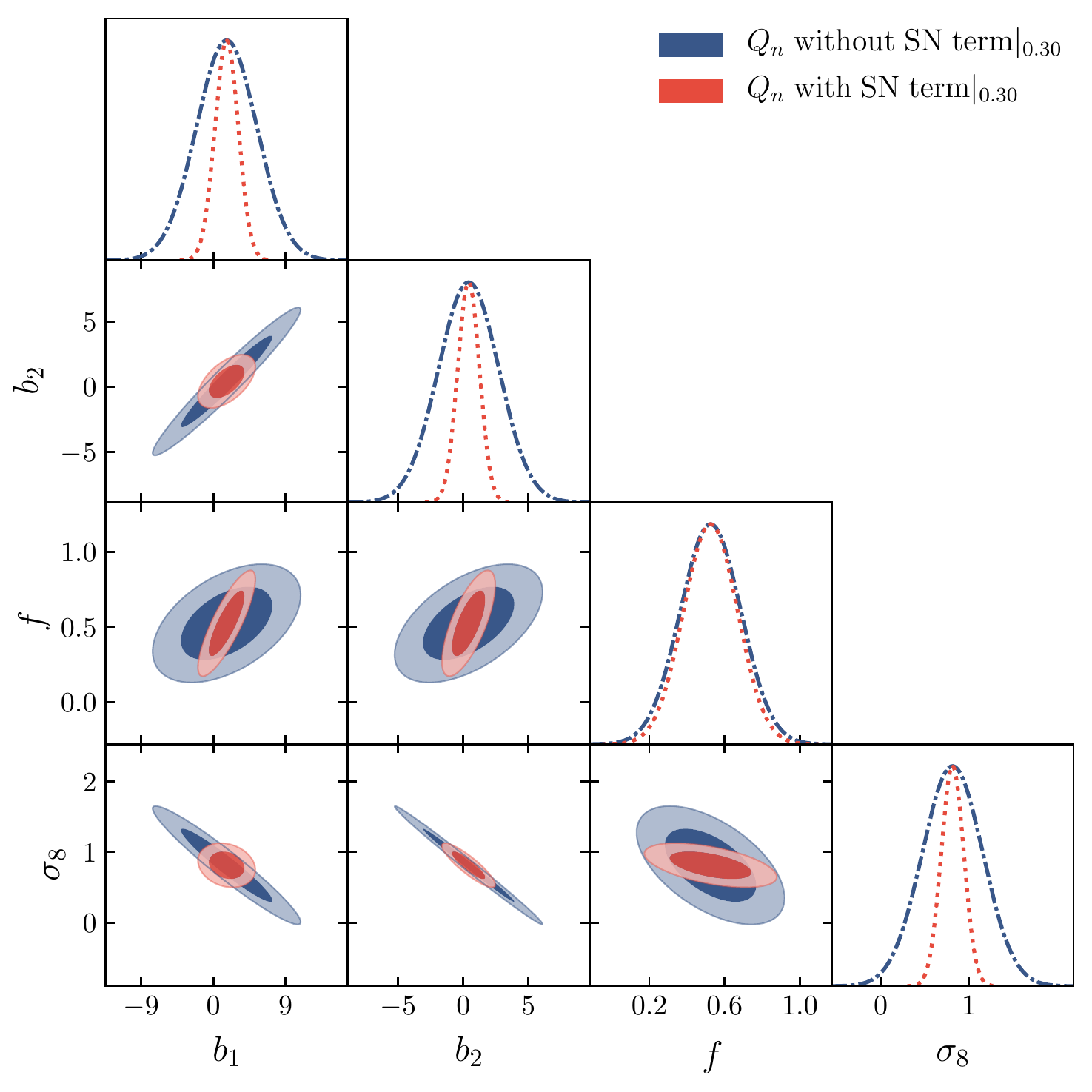}
\caption{Fisher forecasted constraints using the only the LCF multipoles for $k_{\rm max} = 0.30 \,h\Mpc^{-1}$. The blue contours show the constraints when only the effective LCF term is used as the observable, and the shot noise term is not included. The red contours show the constraints when the total LCF, including the shot noise term, is used. The latter case in red gives stronger constraints, because the cosmology-dependence of the shot noise term provides additional constraining power.}
\label{fig:halo_fisher_sn_effect}
\end{figure*}

To determine which multipoles of the LCF are most relevant for breaking the degeneracies between parameters, we compare forecasts where the LCF multipoles are included cumulatively: first, only the monopole of the LCF is combined with the power spectrum multipoles, then the LCF monopole and quadrupole are added, and finally all three LCF multipoles are included. The results in Figure \ref{fig:halo_fisher_Qn_contribs} show that most of the improvement comes from the first two multipoles, $Q_0$ and $Q_2$. In the left panel of the figure, we find that for $k_{\rm max} = 0.15 \,h\Mpc^{-1}$ combining only the monopole of the LCF, $Q_0$, with the power spectrum gives an improvement of 60 to 70 per cent, depending on the parameter, with respect to the power spectrum alone. Further adding the quadrupole of the LCF, $Q_2$, nearly doubles the per cent improvement, while adding the hexadecapole, $Q_4$, changes the constraints by less than 1 per cent. The right panel of Figure \ref{fig:halo_fisher_Qn_contribs} is for $k_{\rm max} = 0.30 \,h\Mpc^{-1}$, where we find again that nearly all of the improvement from the LCF multipoles is contained in $Q_0$ and $Q_2$. In this case, $Q_0$ gives an improvement of 25 to 45 per cent and further including $Q_2$ adds another 10 per cent improvement. Again, including $Q_4$ does not strengthen the constraints any further.

As discussed in Section~\ref{sec:shot noise}, the forecasts can be done either with or without the shot noise term in the LCF. Our benchmark forecast includes the shot noise term, which implicitly assumes that the cosmology-dependence of the shot noise is modelled accurately enough to extract information from it. On the other hand, calculating the forecast \textit{without} the shot noise term requires both assuming that the shot noise is modelled accurately enough to isolate the effective term, and that the true cosmology is already known very accurately. When we compare the forecasts for the joint power spectrum and LCF multipoles with and without the shot noise term in the LCF, we find that the constraints are almost identical. This is due to the fact that most of the constraining power comes from the power spectrum. On the other hand, if only the LCF multipoles are used to constrain the parameters, then the inclusion of the shot noise term makes a noticeable difference, as shown in  Figure~\ref{fig:halo_fisher_sn_effect}. The constraints are stronger for all parameters if the shot noise term is included. This is what we would expect, since when the shot noise term is included, we also get information from the cosmology-dependence in the LCF shot noise.

Finally, we check the impact of including a CMB prior from the Planck 2018 results \citep{Aghanim:2018eyx}. The forecasts with a Gaussian $1\sigma$ prior on $\sigma_8$ of $\Delta\sigma_8 = 0.0060$ are shown in the last three columns of Table~\ref{tab:fisher}.\footnote{We use the data combination called \texttt{base\_plikHM\_TTTEEE\_lowl\_lowE\_lensing\_post\_BAO}, which is the baseline model in Section 2.18 of \href{https://wiki.cosmos.esa.int/planck-legacy-archive/images/4/43/Baseline_params_table_2018_68pc_v2.pdf}{\texttt{https://wiki.cosmos.esa.int/planck-legacy-archive/images/4/43/Baseline\_params\_table\_2018\_68pc\_v2.pdf}}.} In this case, the improvement from the LCF multipoles is very small, less than 4 per cent, because the parameter degeneracies in the power spectrum are broken by the external constraint on $\sigma_8$, such that there is less opportunity for the LCF to further improve the constraints. As discussed in the introduction, however, an important benefit of extracting additional information from the spatial distribution of galaxies to break the parameter degeneracies is that the combination of large-scale structure data is model-independent. In contrast, the combination of large-scale structure data with CMB analyses is model-dependent, since CMB data constrains the primordial amplitude of perturbations, and in order to translate this into a prior on $\sigma_8$, a cosmological model, such as $\Lambda$CDM and general relativity, must be assumed. The growth rate $f$ measured in this way is therefore not model-independent and cannot be consistently used to test models beyond $\Lambda$CDM.\footnote{In this work we consider constraints on the growth in multiple redshift bins to allow for any $z$-dependence. However, CMB data can be used to constrain beyond $\Lambda$CDM models given a model-specific parametrisation of the growth rate.}


\subsection{Theoretical covariance matrices}
\label{sec:theory_cov}

To calculate Fisher forecasts for upcoming surveys, we require theoretical predictions for the forecasting ingredients that we have so far measured in simulations. Previously, we measured the full covariance matrix in simulations and fitted simulation data to find our fiducial values of $b_1$, $b_2$, $\sigma_P$ and $\sigma_B$. Here, and in the rest of Section~\ref{sec:fisher}, we examine how the benchmark forecasts presented in Section~\ref{sec:benchmark} change when these quantities are replaced with theoretical predictions.

\begin{figure*}
\centering
\includegraphics[width=\textwidth]{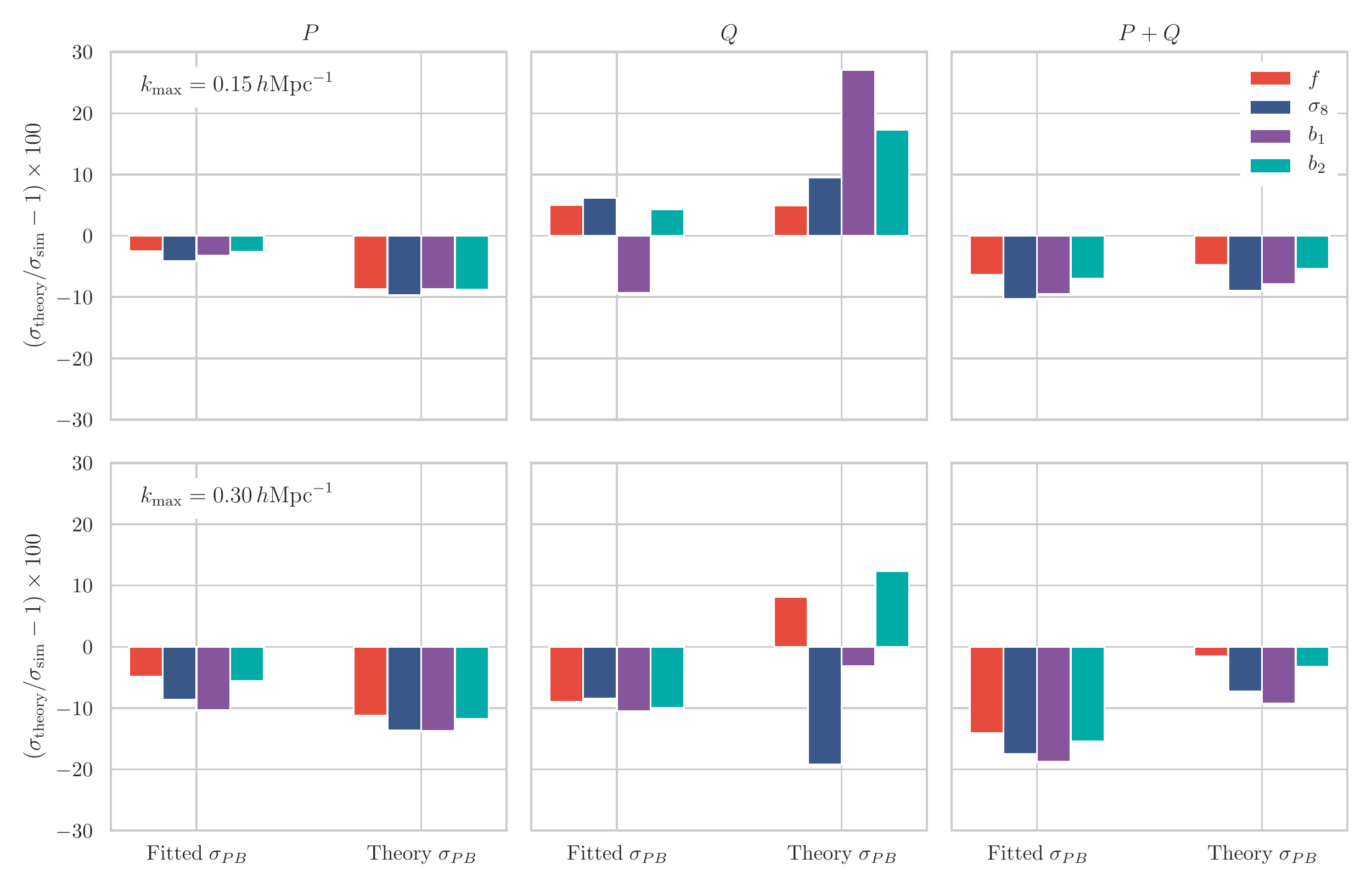}
\caption{Comparison of Fisher forecasted constraints using the simulation covariance matrix vs the theoretical covariance matrix. The bar heights correspond to the per cent change in the forecasted parameter errors on $f$ (red), $\sigma_8$ (blue), $b_1$ (purple) and $b_2$ (turquoise) as a result of using the theoretical covariance matrix. The constraints from the power spectrum only, LCF only and their combination are in the left, center and right columns, respectively. Constraints in the top row are for $k_{\rm max} = 0.15 \,h\Mpc^{-1}$, while those in the bottom row are for $k_{\rm max} = 0.30 \,h\Mpc^{-1}$. Within each panel, the bars labelled ``Fitted $\sigma_{PB}$'' use the fitted values of $\sigma_P$ and $\sigma_B$ from Section~\ref{sec:means}, while the bars labelled ``Theory $\sigma_{PB}$'' use the values of $\sigma_P=\sigma_B$ predicted by linear theory in eq.~\eqref{eq:sigmav2}.}
\label{fig:halo_theorycov}
\end{figure*}

First, we check how well our theoretical model for the covariance matrix can recover the benchmark forecast. We compute the Fisher matrix as in the benchmark forecast, with the only difference that we use the theoretically predicted covariance matrix (shown in the upper right part of Figure~\ref{fig:cov}) instead of the covariance matrix estimated from simulations. The results are shown in Figure~\ref{fig:halo_theorycov} as the bars labelled ``Fitted $\sigma_{PB}$'' (in the left side of each panel) to indicate that the forecast is evaluated at the best-fit values of $\sigma_P$ and $\sigma_B$ from Section~\ref{sec:means}. The height of the bars corresponds to the per cent difference in the forecasted constraints relative to the benchmark forecast with the simulation covariance matrix. For $k_{\rm max} = 0.30 \,h\Mpc^{-1}$ (bottom row in the figure), we find that using the theoretical covariance instead of the simulated one for the power spectrum-only constraints underestimates the forecasted parameter error by up to 10 per cent. A similar difference is seen for the LCF-only constraints. For the power spectrum-LCF joint constraints, using the theory covariance matrix underestimates the constraints by up to 20 per cent. This larger mismatch for the joint constraints does not seem to be due to our model for the cross-covariance; we have checked that ignoring the cross-covariance between the power spectrum and LCF in the theoretical covariance matrix only changes the constraints by less than 3 per cent for both $k_{\rm max}$ values, which is similar to the behaviour we found for the simulation covariance matrix in the benchmark forecast. If $k_{\rm max} = 0.15 \,h\Mpc^{-1}$ (top row of Figure~\ref{fig:halo_theorycov}), the agreement is better: using the theoretical covariance matrix changes the power spectrum-only constraints by 4 per cent, the LCF-only constraints by 10 per cent, and the joint constraints by 10 per cent. 

\begin{figure*}
\centering
\includegraphics[width=\textwidth]{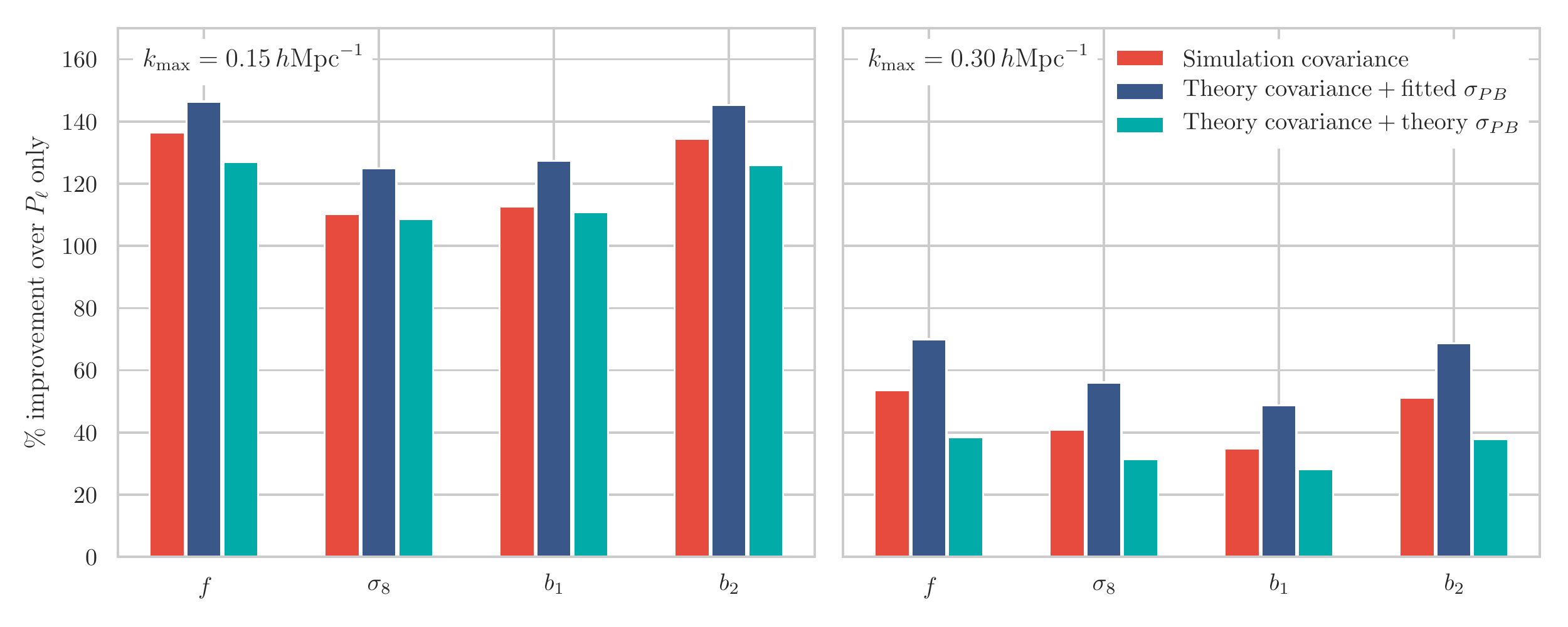}
\caption{Comparison of the improvement brought by the LCF multipoles in the Fisher forecasted constraints using the simulation covariance matrix vs the theoretical covariance matrix. The bar heights correspond to the per cent improvement in the forecasted parameter error from including the LCF multipoles, relative to the case with only the power spectrum multipoles. The left panel is for $k_{\rm max} = 0.15 \,h\Mpc^{-1}$, while the right panel is for $k_{\rm max} = 0.30 \,h\Mpc^{-1}$. For each parameter on the $x$-axis, the constraints are shown using the simulation covariance (red), the theoretical covariance with fitted values of $\sigma_P$ and $\sigma_B$ from Section~\ref{sec:means} (blue), and the theoretical covariance with  values of $\sigma_P=\sigma_B$ predicted by linear theory in eq.~\eqref{eq:sigmav2} (turquoise).}
\label{fig:halo_theorycov_improvement}
\end{figure*}

Figure~\ref{fig:halo_theorycov_improvement} shows the per cent improvement that is gained in the forecasted constraints by adding the LCF to the power spectrum. The red bars correspond to the same per cent improvements that are in Table~\ref{tab:fisher} (without the Planck prior). The blue bars show that when the theoretical covariance is used with the fitted $\sigma_P$ and $\sigma_B$, the per cent improvement is overestimated by 9 to 14 per cent when $k_{\rm max} = 0.15 \,h\Mpc^{-1}$ and by 14 to 17 per cent when $k_{\rm max} = 0.30 \,h\Mpc^{-1}$.

The comparisons between the theoretical and simulation covariance matrices in this section show that the modelling for the covariance matrices presented in Section~\ref{sec:theoretical_covariance} is accurate enough to return forecasted constraints to within $\sim 10$ per cent of the benchmarks for $k_{\rm max} = 0.15 \,h\Mpc^{-1}$ and within $\sim 20$ per cent of the benchmarks for $k_{\rm max} = 0.30 \,h\Mpc^{-1}$. We note that the comparisons in this work are necessarily done at redshift $z=0$ and for the halo catalogs which we are able to identify with the \textsc{l-picola} simulations available to us. The modelling may perform differently for different number densities and halos, but in general we would expect that the theoretical modelling becomes more accurate at the higher redshifts that are more relevant for upcoming surveys, since non-linearities are less important at high redshift. In the next section, we consider the impact of assuming the linear predictions for the fiducial velocity dispersions, in place of the fitted values of $\sigma_P$ and $\sigma_B$ that we have used here.

\subsection{Theoretical velocity dispersions}
\label{subsec:theory_veldisp}

In addition to theoretical predictions for the covariance matrices, the Fisher forecasts for upcoming surveys will also require fiducial values for the velocity dispersions $\sigma_P$ and $\sigma_B$, so we now check the impact of using the linear theory prediction for the velocity dispersions, rather than their fitted values. In Section~\ref{sec:means}, we found that the best-fit $\sigma_P$ and $\sigma_B$ values in the Lorentzian FoG model are consistent with each other and with the linear prediction from eq.~\eqref{eq:sigmav2}, which at $z=0$ is $\sigma_P = 4.5 \,h^{-1}\mathrm{Mpc}$. Therefore, in our forecast with theoretical velocity dispersions, we choose as fiducial values $\sigma_P = \sigma_B = 4.5 \,h^{-1}\mathrm{Mpc}$, but we still treat these as two separate nuisance parameters which vary independently. We note that this requires not only changing where in the parameter space the Fisher derivatives are evaluated, but we also recompute the full theoretical covariance, which depends on the fiducial $\sigma_P$ and $\sigma_B$.

We compare the constraints with the benchmark forecast in Figure~\ref{fig:halo_theorycov} as the bars labeled ``Theory $\sigma_{PB}$'' (right side of each panel) to indicate that the forecast is evaluated at the values of $\sigma_P$ and $\sigma_B$ predicted by linear theory. For $k_{\rm max} = 0.3 \,h\Mpc^{-1}$ (bottom row of the figure), we find that the power spectrum-only constraints with the theoretical covariance underestimates the forecasted parameter error from simulated covariances by up to 15 per cent, and the LCF-only constraints agree to within 20 per cent. For the joint power spectrum-LCF constraints, using the theory covariance makes the constraints match to within 10 per cent. If $k_{\rm max} = 0.15 \,h\Mpc^{-1}$ (top row of the figure), the agreement is roughly similar: within 10 per cent for power spectrum-only constraints, 30 per cent for LCF-only constraints, and 10 per cent for the joint constraints. Figure~\ref{fig:halo_theorycov_improvement} shows that for both $k_{\rm max}$, the theory covariance with the theory prediction for the velocity dispersions returns a per cent improvement that is within a 12 per cent difference with the simulation forecasts. 

These results show a remarkable agreement with the benchmark forecast, considering that they are different by both covariance matrix modelling and fiducial values of $\sigma_P$ and $\sigma_B$. We consider this agreement good enough for our purposes of forecasting constraints from future surveys, and in particular, estimating the benefit of combining the LCF multipoles with the power spectrum, so we proceed to use the theoretical covariance matrix modelling with the linear predictions for the velocity dispersions for the survey forecasts in Section~\ref{sec:surveys}.

We note that we also calculated a forecast where the two velocity dispersions $\sigma_P$ and $\sigma_B$ were treated as if they were the same single nuisance parameter, following the discussion in \cite{Hashimoto:2017klo} and \cite{Yankelevich:2018uaz}. This results in much stronger constraints---the per cent improvement from the LCF is roughly 2 to 6 times larger. However, since this requires a very strong assumption about the relationship between the Fingers-of-God damping factors in the power spectrum and bispectrum, we do not make this assumption in this work.


\subsection{Theoretical galaxy bias}
\label{subsec:theory_bias}

We now briefly discuss how we obtain theoretical predictions for $b_2$ when performing the survey forecasts presented in the next section. For dark matter halos, $b_1$ and $b_2$ depend on the halo mass, and fits for $b_2$ as a function of $b_1$, calibrated to N-body simulations, have been presented in \cite{Lazeyras:2015lgp} and \cite{Hoffmann:2016omy}. In this work, we use the fit from \cite{Lazeyras:2015lgp} to obtain our fiducial values of the quadratic bias: $b_2 = 0.412 - 2.143 b_1 + 0.929 b_1^2 + 0.008 b_1^3$. In general, this relation is not guaranteed to hold for galaxies, and to predict the galaxy $b_1$ and $b_2$ from the halo biases requires modeling how galaxies populate halos using prescriptions such as subhalo abundance matching or a halo occupation distribution. In \cite{Yankelevich:2018uaz}, the latter was used to calculate the galaxy $b_2$ for the $H\alpha$ galaxies that will be observed by Euclid. In that work, the galaxy $b_2$ was very well approximated by the halo $b_2$, which implies that the galaxy bias is insensitive to the details of the HOD. Our forecasts in this work will assume this is also the case for the DESI galaxy samples, and we leave it to future work to include more precise modeling of the galaxy-halo connection for these surveys.

We note that when the fit for $b_2$ is applied to the best-fit $b_1$ from the \textsc{l-picola} simulations in Section~\ref{sec:means}, we obtain a value of $b_2 = -0.55$, which is not in agreement with our best-fit $b_2 = 0.41$. This is most likely due to the fact that our halo catalogs from the \textsc{l-picola} simulations contain all halos above a minimum halo mass of $6.7\times 10^{12}h^{-1}M_{\odot}$, so our fitted values of the halo biases are effective values that cover a large population of halos of different masses, whereas the fit derived in \cite{Lazeyras:2015lgp} has been calibrated on halos that fall within narrow mass bins. 


\section{Forecasts for upcoming surveys}
\label{sec:surveys}

In this section, we apply our forecasting method to the upcoming DESI and Euclid galaxy surveys using the theoretical predictions that were validated in the previous section.  

For DESI, we consider the Bright Galaxy Sample (BGS), Emission Line Galaxies (ELGs), Luminous Red Galaxies (LRGs), and quasars (QSOs) with $14{,}000 \,{\rm deg}^2$ of sky coverage and the redshift bins and galaxy number densities in Tables 2.3 and 2.5 of \cite{Aghamousa:2016zmz}. All redshift bins have width $\Delta z=0.1$, and the bin centers are: $z_{\rm BGS}=0.05-0.45$, $z_{\rm ELG}=0.65-1.65$, $z_{\rm LRG}=0.65-1.15$, and $z_{\rm QSO}=0.65-1.85$. As in that work, we set the fiducial linear bias for each sample by imposing constant $b_1(z)D(z)$, where $D(z)$ is the linear growth factor that is normalized to one at $z=0$: $b_{\rm BGS}(z)D(z)=1.34$, $b_{\rm ELG}(z)D(z)=0.84$, $b_{\rm LRG}(z)D(z)=1.7$, and $b_{\rm QSO}(z)D(z)=1.2$. For Euclid, we assume the survey parameters for the H$\alpha$ emitting galaxies from Table 3 of \cite{Blanchard:2019oqi}, where there are four redshift bins centered around $z_{\rm H \alpha} = 1.00$, 1.20, 1.40, 1.65, and the bin widths are $\Delta z = 0.2$ for the first three bins and $\Delta z = 0.3$ for the highest redshift bin.

In each redshift bin, we forecast the constraints on $f(z)$ and $\sigma_8(z)$, marginalised over $b_1(z), b_2(z), \sigma_P(z)$, and $\sigma_B(z)$. We compute a theoretical covariance matrix for each redshift bin, and we assume that there is no cross-covariance between the power spectrum and LCF multipoles, since we found in Section \ref{sec:benchmark} that the cross-covariance made a difference of less than 3 per cent in the forecasted constraints, compared to the theoretical covariance with cross-covariance included.

\begin{figure}
    \centering
    \subfloat{\includegraphics[width=\textwidth]{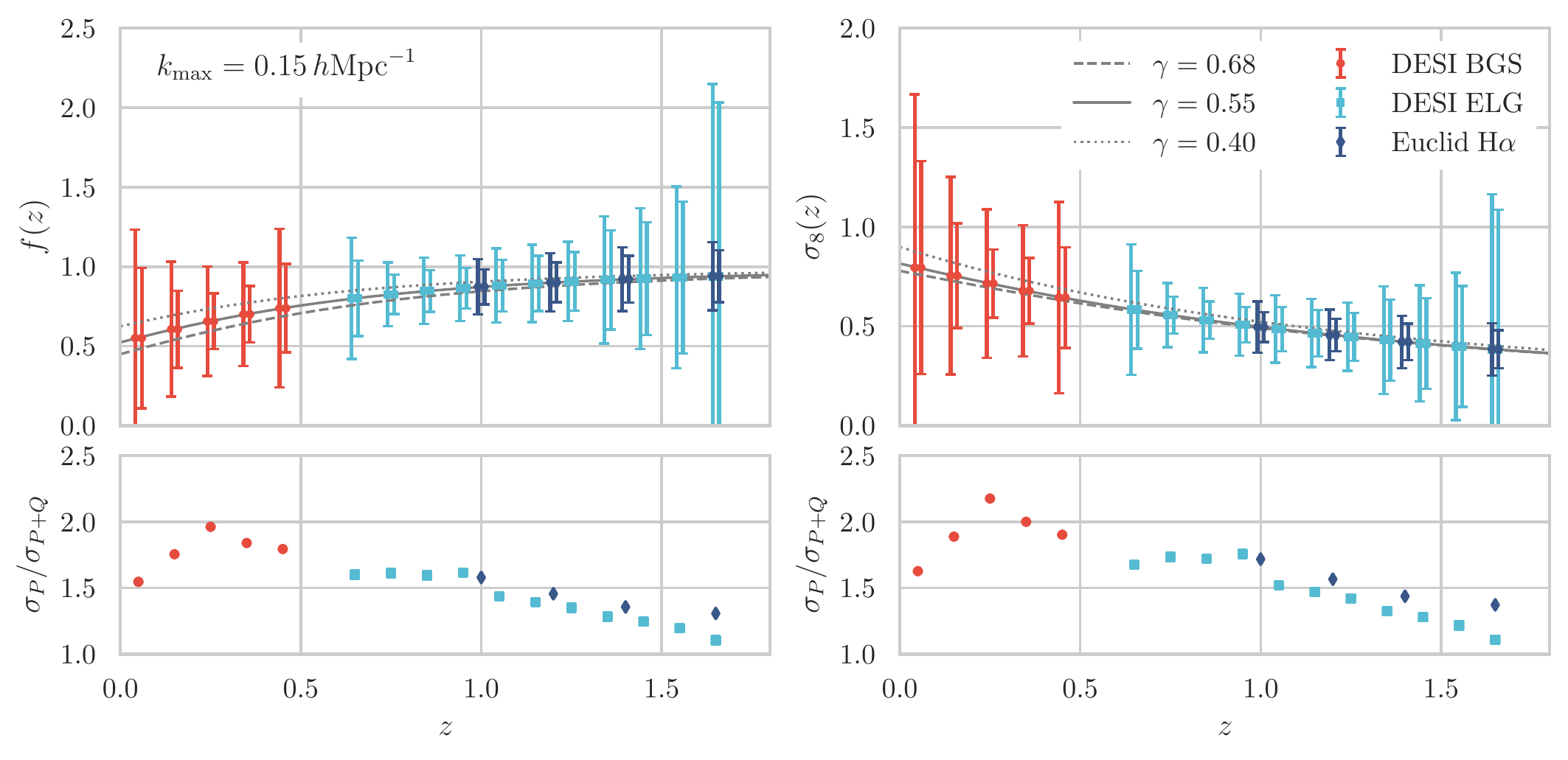}}
    \\
    \vspace{-1.5\baselineskip}
    \subfloat{\includegraphics[width=\textwidth]{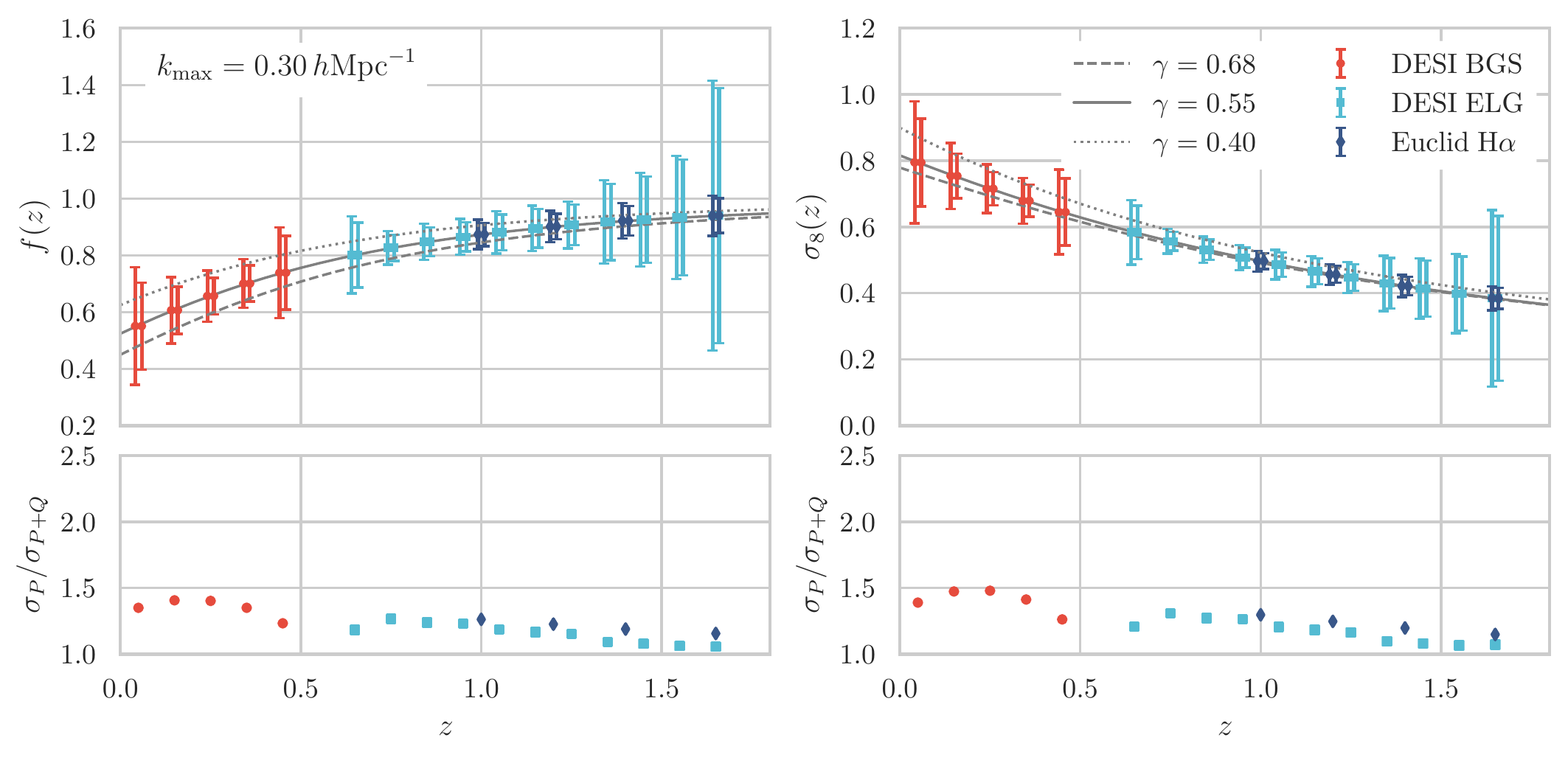}}
    \caption{Fisher forecasted constraints on $f(z)$ (left column) and $\sigma_8(z)$ (right column) from different galaxy samples for $k_{\rm max}=0.15 \,h\Mpc^{-1}$ (top row) and $k_{\rm max}=0.30 \,h\Mpc^{-1}$ (bottom row). Constraints for DESI BGS (red circles), DESI ELG (light blue squares), and Euclid H$\alpha$ galaxies (dark blue diamonds) are shown. Constraints in different redshift bins or different galaxy samples are not combined. Each bin has a pair of error bars, with the left bar indicating the constraint from the power spectrum only and the right bar indicating the joint power spectrum and LCF constraint. The panels showing $\sigma_P/\sigma_{P+Q}$ on the y-axis give the factor of improvement in the constraint from including the LCF multipoles. The gray lines show the theoretical predictions for $f(z)$ and $\sigma_8(z)$ for different values of the growth rate index $\gamma$: 0.40 (dotted), 0.55 (solid), and 0.68 (dashed).}
    \label{fig:surveys1zbinOnly}
\end{figure}

The constraints on $f(z)$ and $\sigma_8(z)$ are shown in Figure~\ref{fig:surveys1zbinOnly} for $k_{\rm max}=0.15 \,h\Mpc^{-1}$ (top row) and $k_{\rm max}=0.30 \,h\Mpc^{-1}$ (bottom row). For DESI, we choose to focus on the forecasts from the BGS and ELG samples, because these two samples together span a large range of redshifts, $z=0.05-1.65$, and where the ELG bins overlap with those of LRGs and QSOs, the ELG forecasts generally give stronger constraints on $f(z)$ and $\sigma_8(z)$ from the power spectrum only. The figure shows a pair of error bars for each redshift bin and galaxy sample, where within each pair, the error bar on the left is obtained from the power spectrum multipoles only and the smaller error bar on the right is from the joint power spectrum-LCF multipoles analysis. We have not combined the constraints from different redshift bins, because doing so would require assuming a model for how $f$ and $\sigma_8$ evolve with redshift. 

To put the size of the error bars into context, we have also plotted the predictions for $f(z)$ and $\sigma_8(z)$ for models with different values for the growth rate index $\gamma$, which is defined by 
\begin{equation}
	f(z) = \Omega_m(z)^\gamma.
\end{equation}
The growth rate index is often used as a simple parametrisation of the growth rate used in searches for modifications to gravity \citep{Linder:2005in,Linder:2007hg}. In general relativity, $\gamma \approx 0.55$, but it can take different values modified gravity model, and the predictions for $\gamma = 0.40$ and 0.68 shown in Figure~\ref{fig:surveys1zbinOnly} roughly approximate the range of values that are consistent with the data to within $\sim 2\sigma$ in recent analyses combining multiple low redshift probes and Planck CMB data \citep{Mueller:2016kpu,Sanchez:2016sas,Grieb:2016uuo,Wang:2017wia,Zhao:2018gvb}. 
The predictions of $\sigma_8(z)$ for different $\gamma$ are calculated as
\begin{align}
	\sigma_8(\gamma,z) = \sigma_8(z) 
	\frac{D_{\rm GR}(z_*)}{D_{\rm GR}(z)}
	\frac{D_{\gamma}(z)}{D_{\gamma}(z_*)}\,,
\end{align}
where 
\begin{equation}
	\frac{D_\gamma(a)}{D_\gamma(a_*)} = \exp \left[ \int_{\ln a_*}^{\ln a} \D{\ln{a'}} \, \Omega_m(a')^\gamma \right].
\end{equation}
We take $z_*=500$ to be a high redshift at which the linear growth factor was very close to that of general relativity.

The panels showing $\sigma_P/\sigma_{P+Q}$ in Figure~\ref{fig:surveys1zbinOnly} indicate the factor of improvement in the constraint from including the LCF measurements, which is equal to the ratio of the error bars. The improvement brought by including the LCF multipoles can be significant. For a fixed $k_{\rm max}$, the DESI BGS sample in the lower redshift bins (red circles) gives a larger improvement than in the higher redshift bins populated by the DESI ELG and Euclid H$\alpha$ galaxies. For a fixed galaxy sample and redshift bin, the relative improvement from the LCF is larger when $k_{\rm max}$ is lower, though the absolute size of the errors is larger. In particular, our forecast for the DESI BGS sample shows that the LCF multipoles can strengthen the constraints on $f(z)$ and $\sigma_8(z)$ by up to $\sim 220$ per cent for $k_{\rm max}=0.15 \,h\Mpc^{-1}$ or $\sim 50$ per cent for $k_{\rm max}=0.30 \,h\Mpc^{-1}$ at redshift $z=0.25$. For $k_{\rm max} = 0.15 \, h\mathrm{Mpc}^{-1}$, the per cent improvements in the constraints on $f$ and $\sigma_8$ averaged over redshift bins are $\sim 90$ per cent for the DESI BGS sample with mean redshift $\overline{z}=0.25$, $\sim 40$ per cent for the DESI ELG sample with $\overline{z}=1.25$, and $\sim 40$ per cent for the Euclid H$\alpha$ galaxies with $\overline{z}=1.3$. For $k_{\rm max} = 0.30 \, h\mathrm{Mpc}^{-1}$, the average improvements are $\sim 40$ per cent for the DESI BGS sample and $\sim 20$ per cent for both the DESI ELG and Euclid H$\alpha$ samples. These forecasts show that the LCF may be very useful for further improving the constraints on the growth rate $f$ with the upcoming generation of galaxy surveys. The fact that the LCF helps to break the degeneracy between $f$ and $\sigma_8$ within each individual redshift bin is highly relevant for performing model-independent analyses that do not need to assume any modelling for the evolution of $f(z)$ and $\sigma_8(z)$ with redshift.

We also consider combining the forecast with the Planck 2018 prior on $\sigma_8(z=0)$, as we did in Section~\ref{sec:benchmark}. Assuming that $\sigma_8(z)$ evolves with redshift according to our fiducial $\Lambda$CDM cosmology, we combine the prior with our forecast in each individual redshift bin. In this case, the improvement from measuring the LCF is minimal, and gives less than 12 per cent reduction in the errors on $f$ and $\sigma_8$ for all redshift bins, galaxy samples, and $k_{\rm max}$. This is because the CMB prior strongly breaks the degeneracy between $\sigma_8(z)$ and $f(z)$, such that there is little degeneracy left for the LCF to break further. However, the resulting $f(z)$ measured in this way is not model-independent and can therefore not be used to consistently test models beyond $\Lambda$CDM.


\section{Conclusions}
\label{sec:conclusions}

The LCF provides a way to harness the cosmological information in phase correlations and is distinct from the power spectrum and bispectrum (or 3-point correlation function) because the LCF in general depends on the full hierarchy of odd cumulants, starting with and going beyond the 3-point function \citep{Wolstenhulme:2014cla}. However, on large scales, $r \gtrsim 10 \, h^{-1}\Mpc$, where the LCF is well-approximated by the Edgeworth expansion, the cosmological information in the power spectrum and LCF is a subset of the information in the power spectrum and bispectrum \citep{Wolstenhulme:2014cla}. Still, the LCF is an attractive alternative to the bispectrum because it acts as a compression of the information in the bispectrum: rather than measuring the full 3-dimensional bispectrum on a large number of triangles in Fourier space, the LCF is a 1-dimensional function of $r$. An additional advantage of the LCF is that, unlike the bispectrum, the LCF has a straightforward geometric interpretation: it measures the prominence of cosmic filaments on different scales \citep{Obreschkow:2012yb}. In this sense, the LCF can be seen as a natural pairing with the 2-point correlation function or power spectrum: the power spectrum is a 1-dimensional function of density \textit{amplitude} correlations that is sensitive to \textit{spherical} clustering, while the LCF is a 1-dimensional function of density \textit{phase} correlations that is sensitive to linear \textit{filamentarity}.

In this work, we have shown that the correlations between phases measured by the LCF provide a powerful way to test general relativity, by improving the constraint on the growth rate of structure, $f$, with the coming generation of galaxy large-scale structure surveys like DESI and Euclid. We have focused on a specific estimator of phase correlations, the line correlation function (LCF), and studied how the multipoles of the LCF can be used in combination with the multipoles of the power spectrum to improve the measurement of $f$. The key property of the LCF which makes it complementary to the power spectrum is the fact that it contains different combinations of $f$ and $\sigma_8$, allowing it to break the degeneracy between these parameters that are present in a power spectrum-only analysis. We have argued that this method has the advantage of not relying on an assumed cosmological model for how the growth of structure evolves with redshift, which is not the case for joint clustering-CMB analyses or joint clustering-weak lensing analyses that require a specific cosmological model to break the degeneracy between $f$ and $\sigma_8$.

We have constructed a model of the LCF multipoles and of their covariance, which goes beyond linear perturbation theory and is valid in the non-linear regime. We have tested this model using a large suite of \textsc{l-picola} realizations, and found that it agrees well with the simulations down to separations of $20 \,h^{-1}\Mpc$. Using Fisher matrices we have forecasted the constraints expected on $f$ and $\sigma_8$ for surveys like DESI and Euclid and found that adding the LCF leads to an improvement of up to 220 per cent for $k_{\rm max}=0.15 \, h\Mpc^{-1}$ and up to 50 per cent for $k_{\rm max}=0.30 \, h\Mpc^{-1}$, depending on the redshift bin and galaxy sample. Averaged over redshift bins, the constraints on $f$ and $\sigma_8$ for $k_{\rm max} = 0.15 \, h\mathrm{Mpc}^{-1}$ are improved by $\sim 90$ per cent for the DESI BGS sample at lower redshifts and $\sim 40$ for both the DESI ELG and Euclid H$\alpha$ galaxies at higher redshifts. For $k_{\rm max} = 0.30 \, h\mathrm{Mpc}^{-1}$, the average improvements are $\sim 40$ per cent for the DESI BGS sample and $\sim 20$ per cent for both the DESI ELG and Euclid H$\alpha$ samples.

Our work has explored the utility of the power spectrum and LCF to access information about the growth rate and amplitude of scalar perturbations. However, as we have noted above, on large scales the power spectrum and LCF together contain a subset of the information in the power spectrum and bispectrum. This naturally raises the question of how our forecasted constraints would compare to those from the power spectrum and bispectrum. Previous works have used the redshift-space bispectrum to break the degeneracy between $f$ and $\sigma_8$ \citep{Gil-Marin:2016wya,Gagrani:2016rfy,Gualdi:2017iey,Gualdi:2020ymf}, but the differences in the details of those works and the present one (such as the data and modelling used, the minimum scale given by $k_{\rm max}$, the specific parameters that were constrained, etc.) prevent us from making quantitative comparisons. A controlled comparison of constraints from the LCF and bispectrum in redshift-space would be an interesting goal for future work.

Further work will also be necessary before the LCF multipole estimator can be applied to more realistic data. For example, there are several observational effects which our modelling and forecast did not include, such as the effect of complex survey window functions and going beyond the plane-parallel approximation. Existing methods for including these effects in the 3-point correlation function may be applicable to the LCF, and indeed the LCF may also benefit from some of the advantages of measuring the 3-point correlation function over the bispectrum (for example, as discussed in \cite{Slepian:2015hca}). Furthermore, the numerical calculation of theoretical predictions for the LCF must be improved before it would be fast enough to be part of a standard MCMC likelihood analysis. In parallel, it would be interesting to study other configurations, beyond correlations restricted to a line, to see if the impact of redshift-space distortions may be enhanced in specific triangular configurations.


\section*{Acknowledgements}

We thank William Wright for useful discussions and Davide Gualdi for comments and suggestions that helped to improve the presentation of our results. JB acknowledges support from the SNSF Sinergia grant No. 173716. FOF and CB acknowledge support from the SNSF. CH was supported by  the  Australian  Government  through  the  Australian Research  Council’s  Laureate  Fellowship  funding  scheme (project FL180100168). DO is a recipient of an Australian Research Council Future Fellowship (FT190100083) funded by the Australian Government. Computations were performed at the University of Geneva on the Baobab computing cluster and on the OzSTAR national facility at the Swinburne University of Technology. OzSTAR is funded by Swinburne University of Technology and the National Collaborative Research Infrastructure Strategy (NCRIS).

We acknowledge the use of the \texttt{emcee} \citep{ForemanMackey:2012ig}, \texttt{GetDist} \citep{Lewis:2019xzd}, \texttt{COLOSSUS} \citep{Diemer:2017bwl}, \texttt{EuclidEmulator} \citep{Knabenhans:2018cng}, and \texttt{nbodykit} \citep{Hand:2017pqn} Python packages.

\section*{Data availability}
The data underlying this article will be shared on reasonable request to the corresponding author.




\bibliographystyle{mnras}
\bibliography{references} 



\appendix

\section{Covariance of the LCF multipoles}
\label{app:Qn_cov}

Given the form of the estimator for the LCF multipoles in eq.~\eqref{eq:Qn_estimator}, we calculate the covariance of the $Q_n$ at lowest order (i.e.\ at Gaussian order) in the Edgeworth expansion. We have
\begin{equation}
	\text{cov}\left[Q_{n_{1}}(r_{i}),Q_{n_{2}}(r_{j})\right]=\Braket{\hat{Q}_{n_{1}}(r_{i})\hat{Q}_{n_{2}}(r_{j})},
	\label{eq:covmult}
\end{equation}
since $\Braket{\hat{Q}_{n}(r)}=0$ at Gaussian order. Eq.~\eqref{eq:covmult} contains the six-point phase correlation
\begin{equation}
	\mathcal{E}_{\ell\ell}=\Braket{\epsilon_{-\mathbf{k}_{1}-\mathbf{k}_{2}}\epsilon_{\mathbf{k}_{1}}\epsilon_{\mathbf{k}_{2}}\epsilon_{-\mathbf{k}_{3}-\mathbf{k}_{4}}\epsilon_{\mathbf{k}_{3}}\epsilon_{\mathbf{k}_{4}}}_{G},
\end{equation}
which can be split into a sum of products of $\Braket{\epsilon_{\mathbf{k}}\epsilon_{\mathbf{q}}}_{G}$ through Wick's theorem. Statistical homogeneity implies that the two-point phase correlation is given by $\Braket{\epsilon_{\mathbf{k}}\epsilon_{\mathbf{q}}}_{G}=\delta_{\mathbf{k}+\mathbf{q}}^{K}$ \citep{Wolstenhulme:2014cla,Eggemeier:2016asq}. Then, neglecting all terms that give rise to background modes ($\bk=0$), $\mathcal{E}_{\ell\ell}$ has six terms
\begin{equation}
	\mathcal{E}_{\ell\ell} = \delta_{\mathbf{k}_{1}+\mathbf{k}_{2}+\mathbf{k}_{3}+\mathbf{k}_{4}}^{K}\delta_{\mathbf{k}_{1}+\mathbf{k}_{3}}^{K}\delta_{\mathbf{k}_{2}+\mathbf{k}_{4}}^{K}+\delta_{-\mathbf{k}_{1}-\mathbf{k}_{2}+\mathbf{k}_{3}}^{K}\delta_{\mathbf{k}_{1}-\mathbf{k}_{3}-\mathbf{k}_{4}}^{K}\delta_{\mathbf{k}_{2}+\mathbf{k}_{4}}^{K}+
	\delta_{-\mathbf{k}_{1}-\mathbf{k}_{2}+\mathbf{k}_{4}}^{K}\delta_{\mathbf{k}_{1}-\mathbf{k}_{3}-\mathbf{k}_{4}}^{K}\delta_{\mathbf{k}_{2}+\mathbf{k}_{3}}^{K}
	+\left(\mathbf{k}_{1}\leftrightarrow\mathbf{k}_{2}\right). 
	\label{eq:kron}
\end{equation}
Inserting eq.~\eqref{eq:kron} into eq.~\eqref{eq:covmult} yields
\begin{align}
	\text{cov}\left[Q_{n_{1}}(r_{i}),Q_{n_{2}}(r_{j})\right] & = \frac{\left(2n_{1}+1\right)\left(2n_{2}+1\right)i^{n_{1}+n_{2}}\left(r_{i}r_{j}\right)^{9/2}}{32\pi^{6}V}\underset{k_{1},k_{2},\left|\mathbf{k}_{1}+\mathbf{k}_{2}\right|\leq\frac{2\pi}{\max(r_{i},r_{j})}}{\iint}\!\!\!\!\D{^3 k_1}\D{^3 k_2}\,
	j_{n_{1}}\left(\kappa_{1}r_{i}\right)L_{n_{1}}\left(\hat{\boldsymbol{\kappa}}_{1}\cdot\hat{\mathbf{n}}\right)\nonumber \\
 	&  \times\left[j_{n_{2}}\left(\kappa_{1}r_{j}\right)L_{n_{2}}\left(\hat{\boldsymbol{\kappa}}_{1}\cdot\hat{\mathbf{n}}\right)+j_{n_{2}}\left(\kappa_{2}r_{j}\right)L_{n_{2}}\left(\hat{\boldsymbol{\kappa}}_{2}\cdot\hat{\mathbf{n}}\right)+j_{n_{2}}\left(\kappa_{3}r_{j}\right)L_{n_2}\left(\hat{\boldsymbol{\kappa}}_{3}\cdot\hat{\mathbf{n}}\right)\right],
 	\label{eq:covQ}
\end{align}
with $\boldsymbol{\kappa}_{1}\equiv \mathbf{k}_{1}-\mathbf{k}_{2}$, $\boldsymbol{\kappa}_{2}\equiv \mathbf{k}_{1}+2\mathbf{k}_{2}$ and $\boldsymbol{\kappa}_{3}\equiv-2\mathbf{k}_{1}-\mathbf{k}_{2}$. Moreover, since $n_{1}$ and $n_{2}$ are even, the third term is equal to the second one through the transformation $\mathbf{k}_{1}\leftrightarrow \mathbf{k}_{2}$. 

This expression contains a six-dimensional integral, over the modulus of $\bk_1$ and $\bk_2$ and over their directions, which we denote respectively by $(\theta_1, \varphi_1)$ and $(\theta_2, \varphi_2)$. Three of these integrals can be done analytically. To do this, we first note that since the multipoles and their covariance do not depend on the line-of-sight direction $\hbn$, we can integrate eq.~\eqref{eq:covQ} over $\hbn$ and divide by $4\pi$. Since the Legendre polynomials are the only contributions that contain $\hbn$, the integral over $\hbn$ reduces to
\begin{align}
	\frac{1}{4\pi}\int d\Omega_{\hbn} L_{n_{1}}\left(\hat{\boldsymbol{\kappa}}_{1}\cdot\hat{\mathbf{n}}\right)L_{n_{2}}\left(\hat{\boldsymbol{\kappa}}_{i}\cdot\hat{\mathbf{n}}\right)&=\frac{4\pi}{(2n_1+1)(2n_2+1)}\sum_{m_1=-n_1}^{n_1}\sum_{m_2=-n_2}^{n_2}Y_{n_1m_1}(\hat{\boldsymbol{\kappa}}_{1})Y^*_{n_2m_2}(\hat{\boldsymbol{\kappa}}_{i})\int d\Omega_{\hbn} Y^*_{n_1m_1}(\hbn)Y_{n_2m_2}(\hbn)\nonumber\\
	&=\frac{\delta_{n_1n_2}}{2n_1+1}L_{n_1}(\hat{\boldsymbol{\kappa}}_{1}\cdot\hat{\boldsymbol{\kappa}}_i) \quad\mbox{for}\quad i=1,2.
\end{align}
We insert this into eq.~\eqref{eq:covQ} and do the following coordinate transformation: $\left\{ \theta_{1},\phi_{1},\theta_{2},\phi_{2}\right\} \rightarrow\left\{ \gamma,\phi,\theta_{2},\phi_{2}\right\} $, where $\cos\gamma\equiv\hat{\mathbf{k}}_{1}\cdot\hat{\mathbf{k}}_{2}$ and $\phi$ is the azimuthal angle of $\mathbf{k}_{1}$ around $\mathbf{k}_{2}$. The Jacobian of this transformation is 1, since it is a rotation. In this coordinate system, the product $\hat{\boldsymbol{\kappa}}_1\cdot \hat{\boldsymbol{\kappa}}_i$ depends only on $k_1, k_2$ and $\gamma$. Therefore the integral over $\phi, \theta_2$ and $\phi_2$ can be performed and gives rise to a factor $8\pi^2$. We obtain
\begin{align}
	\text{cov}\left[Q_{n_{1}}\left(r_{i}\right),Q_{n_{2}}\left(r_{j}\right)\right]  =& \frac{\left(2n_{1}+1\right)(-1)^{n_{1}}\left(r_{i}r_{j}\right)^{9/2}}{4\pi^{4}V}\int_{0}^{2\pi/R}dk_{1}k_{1}^{2}\int_{0}^{2\pi/R}dk_{2}k_{2}^{2}\int_{-1}^{\alpha_{\text{cut}}}d\alpha j_{n_{1}}\left(\kappa_{1}r_{i}\right)\nonumber \\
 	& \times\left[j_{n_{1}}\left(\kappa_{1}r_{j}\right)+2j_{n_{1}}\left(\kappa_{2}r_{j}\right)L_{n_{1}}\left(\hat{\boldsymbol{\kappa}}_{1}\cdot\hat{\boldsymbol{\kappa}}_{2}\right)\right]\cdot\delta^K_{n_1n_2}\,,\label{eq:covfinal}
	\end{align}
where $R\equiv\max\left(r_{i},r_{j}\right)$. $\alpha$ is the cosine of the angle between $\hat{\bk}_1$ and $\hat{\bk}_2$, and $\alpha_{\text{cut}}\equiv\min\{ 1,\max\{ -1,[\left(2\pi/R\right)^{2}-k_{1}^{2}-k_{2}^{2}]/[2k_{1}k_{2}]\} \}$ is imposed to keep $\left|\mathbf{k}_{1}+\mathbf{k}_{2}\right|\leq2\pi/R$ satisfied.
The argument of the Legendre polynomial is given by
\begin{equation}
	\hat{\boldsymbol{\kappa}}_{1}\cdot\hat{\boldsymbol{\kappa}}_{2}=\frac{k_1^2+k_1k_2\alpha-2k_2^2}{\sqrt{\big(k_1^2-2k_1k_2\alpha+k_2^2\big)\big(k_1^2+4k_1k_2\alpha+4k_2^2\big)}}\, .
\end{equation}
From eq.~\eqref{eq:covfinal}, we see that the different multipoles are not correlated.


\section{Cross-covariance between the power spectrum and LCF multipoles}
\label{app:PlQn_crosscov}

We now calculate the cross-covariance between the power spectrum multipoles and the LCF multipoles
\begin{equation}
	\text{cov}\left[P_{n_{1}}\left(k_{i}\right),Q_{n_{2}}\left(r_{j}\right)\right]=\Braket{\hat{P}_{n_{1}}\left(k_{i}\right)\hat{Q}_{n_{2}}\left(r_{j}\right)}-\Braket{\hat{P}_{n_{1}}\left(k_{i}\right)}\Braket{\hat{Q}_{n_{2}}\left(r_{j}\right)}.
	\label{eq:covcross}
\end{equation}
The Gaussian contribution to eq.~\eqref{eq:covcross} exactly vanishes since the first term is a five-point correlation, which is zero for a Gaussian field, and the second term contains a three-point correlation which is also zero for a Gaussian field. Therefore, to account for any non-zero correlation between the LCF and the power spectrum, we need to compute the non-Gaussian contribution to eq.~\eqref{eq:covcross}. 

Eq.~\eqref{eq:covcross} contains the mixed five-point correlator of phases and amplitudes,
\begin{equation}
	\mathcal{E}_{P\ell}=
	\Braket{\Delta(\mathbf{q})\Delta(-\mathbf{q})\epsilon(-\mathbf{k}_{1}-\mathbf{k}_{2})\epsilon(\mathbf{k}_{1})\epsilon(\mathbf{k}_{2})}
	-\Braket{\Delta(\mathbf{q})\Delta(-\mathbf{q})}\Braket{\epsilon(-\mathbf{k}_{1}-\mathbf{k}_{2})\epsilon(\mathbf{k}_{1})\epsilon(\mathbf{k}_{2})}.
\end{equation}
Neglecting the background modes as in Appendix~\ref{app:Qn_cov}, we split this expression into its connected correlators using Wick's theorem. We find two kinds of contributions
\begin{align}
	\mathcal{E}_{PB} & = \Braket{\Delta(-\mathbf{q})\epsilon(\mathbf{k}_{1})}_{c}\Braket{\Delta(\mathbf{q})\epsilon(\mathbf{k}_{2})\epsilon(-\mathbf{k}_{1}-\mathbf{k}_{2})}_{c}+\Braket{\Delta(-\mathbf{q})\epsilon(\mathbf{k}_{2})}_{c}\Braket{\Delta(\mathbf{q})\epsilon(\mathbf{k}_{1})\epsilon(-\mathbf{k}_{1}-\mathbf{k}_{2})}_{c}+\Braket{\Delta(-\mathbf{q})\epsilon(-\mathbf{k}_{1}-\bk_2)}_{c}\Braket{\Delta(\mathbf{q})\epsilon(\mathbf{k}_{1})\epsilon(\bk_2)}_{c}\nonumber\\
	&+\left(\mathbf{q}\leftrightarrow-\mathbf{q}\right),
	\label{eq:EPB}\\
	\mathcal{E}_{P_{5}} & = \Braket{\epsilon(-\mathbf{k}_{1}-\mathbf{k}_{2})\epsilon(\mathbf{k}_{1})\epsilon(\mathbf{k}_{2})\Delta(\mathbf{q})\Delta(-\mathbf{q})}_{c}.
	\label{eq:EP5}
\end{align}
These connected mixed-correlators can be evaluated using the joint PDF of Fourier modes, and at lowest order they are given by \citep{Eggemeier:2016asq}
\begin{align}
	\Braket{\Delta(\mathbf{q})\epsilon(\mathbf{k})}_{c} & = \frac{\left(2\pi\right)^{3}}{V}\frac{\sqrt{\pi}}{2}\sqrt{VP\left(\mathbf{q}\right)}\delta_{D}\left(\mathbf{k}+\mathbf{q}\right),
	\label{eq:del ep}\\
	\Braket{\Delta(\mathbf{q})\epsilon(\mathbf{k}_{1})\epsilon(\mathbf{k}_{2})}_{c} & = \frac{\left(2\pi\right)^{3}}{V}\left(\frac{\sqrt{\pi}}{2}\right)^{2}\sqrt{VP\left(\mathbf{q}\right)}p^{\left(3\right)}\left(\mathbf{q},\mathbf{k}_{1},\mathbf{k}_{2}\right)\delta_{D}\left(\mathbf{k}_{1}+\mathbf{k}_{2}+\mathbf{q}\right),
	\label{eq:del ep ep}\\
	\Braket{\Delta(\mathbf{q}_{1})\Delta(\mathbf{q}_{2})\epsilon(\mathbf{k}_{1})\epsilon(\mathbf{k}_{2})\epsilon(\mathbf{k}_{3})}_{c} & = \left(2\pi\right)^{3}\left(\frac{\sqrt{\pi}}{2}\right)^{3}\sqrt{P\left(\mathbf{q}_{1}\right)P\left(\mathbf{q}_{2}\right)}p^{\left(5\right)}\left(\mathbf{q}_{1},\mathbf{q}_{2},\mathbf{k}_{1},\mathbf{k}_{2},\mathbf{k}_{3}\right)\delta_{D}\left(\mathbf{k}_{1}+\mathbf{k}_{2}+\mathbf{k}_{3}+\mathbf{q}_{1}+\mathbf{q}_{2}\right),
	\label{eq:del del ep ep ep}
\end{align}
where the $N$th order cumulants $p^{\left(N\right)}$ are related to the ordinary $N$th order spectra $P^{\left(N\right)}$ by
\begin{equation}
	p^{\left(N\right)}\left(\mathbf{k}_{1},\ldots,\mathbf{k}_{N}\right)\equiv V^{1-\frac{N}{2}}\frac{P^{\left(N\right)}\left(\mathbf{k}_{1},\ldots,\mathbf{k}_{N}\right)}{\sqrt{P\left(\mathbf{k}_{1}\right)\ldots P\left(\mathbf{k}_{N}\right)}}\,.
	\label{eq:pN cumulants}
\end{equation}
With this the covariance becomes 
\begin{equation}
	\text{cov}\left[P_{n_{1}}\left(k_{i}\right),Q_{n_{2}}\left(r_{j}\right)\right]=\frac{V}{\left(2\pi\right)^{3}}
	\left(\frac{\sqrt{\pi}}{2}\right)^{3}\left(\frac{r_{j}^{3}}{V}\right)^{3/2}\left[\mathcal{C}^{\left(n_{1},n_{2}\right)}_{PB}+\mathcal{C}^{\left(n_{1},n_{2}\right)}_{P_{5}}\right],
\end{equation}
where the contribution from the connected 5-point correlation function is 
\begin{align}
	\mathcal{C}^{\left(n_{1},n_{2}\right)}_{P^{5}} & = \left(2n_{1}+1\right)\left(2n_{2}+1\right)i^{n_2}\int_{k_{i}}\frac{\D{^3 q}}{V_{P}\left(k_{i}\right)}L_{n_{1}}\left(\hat{\mathbf{q}}\cdot\hat{\bn}\right)P\left(\mathbf{q}\right)\nonumber \\
	 & \times\underset{k_{1},k_{2},\left|\mathbf{k}_{1}+\mathbf{k}_{2}\right|\leq\frac{2\pi}{r_{j}}}{\iint}\D{^3 k_1}\D{^3 k_2}\,j_{n_{2}}\left(\kappa_{1}r_{j}\right)L_{n_{2}}\left(\hat{\boldsymbol{\kappa}}_{1}\cdot\hat{\bn}\right)p^{\left(5\right)}\left(\mathbf{q},-\mathbf{q},-\mathbf{k}_{1}-\mathbf{k}_{2},\mathbf{k}_{1},\mathbf{k}_{2}\right),
	 \label{eq:CP5}
\end{align}
and the $\mathcal{C}^{\left(n_{1},n_{2}\right)}_{PB}$ contribution is
\begin{equation}
	\mathcal{C}^{\left(n_{1},n_{2}\right)}_{PB} = \frac{\left(2n_{1}+1\right)\left(2n_{2}+1\right)i^{n_2}}{\pi^{9/2}}\int_{k_{i}}\frac{\D{^3 q}}{V_{P}\left(k_{i}\right)}L_{n_{1}}\left(\hat{\mathbf{q}}\cdot\hat{\bn}\right)\underset{k_{1},k_{2},\left|\mathbf{k}_{1}+\mathbf{k}_{2}\right|\leq\frac{2\pi}{r_{j}}}{\iint}\D{^3 k_1}\D{^3 k_2}\,j_{n_{2}}\left(\kappa_{1}r_{j}\right)L_{n_{2}}\left(\hat{\boldsymbol{\kappa}}_{1}\cdot\hat{\bn}\right)\mathcal{E}_{PB}.
	\label{eq:CPB}
\end{equation}

Since $n_1$ is even, $L_{n_1}(\hat{\mathbf{q}}\cdot\hat{\bn})=L_{n_1}(-\hat{\mathbf{q}}\cdot\hat{\bn})$, and the last three terms in eq.~\eqref{eq:EPB} have the same contribution as the first three terms. Moreover, one can rewrite the first term by relabelling the dummy wavenumbers as
\begin{equation}
	\begin{cases}
	\mathbf{k}_{1} & \rightarrow\mathbf{k}_{2}\\
	\mathbf{k}_{2} & \rightarrow-\mathbf{k}_{1}-\mathbf{k}_{2}\\
	\boldsymbol{\kappa}_{1} & \rightarrow\boldsymbol{\kappa}_{2},
	\end{cases}
\end{equation}
while doing a different relabelling for the third term,
\begin{equation}
	\begin{cases}
	\mathbf{k}_{1} & \rightarrow-\mathbf{k}_{1}-\mathbf{k}_{2}\\
	\mathbf{k}_{2} & \rightarrow\mathbf{k}_{1}\\
	\boldsymbol{\kappa}_{1} & \rightarrow\boldsymbol{\kappa}_{3}.
	\end{cases}
\end{equation}
Then eq.~\eqref{eq:CPB} takes the form 
\begin{align}
	\mathcal{C}^{\left(n_{1},n_{2}\right)}_{PB} & = \frac{2\left(2n_{1}+1\right)\left(2n_{2}+1\right)i^{n_2}}{\pi^{9/2}}\int_{k_{i}}\frac{\D{^3 q}}{V_{P}\left(k_{i}\right)}L_{n_{1}}\left(\hat{\mathbf{q}}\cdot\hat{\bn}\right)\underset{k_{1},k_{2},\left|\mathbf{k}_{1}+\mathbf{k}_{2}\right|\leq\frac{2\pi}{r_{j}}}{\iint}\!\!\!\D{^3 k_1}\D{^3 k_2}\Braket{\Delta(-\mathbf{q})\epsilon(\mathbf{k}_{2})}_{c}\Braket{\Delta(\mathbf{q})\epsilon(-\mathbf{k}_{1}-\mathbf{k}_{2})\epsilon(\mathbf{k}_{1})}_{c}\nonumber\\
	& \times\left[j_{n_{2}}\left(\kappa_{1}r_{j}\right)L_{n_{2}}\left(\hat{\boldsymbol{\kappa}}_{1}\cdot\hat{\bn}\right)+j_{n_{2}}\left(\kappa_{2}r_{j}\right)L_{n_{2}}\left(\hat{\boldsymbol{\kappa}}_{2}\cdot\hat{\bn}\right)+j_{n_{2}}\left(\kappa_{3}r_{j}\right)L_{n_{2}}\left(\hat{\boldsymbol{\kappa}}_{3}\cdot\hat{\bn}\right)\right].
\end{align}
Using eqs.~\eqref{eq:del ep} and \eqref{eq:del ep ep}, one of the Dirac delta functions allows us to integrate over $\D{^3 k_2}$, generating a Theta function $\Theta\left(1-k_{2}r_{j}/2\pi\right)$. The other Dirac delta function is redundant, so it contributes a factor of $\delta_{D}\left(\mathbf{0}\right)=V/\left(2\pi\right)^{3}$. Then relabelling $\mathbf{q}$ as $\mathbf{k}_{2}$, we obtain 
\begin{align}
	\mathcal{C}^{\left(n_{1},n_{2}\right)}_{PB} & = 2\left(2n_{1}+1\right)\left(2n_{2}+1\right)i^{n_2}\int_{k_{i}}\frac{\D{^3 k_2}}{V_{P}\left(k_{i}\right)}P\left(\mathbf{k}_{2}\right)L_{n_{1}}\left(\hat{\mathbf{k}}_{2}\cdot\hat{\bn}\right)\Theta\left(1-\frac{k_{2}r_{j}}{2\pi}\right)\underset{k_{1},\left|\mathbf{k}_{1}+\mathbf{k}_{2}\right|\leq
	\frac{2\pi}{r_{j}}
	}{\int}\D{^3 k_1}p^{\left(3\right)}\left(\mathbf{k}_{2},-\mathbf{k}_{1}-\mathbf{k}_{2},\mathbf{k}_{1}\right)\nonumber\\
	& \times\left[j_{n_{2}}\left(\kappa_{1}r_{j}\right)L_{n_{2}}\left(\hat{\boldsymbol{\kappa}}_{1}\cdot\hat{\bn}\right)+j_{n_{2}}\left(\kappa_{2}r_{j}\right)L_{n_{2}}\left(\hat{\boldsymbol{\kappa}}_{2}\cdot\hat{\bn}\right)+j_{n_{2}}\left(\kappa_{3}r_{j}\right)L_{n_{2}}\left(\hat{\boldsymbol{\kappa}}_{3}\cdot\hat{\bn}\right)\right].
\end{align}

Finally, we rewrite the expression in terms of the power spectrum and 3-point phase correlations. Since the $P^{(3)}$ in $p^{(3)}$ is simply the bispectrum, we use eq.~\eqref{eq:3eps} to write
\begin{align}
	\mathcal{C}^{\left(n_{1},n_{2}\right)}_{PB} & = 2\left(2n_{1}+1\right)\left(2n_{2}+1\right)i^{n_2}
	\left(\frac{2}{\sqrt{\pi}}\right)^3
	\int_{k_{i}}\frac{\D{^3 k_2}}{V_{P}\left(k_{i}\right)}P\left(\mathbf{k}_{2}\right)L_{n_{1}}\left(\hat{\mathbf{k}}_{2}\cdot\hat{\bn}\right)\Theta\left(1-\frac{k_{2}r_{j}}{2\pi}\right) \underset{k_{1},\left|\mathbf{k}_{1}+\mathbf{k}_{2}\right|\leq
	\frac{2\pi}{r_{j}}}{\int}\D{^3 k_1}
	\Braket{\epsilon\left(\mathbf{k}_{2}\right)\,\epsilon\left(-\mathbf{k}_{1}-\mathbf{k}_{2}\right)\,\epsilon\left(\mathbf{k}_{1}\right)}\nonumber\\
	&\times
	\left[j_{n_{2}}\left(\kappa_{1}r_{j}\right)L_{n_{2}}\left(\hat{\boldsymbol{\kappa}}_{1}\cdot\hat{\bn}\right)+j_{n_{2}}\left(\kappa_{2}r_{j}\right)L_{n_{2}}\left(\hat{\boldsymbol{\kappa}}_{2}\cdot\hat{\bn}\right)+j_{n_{2}}\left(\kappa_{3}r_{j}\right)L_{n_{2}}\left(\hat{\boldsymbol{\kappa}}_{3}\cdot\hat{\bn}\right)\right].
\end{align}



\bsp	
\label{lastpage}
\end{document}